\documentclass[final,5p,times]{elsarticle}
\usepackage[dvipsnames]{xcolor}
\usepackage{lineno, amsmath}
\usepackage{graphicx}
\usepackage{algorithm2e}

\expandafter\let\csname ver@natbib.sty\endcsname\relax
\makeatletter
\let\c@author\relax
\makeatother
\usepackage[backend=biber,hyperref=true,doi=false,url=false,isbn=false,uniquename=false,uniquelist=false]{biblatex}
\usepackage{filecontents}
\usepackage[colorlinks=true,linkcolor=blue]{hyperref}
\DeclareMathAlphabet\mathbfcal{OMS}{cmsy}{b}{n}

\journal{Astronomy and Computing}

\begin{document}

\begin{frontmatter}

\title{MAPPRAISER: A massively parallel map-making framework for multi-kilo pixel CMB experiments}

% Group authors per affiliation:
\author[add1,add2,add3]{Hamza El Bouhargani}
\ead{elbouha@lbl.gov}
\author[add1]{Aygul Jamal}
\ead{ajamal@apc.in2p3.fr}
\author[add4,add5]{Dominic Beck}
\ead{dobeck@stanford.edu}
\author[add1]{Josquin Errard}
\ead{josquin@apc.in2p3.fr}
\author[add6]{Laura Grigori}
\ead{Laura.Grigori@inria.fr}
\author[add1,add2]{Radek Stompor}
\ead{radek.stompor@in2p3.fr}

\address[add1]{Université Paris Cité, CNRS, Astroparticule et Cosmologie, F-75013 Paris, France}
\address[add2]{CNRS-UCB International Research Laboratory, Centre Pierre Bin{\'e}truy, IRL2007, CPB-IN2P3, Berkeley, CA 94720, USA}
\address[add3]{Lawrence Berkeley National Laboratory, Berkeley, CA 94720, USA}
\address[add4]{Kavli Institute for Particle Astrophysics and Cosmology, SLAC National Accelerator Laboratory, 2575 Sand Hill Rd, Menlo Park, California 94025, USA}
\address[add5]{Department of Physics, Stanford University, Stanford, California 94305, USA}
\address[add6]{INRIA Paris, Sorbonne Université, Université Paris-Diderot SPC, CNRS, Laboratoire Jacques-Louis Lions, ALPINES Team, France}

\begin{abstract}
Forthcoming cosmic microwave background (CMB) polarized anisotropy experiments have the potential to revolutionize our understanding of the Universe and fundamental physics. The sought-after, tale-telling  signatures will be however distributed over voluminous data sets which these experiments will collect. These data sets will need to be efficiently processed and unwanted contributions due to astrophysical, environmental, and instrumental effects characterized and efficiently mitigated in order to uncover the signatures. This poses a significant challenge to data analysis methods, techniques, and software tools which will not only have to be able to cope with huge volumes of data but to do so with unprecedented precision driven by the demanding science goals posed for the new experiments. 

A keystone of efficient CMB data analysis are solvers of very large linear systems of equations. Such systems appear in very diverse contexts throughout CMB data analysis pipelines, however they typically display similar algebraic structures and can therefore be solved using similar numerical techniques. Linear systems arising in the so-called map-making problem are one of the most prominent and common ones.

In this work we present a massively parallel, flexible and extensible framework, comprised of a numerical library, MIDAPACK, and a high level code, MAPPRAISER, which provide tools for solving efficiently such systems. The framework implements iterative solvers based on conjugate gradient techniques: enlarged and preconditioned using different preconditioners. We demonstrate the framework on simulated examples reflecting basic characteristics of the forthcoming data sets issued by ground-based and satellite-borne instruments, executing it on as many as 16,384 compute cores. The software is developed as an open source project freely available to the community at: \url{https://github.com/B3Dcmb/midapack}. 
\end{abstract}


\end{frontmatter}

%\linenumbers

\section{Introduction}

Observations of cosmic microwave background (CMB) ani\-sotro\-pies have played a major role in shaping up our current view of the Universe, its composition and evolution. This has been epitomized by the results obtained by the Planck satellite~\cite{Planck2018_legacy, Planck_2018_cosmo}, which, with complementary information provided by other probes, have  established the current standard cosmological model determining its parameters with a few percent precision. The scientific potential of the CMB has not been exhausted yet and only experiments forthcoming on the timescale of this decade are expected to exploit the full scientific information contained in the polarization properties of the CMB anisotropies. This is expected to open a new window on the evolution and physics of the very early Universe providing unique insights about physical laws at extremely high energies exceeding by many orders of magnitude energies accessible by any current and projected human-made experiments. The potential impact of these new experiments is therefore difficult to overestimate. 

Given the minute amplitudes of the sought-after signatures, 
data sets collected by these experiments and containing the cos\-mo\-lo\-gi\-cal\-ly-relevant information
will have to surpass in volume and complexity any of the CMB data sets available to date by orders of magnitude. CMB anisotropies constitute a stochastic, weakly correlated, sky-stationary signal, which, in the CMB data analysis process, has to be differentiated from other stochastic and usually correlated contributions. This typically disfavors any \emph{divide-and-conquer} approaches and instead huge volumes of the CMB data have to be processed simultaneously. This sets demanding requirements on computer memory which can be only met by the largest massively-parallel, distributed-memory supercomputers. A key operation of this type is referred to as a map-making procedure and it attempts to compress the raw, time-domain data as obtained by the experiments, into smaller, and potentially more manageable, map-like objects. In fact, actual map-making procedures are more general and map-making-like operations are ubiquitous in the CMB data analysis, implicitly or explicitly, on other processing stages emphasizing the key role map-making solvers play in the entire analysis process. Depending on a specific application, a solution to the same map-making system may need to be derived only once or multiple times for many different right hand sides, potentially requiring different application-specific solvers. 

The anticipated data volumes and the diversity of the contexts and applications define the challenge for any successful map-making software. In this work, we present a unified numerical framework designed to meet this challenge. The framework is massively parallel, flexible and extensible to permit processing very large data volumes while accounting for the specificity of different data sets leading to different models of the processed data.

This paper is organized as follows. In Sect.~\ref{sect:formalism} we introduce the map-making problem, and provide a general approach for reconstructing sky maps while deprojecting several time-domain systematics that may be present in the data. In Sect.~\ref{sect:framework} we discuss the implementation details of the framework proposed in this work, and in Sect.~\ref{sect:applications} we present some applications on simulated data from ground-based and satellite-borne experiments demonstrating the methods implemented in the framework.

\section{Map reconstruction}
\label{sect:formalism}

In this section, we lay out the basic formalism of sky maps reconstruction. We start from a generic data model of CMB experiments, from which we derive a general expression for unbiased sky maps estimators. We then discuss different types of these unbiased map estimators which employ different models of the noise and are adapted to deal with selected time-domain systematics.

\subsection{Data model}

In the map-making problem we typically assume that the raw data of the detectors, have been properly calibrated and cleaned in the pre-processing stage. The data are taken at a fixed time step and for each detector we order them by the time at which they were collected. Conceptually, the data from different detectors are typically assumed to be concatenated together in a single data vector. We denote this data vector as  $\mathbf{d}$. We use a linear model to describe the data representing them therefore as,
\begin{eqnarray}
\mathbf{d} & = & \sum_i \mathbf{T}_i\,\mathbf{x}_i \, + \, \mathbf{n}.
\label{eqn:dataModel0}
\end{eqnarray}
Here, the index $i$ goes over different identified contributions to the measurement, whatever their origins: these can describe for example sky stationary signals, scan synchronous signals, HWP spin synchronous signals, or long term drifts. The columns of each matrix $\mathbf{T}_i$ define a time-domain template describing the effect.  $\mathbf{n}$ denotes then a stochastic contribution, which is characterized only statistically and assumed for most of the application to be piece-wise time-stationary. We refer to this term generically as noise. 

A key assumption behind the model is that all relevant contributions can be described by a limited number of amplitudes stored in the vectors $\mathbf{x}_i$, and the corresponding templates, i.e., columns of the matrices $\mathbf{T}_i$, which are assumed to be known. For continuous signals, such as the sky signal itself, it implies that they can be discretized with sufficient precision.

For definiteness we assume that the $i=0$ term is the actual sky signal. The corresponding template matrix, $\mathbf{T}_0$, is then typically referred to as a pointing matrix and denoted as $\mathbf{P}$. The vector $\mathbf{x}_0$ stands then for a discretized sky signal, typically a pixelized sky map, which we denote as $\mathbf{m}$, ($=\mathbf{x}_0$). On collecting all the remaining template matrices, $\mathbf{T}_i$, in a single matrix $\mathbf{T} \equiv [ \mathbf{T}_1, ..., \mathbf{T}_i, ...]$ and concatenating all the amplitude vectors, $\mathbf{x}_i$, together, in a single vector, $\mathbf{x}$, we can rewrite Eq.~\eqref{eqn:dataModel0} as,
\begin{eqnarray}
\mathbf{d} & = & \mathbf{P}\,\mathbf{m} \, + \, \mathbf{T}\,\mathbf{x} \, + \, \mathbf{n}
\; \equiv \;
\boldsymbol{\mathcal{P}}\,\boldsymbol{\mathit{y}} \, + \, \mathbf{n}.
\label{eqn:dataModel}
\end{eqnarray}
Here, for future convenience we have introduced a generalized pointing matrix, $\boldsymbol{\mathcal{P}}$, and a generalized map, $\boldsymbol{\mathit{y}}$, which combine all the template matrices and amplitude vectors, respectively, including those describing the sky signal.

In the case of modern multi-kilo pixel experiments the size of the time ordered data can reach $\mathcal{N}_t=\mathcal{O}(10^{12-15})$, while the number of the amplitudes is typically much smaller. In particular, the number of sky amplitudes, that is the total number of Stokes parameters we are solving for, is usually on order of $\mathcal{N}_\mathrm{pix} = \mathcal{O}(10^{5-8})$. The data are therefore hugely redundant reflecting the fact that features of the actual sky signal of current interest are very subdominant as compared to the noise of the measurements.

\subsection{Map-making solution}
\label{sect:formalism:map-making}

The map-making problem consists of estimating the sky signal, $\mathbf{m}$, by compressing the redundant information contained in the raw data and deprojecting the undesirable effects defined by the combined template matrix $\mathbf{T}$. This is done given some assumptions about the statistical properties of the noise, $\mathbf{n}$. The most general estimator achieving all these goals can be written down in a concise form as~\cite{Stompor_2001, Poletti_2017},
\begin{eqnarray}
\hat{\mathbf{m}} & = & \left(\,\mathbf{P}^\mathrm{T}\,\mathbf{F_T}\,\mathbf{P}\,\right)^{-1}\,\mathbf{P}^\mathrm{T}\,\mathbf{F_T}\,\mathbf{d},
\label{eqn:genMapEstimator}
\end{eqnarray}
where 
\begin{eqnarray}
\mathbf{F_T}&\equiv&\mathbf{W}^{-1}\,\left(\,\mathbf{I}-\mathbf{T}\,\left(\,\mathbf{T}^\mathrm{T}\,\mathbf{W}^{-1}\,\mathbf{T}\,\right)^{-1}\,\mathbf{T}^\mathrm{T}\,\mathbf{W}^{-1}\,\right)
\label{eqn:defFT}
\end{eqnarray}
is a filtering and weighting operator, which simultaneously filters all the contributions contained in the subspace spanned by the columns of the template matrix, $\mathbf{T}$, i.e., $\mathbf{F_T} \, \mathbf{T} \, = \, 0$, and weights modes orthogonal to it by the weight matrix, $\mathbf{W}$.
Consequently, the estimator given by Eq.~\eqref{eqn:genMapEstimator} is explicitly unbiased as long as the system matrix, $\mathbf{P}^\mathrm{T}\,\mathbf{F_T}\,\mathbf{P}$, is non-singular and the assumed templates span the subspace affected by the unwanted signals. Indeed,
\begin{eqnarray}
\hat{\mathbf{m}} & = & \left(\,\mathbf{P}^\mathrm{T}\,\mathbf{F_T}\,\mathbf{P}\,\right)^{-1}\,\mathbf{P}^\mathrm{T}\,\mathbf{F_T}\,(\mathbf{P} \, \mathbf{m} \,+\,\mathbf{T}\,\mathbf{x}\,+\,\mathbf{n}),
\nonumber\\
& = & 
\mathbf{m} \, + \, \left(\,\mathbf{P}^\mathrm{T}\,\mathbf{F_T}\,\mathbf{P}\,\right)^{-1}\,\mathbf{P}^\mathrm{T}\,\mathbf{F_T}\, \mathbf{n},
\end{eqnarray}
and the last term vanishes when averaged over the statistical ensemble of noise realizations as $\langle \mathbf{n}\rangle_\mathrm{noise} \, = \, 0 $.

We note for completeness that if some prior information concerning the sought-after amplitudes of the templates is available, assuming those are given by a multivariate Gaussian distribution, it can be incorporated in the map estimation modifying the filtering operator, which in such a case reads,
\begin{eqnarray}
\mathbf{\tilde{F}_T} \;=\; \mathbf{W}^{-1}\,\left(\,\mathbf{I}-\mathbf{T}\,\left(\boldsymbol{\Theta}^{\;-1}\,+\,\mathbf{T}^\mathrm{T}\,\mathbf{W}^{-1}\,\mathbf{T}\,\right)^{-1}\,\mathbf{T}^\mathrm{T}\,\mathbf{W}^{-1}\,\right),
\label{eqn:defFtPrior}
\end{eqnarray}
where $\boldsymbol{\Theta}$ is the covariance matrix of the sought after amplitudes, $\mathbf{x}$. This matrix may be rank deficient if the prior information is available only for some of the amplitudes and not others. No prior information at all corresponds to $\boldsymbol{\Theta} \rightarrow \infty$ and $\boldsymbol{\Theta}^{\;-1} \rightarrow 0$ and we recover Eq.~\eqref{eqn:defFT}. We note that the templates for which the prior is available are not completely filtered out from the data during each map-making procedure, i.e., $\mathbf{\tilde{F}_T}\,\mathbf{T} \, \ne 0$, rather they are Wiener-filtered instead. The sky signal estimates are therefore not unbiased when averaged over the noise ensemble, however, their overall uncertainty over the ensemble of the noise and template amplitudes is minimized given the assumed priors~\cite{bunn_etal_1996}.

Eq.~\eqref{eqn:genMapEstimator} can be implemented implicitly, for instance, by solving for the full amplitude vector, $\boldsymbol{\mathit{y}}$, weighting the data with help of the weight matrix, $\mathbf{W}$, i.e.,
\begin{eqnarray}
\hat{\boldsymbol{\mathit{y}}} & = & \left(\,\boldsymbol{\mathcal{P}}^\mathrm{T}\,\mathbf{W}^{-1}\,\boldsymbol{\mathcal{P}}\,\right)^{-1}\,\boldsymbol{\mathcal{P}}^\mathrm{T}\,\mathbf{W}^{-1}\,\mathbf{d},
\label{eqn:fullEstimator}
\end{eqnarray}
and truncating it afterwards to retain the sky signal part only~\cite{Stompor_2001}.

Alternately, one can opt for a two step approach where the amplitudes of the undesirable signals, $\mathbf{x}$, are first estimated as,
\begin{eqnarray}
\hat{\mathbf{x}}=\left(\,\mathbf{T}^\mathrm{T}\,\mathbf{F_P}\,\mathbf{T}\,\right)^{-1}\,\mathbf{T}^\mathrm{T}\,\mathbf{F_P}\,\mathbf{d},
\label{eqn:tempAmpEstimator}
\end{eqnarray}
and the sky map is then recovered on the second step via,
\begin{eqnarray}
\hat{\mathbf{m}} & = & 
(\mathbf{P}^\mathrm{T}\,\mathbf{W}^{-1}\,\mathbf{P})^{-1}\,\mathbf{P}^\mathrm{T}\,\mathbf{W}^{-1}\,(\mathbf{d}\,-\,\mathbf{T}\,\hat{\mathbf{x}}).
\end{eqnarray}
This approach is then essentially equivalent to the destriping technique originally proposed for Planck~\cite{Burigana_1999, Delabrouille_1998, Maino_1999, Planck18_HFI_dataproc, Planck18_LFI_dataproc, NPIPE_dataproc}, and later studied also in the context of ground-based experiments~\cite{Sutton_2009, Sutton_2010, Poletti_2017}.

Both these approaches have been successfully employed in the past. In the realm of the multi-kilo pixel experiments, the number of instrumental, detector-specific contributions in the data is typically exceeding by far the number of the observed sky pixels. Thus, the system matrices in Eq.~\eqref{eqn:fullEstimator} and~\eqref{eqn:tempAmpEstimator} are significantly larger than the one appearing explicitly in Eq.~\eqref{eqn:genMapEstimator}. Consequently, solving directly Eq.~\eqref{eqn:genMapEstimator} may look as potentially more computationally attractive. For this to be indeed the case, there must be a computationally-efficient way to construct the template orthogonalization kernel $(\mathbf{T}^\mathrm{T}\,\mathbf{W}^{-1}\,\mathbf{T})^{-1}$, which in general can be rather cumbersome to calculate. Fortunately, in many relevant applications the kernel is usually very structured and sparse. This stems from the fact that most of the templates of interest have compact and disjoint support, which is moreover known a priori. This can be capitalized on to save on the calculations making the estimator in Eq.~\eqref{eqn:genMapEstimator} not only computationally feasible but also preferred in many applications. 
We therefore implement this approach in the proposed framework.

Eq.~\eqref{eqn:genMapEstimator} generalizes and includes as special cases many of the standard map-making techniques. Those correspond to some specific choices of either the weight matrix  or the template matrix or both. These particular cases include: (1) the standard maximum likelihood (minimum variance) map-making, where the templates are often dropped and the weights are taken to be given by the inverse covariance of the noise term, $\mathbf{n}$, which is assumed to be Gaussian; (2) the binned map-making where the weights are taken to be diagonal and equal to the noise RMS and the templates neglected; (3) the standard destriper, where the templates are taken to be piece-wise constant offsets and the weights diagonal and set by the noise RMS; (4) the generalized destriper, where the templates can be more general and the covariance of their amplitudes could be included as prior information.

Eq.~\eqref{eqn:genMapEstimator} is very flexible allowing for diverse choices for the pointing, the template, and the weight matrices with the only practical constraint being numerical considerations related to their application. As we discuss in the following section this permits applications of this equation in many cases of practical interest.

\subsection{Pointing matrix}
\label{sect:formalism:pointing}

The sky signal measured by an instrument is a convolution of the actual sky signal, expressed with the three Stokes parameters $I, Q$ and $U$, with an instrument response matrix, $\boldsymbol{\mathcal{R}}(\delta\boldsymbol{\gamma},  \psi, \nu)$.  
This matrix will in general depend on both the sky angle, $\delta\boldsymbol{\gamma}$, measured with respect to the observation direction, $\boldsymbol{\gamma}$, the orientation of the instrument with respect to the sky coordinates, $\psi$, and the frequency, $\nu$. Therefore, a measurement made in $\nu_c$ frequency band, defined by a bandpass, $\mathcal{W}(\nu, \nu_c)$, can be written in general as,
\begin{eqnarray}
\mathbf{d}_{\nu_c}(t) & = & \int d\nu \int d\delta\boldsymbol{\gamma}\;
\mathbf{e_i}^\mathrm{T}\, \boldsymbol{\mathcal{R}}(\delta\boldsymbol{\gamma}, \psi(t), \nu)\,\times\nonumber\\
&&\ \ \ \ \ \ \ \ 
\times\;
\mathcal{W}(\nu, \nu_c)\,
 \, \mathbf{s}(\boldsymbol{\gamma}(t)+\delta\boldsymbol{\gamma}, \nu) \, + \, \mathbf{n}(t),
\label{eqn:instRespDataModel}
\end{eqnarray}
where $\boldsymbol{\gamma}(t)$ is the observation direction at time $t$ and $\mathbf{e_i}$ is a unit vector which defines whether we ultimately measure total power only, $i=0$, or Stokes $Q$, $i=1$, or $U$, $i=2$, parameter. $\mathbf{s}(\boldsymbol{\gamma}, \nu)$ is a vector of the three Stokes parameters as defined in the direction $\boldsymbol{\gamma}(t)$ and at frequency $\nu$.

It is frequently possible to represent the general instrument response matrix as a product of a Mueller matrix describing the detection chain of the instrument and a matrix defining the beam effects, 
\begin{eqnarray}
 \boldsymbol{\mathcal{R}}(\delta\boldsymbol{\gamma}, \psi(t), \nu) \; \equiv \; 
 \boldsymbol{\mathcal{M}}(\boldsymbol{\iota}(t), \nu)\,
 \boldsymbol{\mathcal{B}}(\delta\boldsymbol{\gamma}, \nu)\,\mathbf{R}(\psi(t)),
\end{eqnarray}
where the matrices, $\boldsymbol{\mathcal{M}}$ and $\boldsymbol{\mathcal{B}}$, are defined with respect to instrument coordinates and the matrix $\mathbf{R}(\psi(t))$, denotes a rotation between the sky and instrument coordinates. We note that the Mueller matrix, $\boldsymbol{\mathcal{M}}$, may in general depend on time via some of the instrument parameters, such as, phase of a polarization modulator, as denoted by parameter $\boldsymbol{\iota}(t)$.

Assuming that the beam matrix is diagonal, axially symmetric, i.e., depending only on the magnitude of the angle, $\delta\boldsymbol{\gamma}$, and not on its orientation, and that the beams for the Stokes parameters $Q$ and $U$ are identical, we can commute the rotation operator and the beam matrix, rewriting Eq.~\eqref{eqn:instRespDataModel}, as,
\begin{eqnarray}
\mathbf{d}_{\nu_c}(t) & = & \int d\nu\, \mathcal{W}(\nu, \nu_c)\, \mathbf{e_i}^\mathrm{T}\, \boldsymbol{\mathcal{M}}(\boldsymbol{\iota}(t), \nu)\,\mathbf{R}(\psi(t)) 
\label{eqn:modelBeamConv}
\\
&& \times \underbrace{\int\, d\delta\boldsymbol{\gamma}\,\boldsymbol{\mathcal{B}}(\delta\boldsymbol{\gamma}, \nu)\, \mathbf{s}(\boldsymbol{\gamma}(t)+\delta\boldsymbol{\gamma}, \nu)}_{\displaystyle \equiv \bar{\mathbf{s}}(\boldsymbol{\gamma}(t), \nu)} \, + \, \mathbf{n}(t),\nonumber
\end{eqnarray}
where $\bar{\mathbf{s}}(\boldsymbol{\gamma}(t), \nu)$ stands for a beam-smoothed sky signal, which can be therefore characterized by a finite set of pixelized sky signal amplitudes, $\bar{\mathbf{s}}(p(t), \nu)$, where $p(t)$ is a sky pixel observed at time $t$.

For the frequency bandpass convolution, the simplest case is when the bands defined by $\mathcal{W}(\nu, \nu_c)$ are sufficiently narrow so the beam-convolved sky signal is essentially constant across the band. This allows us to write,
\begin{eqnarray}
\mathbf{d}_{\nu_c}(t) & \approx & \left[\int d\nu\, \mathcal{W}(\nu, \nu_c)\, \mathbf{e_i}^\mathrm{T}\, \boldsymbol{\mathcal{M}}(\boldsymbol{\iota}(t), \nu)\,\mathbf{R}(\psi(t))\right] \nonumber\\ && \times\,\bar{\mathbf{s}}(p(t), \nu_c) \, + \, \mathbf{n}(t), \nonumber \\
&\equiv & \mathbf{w}(t) \, \bar{\mathbf{s}}(p(t), \nu_c) \, + \, \mathbf{n}(t).
\label{eqn:pointMatWghts}
\end{eqnarray}
 The vector, $\mathbf{w}(t)$, defines the pointing weights with which the different Stokes parameters at the band-center frequency, $\nu_c$, are coadded together to give the measured signal, while $p(t)$ defines a sky pixel number observed at time $t$.
This equation also holds when the bands are arbitrary but the Mueller matrix is not dependent on the frequency, $\nu$. However, the recovered sky signal represents then the actual sky signal convolved with the experimental bandpass.

In a more general case when the bands are arbitrary and the Mueller matrix depends on the frequency, then as long as the time-dependence of all the relevant elements of the Mueller matrix is known~\cite{Verges_2021}, the algebraic structure of Eq.~\eqref{eqn:pointMatWghts} is preserved.
In such cases the recovered sky signal will be however in general a linear combination of Stokes parameters convolved with frequency response functions defined by the bandpasses and the Mueller elements. Moreover, for each pixel the number of the amplitudes, required to describe the data may be significantly larger than the number of the Stokes parameters themselves~\cite{Verges_2021}.

In all these cases the information about the pointing weights and the pixel as observed at time $t$ can be encoded in a single matrix, the pointing matrix $\mathbf{P}$, which is tall and skinny with the number of rows given by the total number of measurements and the number of columns given by the total number of observed pixels multiplied by the number of observables assigned to each pixel. The data model can be then expressed conveniently in a vector form as,
 \begin{eqnarray}
\mathbf{d} \; = \; \mathbf{P}\,\mathbf{m} \, + \, \mathbf{n}.
\label{eqn:dataModelVect}
\end{eqnarray}
In a row of the matrix $\mathbf{P}$ corresponding to time $t$, the only non-zero elements are those in columns corresponding to pixel $p$ observed at the time $t$. These elements are given by the respective elements of the pointing weight vector $\mathbf{w}$ and are known given our knowledge of the instrument and the observation. In the most straightforward applications, there are therefore either $1$ -- for total intensity measurements, or $2$ -- for $Q$ and/or $U$ polarization measurements, or $3$ -- for polarization-sensitive measurements, non-zero elements in each row. Consequently, the pointing matrix is  typically very sparse and thus manageable in spite of its huge size.

Important examples include the case of a fixed single-slot polarizer, when the measurements can be expressed as,
\begin{eqnarray}
d_t = I_{p_t} + \cos(2\varphi _{t})Q_{p_t} + \sin(2\varphi_{t})U_{p_t} + n_t,
\label{eqn:dataModelSingleSlot}
\end{eqnarray}
where $\varphi$ denotes an angle of the polarizer with respect to the sky coordinates. Similarly, for an idealized frequency-independent, half-wave plate, used to modulate the incoming signal, the data model reads,
\begin{eqnarray}
d_t = I_{p_t} + \cos(2\varphi _{t} + 4\phi_t)Q_{p_t} + \sin(2\varphi_{t}+4\phi_t)U_{p_t} + n_t,
\label{eqn:dataModelHwp}
\end{eqnarray}
where $\phi_t$ stands for the HWP orientation with respect to the instrument coordinates. 
In both these cases, there are three non-zero entries per row of the pointing matrix, $\mathbf{P}$. As mentioned earlier this number can be lower, i.e., for total intensity measurement, or in the case of pair-differencing approaches. It can also be larger, for instance, in the case of more realistic models of the half-wave plate, where the induced rotation of the incoming polarization depends on the frequency and one may need to introduce as many as $5$ different amplitudes to characterize the signal in each sky pixel. This results therefore in $5$ non-zeros per row of $\mathbf{P}$~\cite{Verges_2021}. The number of non-zeros per row may be even larger in the cases of the generalized pointing matrix, $\boldsymbol{\mathcal{P}}$, as it has to incorporate information about the non-sky degrees of freedom stored in the matrix $\mathbf{T}$. In all these cases  the pointing matrix is  nevertheless extremely sparse and the number of non-zero values per row is very limited.

A notable exception is the case of asymmetric beams, i.e., when $\boldsymbol{\mathcal{B}}$ depends explicitly on a vector $\delta\boldsymbol{\gamma}$ and not just its length. If beams happen to be strongly asymmetric to the extent that it is necessary to correct for it on the map-making stage, the data model in Eq.~\eqref{eqn:dataModelVect} would still apply and though the pointing matrix would be in such a case dense, it would be also structured, permitting specialized algorithms for the map-making problem as have been indeed proposed~\cite{Armitage_2004, Harrison_2011, Keihanen_2012}. In the framework discussed here the focus to date has been specifically on the cases with an arbitrary but sparse pointing matrix. And while the framework readily allows for extensions which would accommodate also cases with asymmetric beams, this requires however further development that is left for the future and is not considered in this work.

\subsection{Weights}
\label{sect:formalism:weights}

The estimator in Eq.~\eqref{eqn:genMapEstimator} is unbiased whenever the templates are appropriately chosen, given the anticipated contaminants, and the system matrix is non-singular. This property is independent of the choice of the assumed weighting. The weighting is however key in obtaining an estimate with as high signal-to-noise ratio as possible.

A maximum likelihood estimate (which is as well a minimum variance one) is indeed given by Eq.~\eqref{eqn:genMapEstimator} with the weights corresponding to the noise covariance,
\begin{eqnarray}
\mathbf{W} = \mathbf{N} \equiv \langle \mathbf{n}\,\mathbf{n}^\mathrm{T}\rangle.
\end{eqnarray}
Given the sizes of the current and forthcoming data sets, such covariance matrices are readily unmanageable in their fully general form. However, the noise properties of any single detector data can be typically adequately characterized as piece-wise stationary. The corresponding noise covariance is then block-diagonal where each block corresponds to a different stationary interval and has the structure of a Toeplitz matrix. This means that
for the $i$th interval, the correlation between the noise at times $t$ and $t'$ depends merely on the time interval, $|t-t'|$, and not actual values of $t$ and $t'$. Consequently, the $i$th block of the covariance matrix satisfies the condition, 
\begin{eqnarray}
\mathbf{N}^{\left(i\right)}(t,t') \; = \; \mathbf{N}^{\left(i\right)}(0, |t-t'|).
\end{eqnarray}

The noise in the CMB measurements is typically correlated all the way to very long timescales and the Toeplitz blocks of the noise covariance are in general dense. However, at least for well-optimized scanning strategies, accounting for the correlations on the largest time scales is usually not necessary, as they affect predominantly only signals varying on angular scales too large to be well-constrained by the experiment anyway and therefore manifest themselves merely as an offset of the map (or a submap)~\cite{Szydlarski_2014}. Consequently, it is justifiable, and desirable, to assume that the correlations are band limited and set to zero beyond some characteristic correlation length parameter, $\lambda$ (also called half-bandwidth). The value of $\lambda$ should be fixed according to properties of the noise and the adopted scanning strategy~\cite{Szydlarski_2014}.

In the map-making problem, Eq.~\eqref{eqn:genMapEstimator}, we need an inverse of the weight matrix. The inverse of a general  Toeplitz matrix is not Toeplitz and it is not easily computable. The inverse Toeplitz blocks once computed would have to be stored in the computer memory, which would quickly undermine the numerical feasibility of the procedure. However, for a banded-Toeplitz matrix, in particular with a rather narrow band, its inverse can be very well-approximated by a Toeplitz matrix. The approximate inverse is given by,
\begin{eqnarray}
{\mathbf{N}^{\left(i\right)}}^{-1}(t,t') \simeq \left\{\begin{matrix}
{\mathbf{C}^{\left(i\right)}}^{-1}(t,t'),& \text{if } |t-t'| \; \leq \lambda \\ 
 0, & \text{otherwise} 
\end{matrix}\right.
\label{eqn:MADCAP}
\end{eqnarray}
Here $\mathbf{C}^{\left(i\right)}$ is a circulant matrix embedding the Toeplitz block $\mathbf{N}^{\left(i\right)}$. The inverse of a circulant matrix is also circulant and can be easily calculated with the help of Fast Fourier Transforms (FFTs) at low numerical cost~\cite{golub13} as discussed later in Sect.~\ref{sect:framework}.

The noise covariance has to be derived typically from the same data as the data used later for the map-making~\cite{Ferreira_2000, Stompor_2001, Wehus_2012}. As a consequence, even for the case of Gaussian noise in the time-domain, the noise of the resulting map is not strictly Gaussian and some care may need to be taken in minimizing or accounting for it in particular at the largest angular scales. For most of the applications this effect is however thought to be largely irrelevant.

One specific choice which is of interest, in particular because of its numerical efficiency, is that of a diagonal weight matrix. This is a limiting case of the block-Toeplitz case discussed above with $\lambda = 0$, however, the application of the weights in such cases can be done directly in the time-domain without any need for Fast Fourier Transforms. From the point of view of the quality of the estimated maps this is however often a rather bad choice which will frequently result in the estimates being dominated by the low frequency noise modes. These modes can be often treated as templates as it is for instance the case of the destriper techniques~\cite{Burigana_1999, Delabrouille_1998, Maino_1999, Revenu_2000, Keihanen_2004}. We discuss choices of templates suitable for this task in \ref{sect:appA}. 

Whenever a prior can be imposed on the amplitude of the templates, as in Eq.~\eqref{eqn:defFtPrior}, introducing the templates is equivalent to adding a low-rank correction of the weight covariance, so in such a case the weights, $\mathbf{W}$, are replaced by~\cite{Stompor_2001},
\begin{eqnarray}
\mathbf{W} \; \rightarrow \; \mathbf{W} \, + \, \mathbf{T}\,\mathbf{\Theta}\,\mathbf{T}^\mathrm{T}.
\end{eqnarray}
Conversely, any low rank correction to the weight matrix can be rephrased as a set of templates and implemented via the projection operator, $\mathbf{F_T}$. A successful set of templates will correspond to a low rank approximation of the full noise covariance, which reflects all of its characteristics which are relevant for the problem at hand. The challenge is therefore in finding out what those characteristics are. 

The need for templates is not limited to the cases when the weight matrix is diagonal. In particular, it is often desirable to use templates to mitigate some systematic effects in the maximum likelihood map-making, when the weights are block-Toeplitz. In practice, the number of such templates is expected to be lower than the number of templates needed to model the low-frequency noise correlations in the case of diagonal weights. It is therefore computationally advantageous to extend the solution vector by adding the amplitudes of the required templates and solve this extended system using standard maximum likelihood technique with $\mathbf{F_T} = \mathbf{W}$ as in Eq.~\eqref{eqn:fullEstimator}. This is equivalent to the meta-pixel approach of \cite{Stompor_2001}. Consequently, the proposed framework implements two options: (1) a maximum likelihood map-making with an arbitrary (but sparse) pointing matrix and block-Toeplitz weights, and, (2) cases of diagonal weights with templates which are directly included in the operator $\mathbf{F_T}$, as in Eq.~\eqref{eqn:defFT} or Eq.~\eqref{eqn:defFtPrior}. However, other cases could be implemented using the functionality provided by the proposed libraries.

All the considerations above apply directly to data of a single detector. In the realm of multi-kilo pixel arrays, we have to deal with data  collected concurrently by many detectors. It is therefore overly optimistic to assume no correlations between the measurements of different detectors. As pointed out earlier, neglecting those will not bias our sky signal estimates however it may easily render them very noisy. Sources of the correlations can be diverse. They could be either due to some common noise source seen by multiple detectors, such as atmospheric signal unavoidably present in the case of the ground-based experiments, or can be induced in the detection chain of the instrument, for instance, in its readout system. In the case of the maximum likelihood map-making such correlations could be in principle included as an off-diagonal block of a full weight (noise correlation) matrix, which includes weights for all detectors simultaneously. However, this would lead to significant numerical overhead rendering such a map-making hardly useful in practice. Consequently, the proposed framework does not provide this functionality at this time as a default.

Some workaround approaches which are more manageable but can render comparable performance are however frequently available. One is to use templates to account for those correlations or to capitalize on differencing scheme for data of some of the detectors devised to remove the correlated component.

\subsection{Templates}
\label{sect:formalism:templates}

The templates can be in principle arbitrary reflecting the diversity of possible non-cosmological contributions which may be present in the data. Whether those can be deprojected from the data on the map-making stage will in general depend on three facts: (1) whether we can define efficiently a subspace, as spanned by the columns of the matrix $\mathbf{T}$, which contains all possible undesirable contributions; (2) whether this can be done while retaining limited dimensionality of the subspace; and (3) whether this subspace overlaps with the subspace spanned by the sky signal. If (1) is not met then some unaccounted for residuals will be present in the estimated sky signal. Failing on (2) may affect the quality, signal-to-noise, of the map, but the estimate will not be biased. Higher the dimensionality of the deprojected subspace increases typically the chance that (3) is not met either. If (3) is not fulfilled the system matrix in Eq.~\eqref{eqn:genMapEstimator} is singular and therefore some of the modes of the sky signal may not be recoverable. We note that if such modes are properly identified the map-making procedure can still be often applied in a meaningful way estimating all the modes for which this can be achieved. The information about the non-recovered modes would have to be then passed on and accounted for on all subsequent stages of the analysis~\cite{Poletti_2017}. 

As mentioned earlier, from the computational point of view some templates may be difficult to implement as they may lead to significant computational overhead, in particular in the construction of the orthogonalization kernel. This issue can be often sidestepped by introducing multiple independent copies of the same template specific to different, disjoint time intervals. Crucially, the procedure makes all the computations more local in the computer memory, thus avoiding extra need for extensive global communication between the compute nodes. At the same time the increase of the size of the template orthogonalization kernel is mitigated by the fact that it is limited to additional diagonal blocks. There is a clear downside to this procedure as it unavoidably increases the dimensionality of the deprojected subspace, therefore leading to a loss of precision. This however may be often acceptable. Moreover, in actual applications many templates of interest are naturally specific to limited ranges of the observation time. As we comment later, the proposed framework is therefore restricted to this kind of templates.
Within such limits the framework allows in principle for any user-defined templates. In addition, some templates are implemented directly in the package. They are described in~\ref{sect:appA}. 

\subsection{Degeneracy issues}
So far our pointing or generalized pointing matrices were considered to be full column rank, therefore ensuring that the system matrices are invertible. In practice, this assumption can be violated for different reasons which we will explore in the following subsections.
\subsubsection{Scanning strategy sourced degeneracies}
In order to solve for three Stokes parameters $I$, $Q$ and $U$, in each pixel in the sky, the scanning strategy should ensure that each pixel is observed at least three times with sufficiently different orientations. The polarization modulation induced by the half-wave plate helps in this regard as it ensures that in each pixel crossing, sufficient samples are taken with different orientations to allow for the reconstruction of the pixel Stokes parameters. However, for some pixels, particularly in the periphery of the observed sky patch, it may happen that the number of revisits of the pixel is very low and on each revisit the pixel crossing time is very short therefore not allowing for sufficient sampling of the polarization angles.

\sloppy
In practice this issue is solved by examining the condition numbers of each block in the block diagonal matrix $\mathbf{P^\mathrm{T}\, \mathrm{diag\,}W^{-1}\, P}$. Each block corresponds to an observed sky pixel, which if observed with sufficient redundancy, should ideally have a condition number close to 2. The higher the condition number, the more the pixel noise gets boosted. We therefore set a threshold on the condition number of each pixel, above which the pixel, and its corresponding TOD samples, get effectively removed from the data.

\subsubsection{Templates linear dependencies}
The second type of degeneracies stems from linear dependencies between templates. One prominent example, is the 0th order polynomial template and the ground template which both filter the global offset of the TOD (see~\ref{sect:appA}, and~\cite{Poletti_2017}).

Such degeneracies prevent the kernel matrix $\mathbf{T}^\mathrm{T}\,\mathbf{W^{-1}}\,\mathbf{T}$ from being invertible. In practice, this does not pose any problem for the sky signal recovery and is easily addressed: we orthogonalize the templates by computing the Moore-Penrose pseudo-inverse of the kernel via a singular value decomposition (SVD). This is made possible because of the simple structure of the kernel matrix as discussed earlier in this section. See also~\ref{sect:appA} for more details in the case of specific templates.

\subsubsection{Templates and sky degeneracies}
This last type of degeneracies is due to linear dependencies between the filtered templates and some sky modes. When such degeneracies are present, the concerned sky modes are impossible to reconstruct and would therefore be missing from the maps, because they are filtered from the data by the deprojection operator $\mathbf{F_T}$.

In practice, this means that the matrix $\mathbf{P}^\mathrm{T}\,\mathbf{F_T}\,\mathbf{P}$ would have singularities. Given the size, and the fact that this matrix is dense in general, the direct computation of these singular modes to regularize the inversion is not feasible. In an iterative approach such as the one we adopt in this work, the presence of such degeneracies manifests itself as a significant slow down in the convergence rate. This is caused by the small eigenvalues in the eigenspectrum of the system matrix introduced by the singular or nearly singular modes. One way to address this issue, is to approximate these modes with an iterative method such as the Lanczos procedure, and deproject them from the system. This provides the basis for the \textit{a posteriori} two-level preconditioning technique, that we will address in more detail in Sect.~\ref{sect:framework}.

\section{Parallel mapping framework}

\label{sect:framework}

In this section, we provide a detailed description of the software framework and highlight key aspects of its implementation. The software is publicly accessible at the following link: \url{https://github.com/B3Dcmb/midapack}. We start from an overview of the code with its overall architecture, then describe each of its building blocks and how they fit into the global framework.

\subsection{Code overview}
\begin{figure}
\centering
\includegraphics[width=0.5\textwidth]{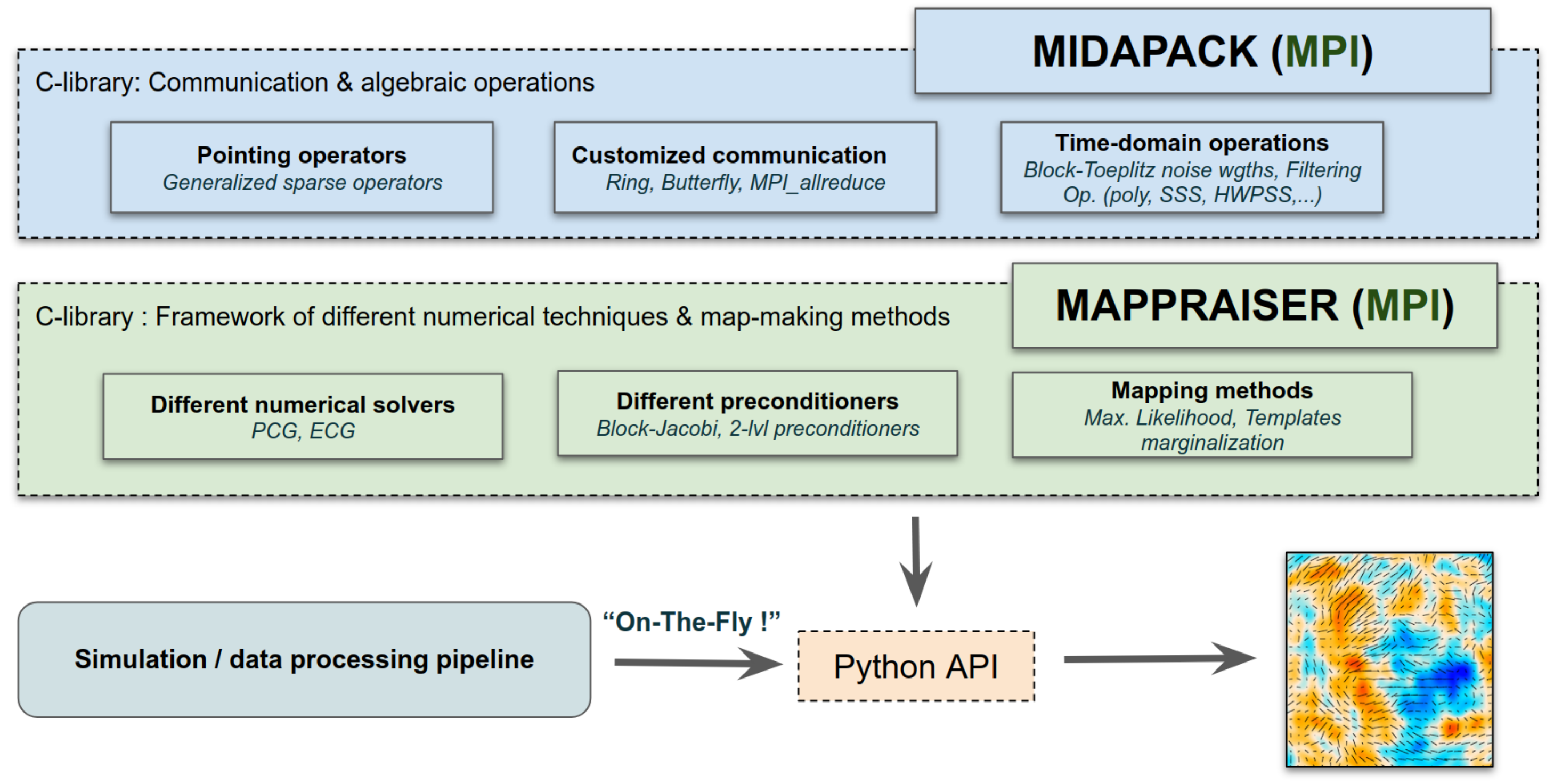}
\caption{\label{code_overview} 
Overview of the software top-level architecture.}
\end{figure}

A summary of the high-level code architecture, is given in figure \ref{code_overview}. The software is comprised of two main C-libraries:
\begin{itemize}
    \item The first library, called MIcrowave Data Analysis PACKage (MIDAPACK), provides low-level operations needed to perform high-level procedures in the data analysis pipe\-lines. Such operations include, sparse linear algebraic operations for generalized pointing and template operators, customized communication schemes for distributed data reduction, operators for structured matrices, such as Toeplitz matrices, or block-diagonal templates kernels, necessary to perform some time-domain operations such as templates filtering or noise weighting. We note that these operations can be relevant for all main stages of the data analysis pipeline, but in accordance with the scope of this paper we mainly focus on map-making applications.
    \item The second library, called MidAPack PaRAllel Iterative Sky EstimatoR (MAPPRAISER), builds on the first library to provide different numerical techniques, solvers and map-making methods to compute unbiased maps, and filter time-domain systematics concurrently with the map projection.
\end{itemize}

The routines conform with the MPI programming model and are therefore able to use massively parallel platforms with memory distributed between a large number of compute nodes. Some routines also make use of multithreading, but this option still needs to be optimized to be fully functional in the code. However these routines may provide the basis for a future extension of the framework to a full hybrid OpenMP/MPI model.

On top of this two-level core layer of the code, there is a Python wrapper, which allows the map-making code to interface easily with simulation or real data analysis pipelines. In particular, the Python wrapper can be used to perform the map-making run ``on-the-fly'' chaining the map-making code with simulation software or with a previous step of an actual data processing pipeline, therefore reducing I/O cost.

\subsection{The microwave data analysis package (MIDAPACK)}
\label{sect:framework:package}
The MIDAPACK library consists of three main sub-packages:
\begin{itemize}
    \item \textbf{The pointing operations package MAPMAT}: this package contains sparse matrix vector products operations to perform the projections from time-domain data to map-domain data and vice-versa. In their most general form, these matrices are rectangular with considerably more rows than columns, given the relative sizes of the time-domain and map-domain data and are referred to as \textit{tall-and-skinny} matrices. Their most important property however is their sparsity as they feature at most few non-zero elements per row.
    \item \textbf{The Toeplitz algebra package}: this package provides routines to perform the product of Toeplitz-structured matrices with an arbitrary data matrix. We refer hereafter to  matrices which are symmetric, block-diagonal with banded Toeplitz blocks, or matrices with all of the above properties along  some rows and corresponding columns missing as ``Toeplitz-structured''. These special matrices are common in CMB data analysis, as they encode the correlation properties of stationary time-domain processes such as instrumental 1/f-noise or atmosphere extending up to the correlation length corresponding to the band width of the matrix. This can be adjusted to provide the best trade-off between the computational time and quality of the result. 
    These routines are used in noise weighting operations, i.e., weighting the different frequency modes in the time-domain data with their corresponding inverse noise power. 
    \item \textbf{The templates algebra package}: this package provides routines to handle multiple sets of templates modeling different sorts of systematics. The routines allow to perform deprojections of templates, i.e., time-domain filtering operations as well as to orthogonalize a set of different templates to regularize the deprojection operation.
    The package allows for and seamlessly manipulates heterogeneous templates which may be defined by different means. This is an important functionality given the diversity of effects which the template formalism aims at describing.
    
    This is facilitated by grouping the templates into template classes. A general template corresponds to a single column of any of the matrices $\mathbf{T}_i$, Eq.~\eqref{eqn:dataModel0}. Templates corresponding to columns of the same matrix $\mathbf{T}_i$ aim at describing the same physical effects and often share some properties. Hereafter, we uniquely assign, a template class to each matrix $\mathbf{T}_i$ and all columns of the matrix define therefore templates which belong to the same template class.
    
    The templates can be defined individually or as a class. They can be defined explicitly via explicitly stored elements of the corresponding column of the matrix $\mathbf{T}_i$. They can also be defined implicitly and computed only as needed using some predefined analytic prescription for a set of parameters. This is for instance the case of the polynomial templates, which are uniquely defined by the order of the polynomial.
    The templates can be also often stored in a compressed form. This is the case when the modelled contribution is synchronous with some instrumental or observation parameter (or parameters), which in turn is a function of the observation time. Such contribution are common including, for example, scan synchronous or ground-pickup as discussed in~\ref{sect:appA}. To be manageable such contributions are discretized by binning (or pixelizing) the synchronous parameter into disjoint bins. This procedure introduces as many independent signals as bins. In the template formulation, this means that there is a separate template needed for each bin. Given that each template is a time-domain object, the memory needed to store all such templates becomes quickly prohibitive and at the very least restricting importantly the number of bins we can consider.

    However, such templates can be defined as a class in a memory efficient fashion capitalizing on the fact that there is only one bin (or more generally, a limited number of bins) contributing to a measurement at any given time. The contribution of such signal to a measurement made at some time $t$ depends on a uniquely assigned bin number $b(t)$, which corresponds to this time. This information can be readily stored as a single time-domain vector, $b(t)$. In addition we may need at most another time-domain vector which defines with what weight the signal amplitude corresponding to that bin should be added to a measurement at time $t$. These two objects are sufficient to characterize any binned contributions independently on the number of assumed bins, therefore providing a convenient and common compressed form encoding information relevant to the template class.

    A template is in general an arbitrary time-domain-like vector. However, in the proposed package we restrict ourselves to templates which have a compact support fully contained within the length of a single full scan of one of the detectors. The templates considered here are therefore specific to scans of each detector.

    We note that while more complex situations may arise, scan specific templates are indeed the most typical occurrence in actual applications, and are sufficient to model most, if not all, relevant systematic effects. This restriction allows to significantly reduce the computational and communication overhead.

    One consequence of it is that while templates assigned to a given process may or may not be orthogonal, templates assigned to two different processes, even if belonging to the same template class are always mutually orthogonal.

    While the package allows for a broad set of user-defined templates it also implements directly some of them. These are described in detail in~\ref{sect:appA}.
\end{itemize}
    
In the following subsections, we will discuss the implementation details of each one of these sub-packages. Before doing so we need to define the distribution scheme of the time domain data. A typical, modern experiment features multiple detectors observing the sky over en extended period of time. The full observation time can be broken into a number of independent time-intervals common for all detectors in a way that for each detector the noise can be considered stationary within each interval and uncorrelated between them. We hereafter refer to these time-intervals as ``scans''. The timestreams of all detectors are stacked together vertically in a two dimensional array with columns corresponding to the detectors and rows to times at which data were taken. This full data array can be broken down into  disjoint sub-arrays each made of full-length rows corresponding to one of the scans.

\begin{figure}
\centering
\includegraphics[width=0.48\textwidth]{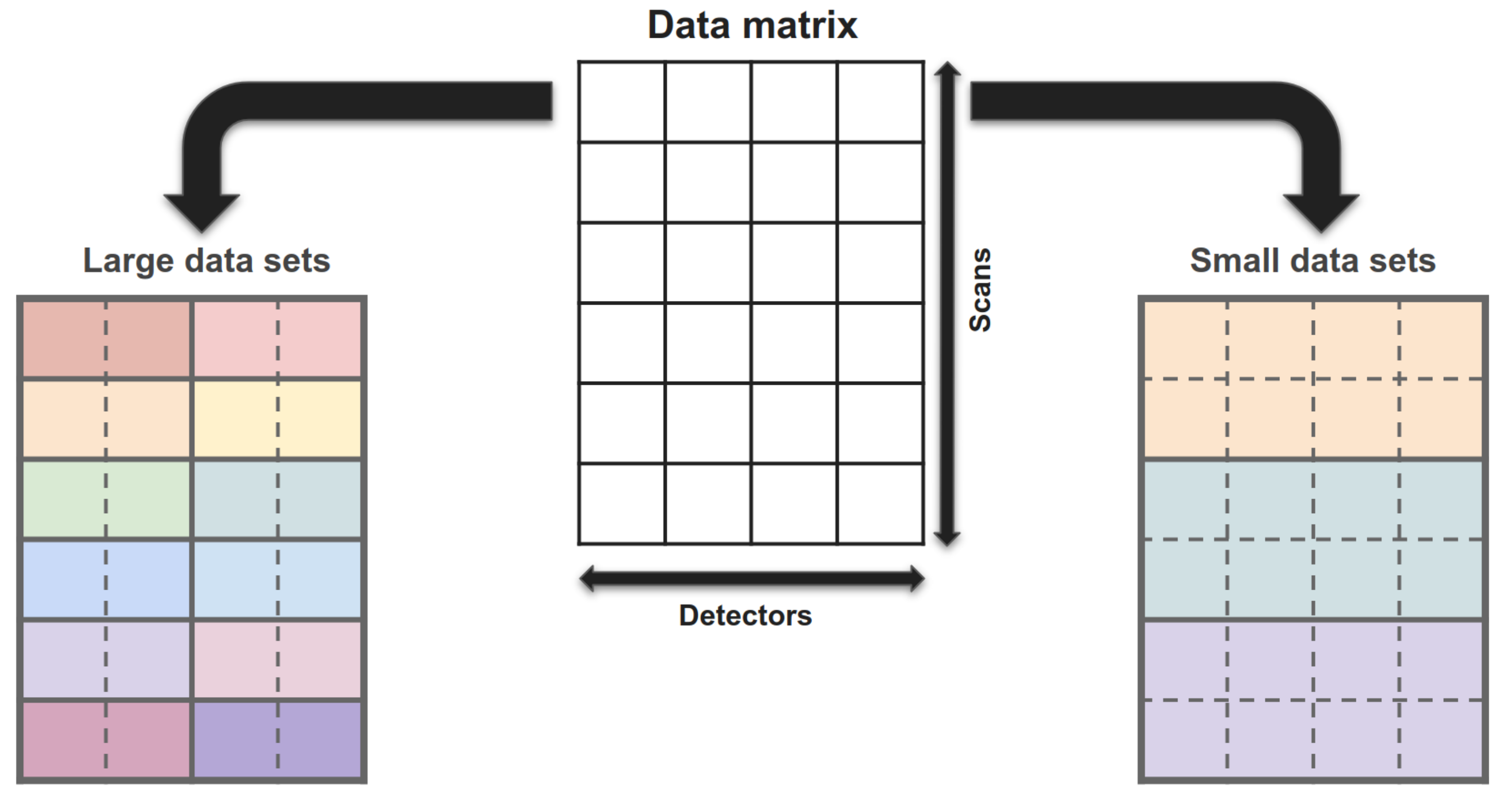}
\caption{\label{data_distri} 
Distribution scheme of the TOD depending on the processed data set size. Each rectangular block corresponds to one detector time stream for the full duration of a scan. The colors represent the assignment of each block of data to a given MPI process. }
\end{figure}

Typically, we can consider two regimes depending on the size of the data. In the small data sets regime (small number of detectors, short scans, low sampling rate), each MPI process can hold the timestreams of all detectors for a given number of scans. In the large data sets regime (large number of detectors, long scans, high sampling rate), each MPI process can hold only the timestreams of a single scan and only for a subset of detectors. To illustrate these schemes, we show a toy example of this data distribution in figure~\ref{data_distri}. The choice of not splitting scans of a single detector between multiple processes avoids unnecessary communication between neighbouring processes either during noise weighting or time-domain filtering operations.

\subsubsection{MAPMAT: pointing algebra package}
\noindent \textit{\textbf{Pointing data distribution \& pixelization schemes:}}
The pointing data distribution follows from the distribution scheme adopted for the timestreams: each process is assigned the pixels and pointing weights corresponding to the detector timestreams it is processing. This also defines the distribution of pixel-domain objects, such as maps. As a result, given the high redundancy of the typical scanning strategy, information concerning a given pixel may be stored in the memory of multiple processes. The pixels indices are defined via two indexation schemes. A global pixel indexation, corresponding to the global pixel index on the sky, and a local pixel indexation, corresponding to a reordering of the pixels observed at least once by the local timestreams of each MPI process.
The package is independent of the particular choice made for the sky pixelization.\\

\noindent \textit{\textbf{Data structure:}}
The pointing matrix is defined with a C-struct called \texttt{Mat}. Each MPI process holds an instance of this structure in memory. One integer,  \texttt{nnz}, specifies the number of non-zero elements per row. The values and columns locations of these elements are specified with two arrays as is done in the ELL\footnote{\scriptsize a storage format originally used by the ELLPACK package, storing sparse matrices only via their non-zero entries and their corresponding column numbers. } storage format, for efficient memory consumption. It also contains the ordered set of locally observed pixels, and a precomputed set of arrays and parameters necessary to conduct the customized collective communication operations in an efficient way, along with a flag specifying the particular communication scheme chosen for these operations.\\

\noindent \textit{\textbf{Projection operations \& collective communication:}}
The package features two main core operations:
\begin{itemize}
    \item \textbf{The un-pointing operation:} This operation generates a time-domain vector, $\mathbf{t}$, from a given map-domain object, $\mathbf{m}$, and a pointing matrix $\mathbf{P}$: $\mathbf{t}=\mathbf{P}\,\mathbf{m}$. Given the data distribution scheme, this operation can be performed locally in each MPI process, as a simple sum of the products of the pointing weights with the corresponding map entries.
    \item \textbf{The pointing operation:} This is a projection operation, where given a time-domain vector, $\mathbf{t}$, and a pointing matrix $\mathbf{P}$, we compute a map vector $\mathbf{m}$, where each pixel is the sum of the time samples observing it: $\mathbf{m} = \mathbf{P^\mathrm{T}}\,\mathbf{t}$. Given that the data are distributed in time-domain, the time samples observing a given pixel, are typically distributed over many MPI processes. Hence this operation is performed in two steps: we first compute local products of the pointing weights with the local time samples, $\boldsymbol{m_i}$. This results in submaps of the full map which contain only a subset, $\mathcal{P}_i$, of pixels, which is relevant to the time-domain data assigned to this process. We then compute the sum of these local submaps over all processes obtaining the global map projection, i.e., 
    $$
    \mathbf{m}(p)\, = \, \sum_{i: \; p \, \in\,  \mathcal{P}_i}\, \boldsymbol{m_i}(p),
    $$ 
    where for a pixel $p$ of the full map, the sum goes over those processes, $i$, for which $p$ belongs to the local set of pixels, $\mathcal{P}_i$.
    
    The second step requires a global, irregular communication operation. Given that these types of operations are not scalable, this represents a bottleneck in the map-making procedure. In addition to the standard \texttt{MPI$\_$Allreduce}, the MAPMAT package implements several customized communication algorithms such as Ring and Butterfly~\cite{pachenko11}.
    
\end{itemize}

\subsubsection{Toeplitz algebra package}
\noindent \textit{\textbf{Data distribution:}}
For each detector timestream, we build a Toeplitz block that is completely defined by one array containing a number of elements equal to the half-bandwidth $\lambda$. These blocks are therefore distributed following the time-domain data layout: each MPI process holding in memory the Toeplitz blocks corresponding to the detectors and the scans it has been assigned. In case, we want to include correlations between detectors, we would need to also build cross-correlation Toeplitz blocks, and hold them in memory, and eventually need a significant amount of communication depending on the spatial scale of the cross-correlations.\\

\noindent \textit{\textbf{Data structure:}}
The Toeplitz-structured noise covariance is defined with a C-struct called \texttt{Tpltz}. Each MPI process holds an instance of this structure in memory. It is parametrized with the total number of rows, the number of Toeplitz blocks both in total and in the local memory of the MPI process, the localization of the local segment in the global matrix, in addition to the size of the local timestreams, and an array of instances of a C-struct called \texttt{Block}. This latter structure defines each Toeplitz block, and contains the actual elements of the first row, the half bandwidth, the total size of the block and its localization in the global data structure. The \texttt{Tpltz} structure is also assigned an MPI communicator, since the product operations can also handle data of a single scan that is split between many MPI processes, hence requiring some point-to-point communication. However, in practice, it is the best to avoid distributing the data in such a way, reducing the communication overhead.\\

\noindent \textit{\textbf{Toeplitz product operations \& numerical algorithms:}}
The package provides routines to perform the product of a Toeplitz-struc\-tured matrix with a general data matrix, which is stored as vectors in the column-wise order. Three product operations can be performed:
\begin{itemize}
    \item \textbf{stmm routines:} these concern the multiplication of a symmetric banded Toeplitz matrix by a data matrix of the same size. If the data are distributed over several MPI processes, then each process should have the same and full Toeplitz matrix stored in memory.
    \item \textbf{stbmm routines:} for the multiplication of a symmetric, block-diagonal matrix, with Toeplitz banded blocks by a data matrix. As explained above, such structure is defined with a list of instances of the C-struct \texttt{Block}, each representing a Toeplitz block in the block-diagonal matrix. Each block is then multiplied with the corresponding subblock of the data matrix. We also note that the blocks need not be contiguous, i.e., we can have rows in the data matrix which correspond to no given block in the Toeplitz structure, such rows are copied unchanged implicitly assuming the corresponding entries in the Toeplitz structured matrix to be equal to $1$. Given the data layout adopted in practice, each MPI process will have a number of full detector timestreams, and therefore all the corresponding Toeplitz blocks. However, in general, as was discussed before the package also handles, detector timestreams being split between multiple MPI processes, in this case each of those MPI processes should have the corresponding full Toeplitz blocks stored in memory, and communication is used to copy necessary data to enable the product operation locally.
    \item \textbf{gstbmm routines:} these are similar to stbmm routines, but with some sets of rows and corresponding columns removed from the Toeplitz-stuctured matrix, we refer to these rows and corresponding columns as gaps. The routines perform a product in a similar way to the stbmm routines but neglecting the contributions from the columns in the gaps and setting the corresponding rows in the output to zero.
\end{itemize}

So far we have discussed, which operations may be called according to the data structure, but we have not discussed how the product operation with the noise covariance blocks is performed in itself. The algorithm is based on a \textit{shift-and-overlap} approach \cite{num_recipes_3rd}, where a product of a single band-diagonal Toeplitz matrix by a data matrix is done by dividing them into a set of overlapping blocks, and performing a sequence of products of the overlapping Toeplitz subblocks by the corresponding overlapping segments of the data matrix. Each of the latter products is performed in turn by embedding the Toeplitz subblock in a minimal circulant matrix and performing the multiplication via Fast Fourier transforms. The size of the subblock is set appropriately to optimize the computation and is typically a multiple of the half-bandwith $ \lambda$. The overall complexity of the operation is ${\cal O}( n \ln \lambda)$, where $ n $ is the size of the initial Toeplitz matrix. Performing the operation directly by embedding the full Toeplitz matrix into a circulant one that is twice as large and computing the product via FFTs would yield a complexity of ${\cal O}( 2n \ln{2n})$.

When communication is needed as in the cases we discussed before, it remains local, involving only neighboring processes. Each process needs to send to and receive from a neighboring process a vector of data of the length defined by the half bandwidth of the Toeplitz block, $ \lambda$, shared between them. This is sufficient to compute a part of the Toeplitz-vector product corresponding to the input data of each process locally without any need for further data exchanges. In particular we note that all the FFT calls used by the package are either sequential or threaded. The point-to-point communication pattern enabled by the package can be either \textit{non-blocking}, using \texttt{MPI$\_$Isend} and \texttt{MPI$\_$Irecv} calls, or \textit{blocking} with \texttt{MPI$\_$Sendrecv} calls. The choice is specified via a flag \texttt{FLAG$\_$BLOCKINGCOMM} determined when constructing the \texttt{Tpltz} structure, and set by default to 0 for non-blocking communication.

\subsubsection{Templates algebra package}
\noindent \textit{\textbf{Data distribution:}}
As described in Sect.~\ref{sect:framework:package} the templates considered here are specific to scans and detectors. Consequently each MPI process stores only templates relevant to the scans and detectors assigned to it as a result of the overall time-domain distribution scheme described earlier. This data distribution is both memory and computation efficient.\\

\noindent \textit{\textbf{Data structure:}}
At the onset of the computation, each MPI process is assigned its specific set of template classes. For each template class, each process stores information only about the templates corresponding to the detector time streams it is processing. 

Different template classes are defined with a C-struct called \texttt{TemplateClass}. The structure defines the time-domain interval for which this class is relevant as well as flags defining the type of templates in the template class as well as a stationary interval to which it corresponds.

For the templates defined implicitly, additional flags store the information about the parameters necessary to calculate the templates. For the templates defined in a compressed form, the structure defines the range of respective bins or pixels as well as two time-domain arrays, defining the mapping between the time indices and bin/pixel numbers and the corresponding weights.
 Given that templates are time-domain objects, they can have a significant memory footprint. Therefore, we implement two operating modes for the templates: the weights and bins can either be fully stored in memory at the moment of templates construction, if the processed data set is sufficiently small with a manageable memory footprint, or they can be computed on-the-fly while conducting the product operations with the templates matrix, for large data sets, hence trading speed for memory efficiency. Each MPI process holds an array of instances of \texttt{TemplateClass} in memory, and as such would have all necessary information to perform the time-domain filtering operations locally, without any need for communication with other processes.\\

\noindent \textit{\textbf{Templates algebraic operations:}}

The main utility of the routines provided by this package is to allow the application of the filtering operator, $\mathbf{F_T}$, to a time-domain vector. Three core routines form the basis for such an operation:
\begin{itemize}
    \item \textbf{Templates projection in time-domain:} This operation generates a time-domain vector, $\mathbf{t}$, from a given set of templates amplitudes, $\boldsymbol{x}$, and a templates matrix, $\mathbf{T}$: $\mathbf{t}=\mathbf{T}\,\boldsymbol{x}$. Each process loops over the set of scans it is assigned, and for each scan, loads a set of templates weights, which typically depend only on boresight pointing parameters and therefore do not vary for individual detectors in the same scan. Higher order templates weights are constructed recursively, by using for example recursive relations between Lengendre polynomials, or trigonometric functions. This allows us to avoid repeating the recursive computation for each detector time stream, which can significantly slow down the operation. Once the weights are stored in memory, we loop over the array of template classes: for each template we construct the array of bins on-the-fly, point to the already loaded weights, perform the series of products with the corresponding segment of templates amplitudes, and finish with freeing the memory segment occupied by the bins, and move on to the next template class. Once all the templates of the scan have been spanned, we remove the loaded weights from memory, and repeat the process for the next scan, until all scans are processed.
    \item \textbf{Time-domain projection in templates space:} This operation projects a time-domain vector, $\mathbf{t}$, in the templates amplitudes space, using a templates matrix $\mathbf{T}$: $\boldsymbol{y}=\mathbf{T^\mathrm{T}}\,\mathbf{t}$. The operation is performed in a similar way to the one above, with loops over scans and template classes, and templates bins and weights constructed on-the-fly before computing the products with the relevant segments of data.
    \item \textbf{Templates kernel construction and inversion:} What we refer to as templates kernel here, is the block-diagonal matrix $\mathbf{T\,W^{-1}\,T}$. Each block of this matrix is constructed solely from the set of templates that apply to a single detector time stream, as long as the templates do not model any common detector modes. As such the construction of this kernel as well as its inversion are considerably simplified. We loop over the scans, and the detectors, and for each detector in a given scan, we perform a double nested loop on the template classes associated with the detector time stream, weights and bins are computed on-the-fly depending on the template class, and we compute the kernel entries with a series of products between the different templates weights. The inverse of the kernel is simply the inverse of each block. The blocks are inverted with a singular value decomposition (SVD), where singular values below a certain threshold, typically $10^{-12}$, are set to zero, effectively regularizing the inversion. This is equivalent to an orthogonalization procedure, and therefore, spurious effects due to linear dependencies between templates are neutralized. Note that it is possible to explicitly exploit the inner structure of the kernel blocks to speed-up both their inversion and construction. For example, we know that Legendre polynomials or cosine and sine harmonics over a full period are orthogonal by definition, and would therefore yield a diagonal kernel, therefore we can ignore some off-diagonal entries while constructing the kernel, and exploit the structure of the kernel when performing the inversion. As in the current applications the kernel inversion did not constitute a significant fraction of the total runtime such optimizations are not included in the current release and are left for future work.
\end{itemize}

\subsection{The MIDAPACK parallel iterative sky estimator (MAPPRAISER)}
\subsubsection{Mapping methods}

As discussed in Section~\ref{sect:formalism}, Eq.~\eqref{eqn:genMapEstimator} represents a broad class of plausible map-making approaches which result in an unbiased estimate of the sky signal. However, implementing it in its full generality is cumbersome and the resulting code likely to be rather inefficient.
Instead, in the proposed framework we have implemented two highly optimized, specific cases of this general estimator, which taken together cover most of the cases of practical interest.

These two specific map-making methods are, 
\begin{enumerate}
\item the case where the weights $\mathbf{W}^{-1}$ are assumed to have the form of a Toeplitz block-diagonal matrix and the deprojection/weighting operator, $\mathbf{F_T}$, contains no templates, i.e., $\mathbf{F_T}=\mathbf{W}^{-1}$. The templates can be included as part of the generalized pointing matrix, see  Eq.~\eqref{eqn:fullEstimator}. We note that this method  allows for a (nearly) maximum likelihood solution for the map in cases when the time-domain noise is piece-wise stationary and uncorrelated between different detectors. Indeed, the imposed structure of the weights is flexible enough to allow for a very good approximation to the actual inverse noise covariance as discussed in Section~\ref{sect:formalism:weights}. The MAPPRAISER package computes the weights using information derived directly from the collected data as in~\cite{Stompor_2001, Prunet_2001, Wehus_2012}. To do so, we consider that over each scan, every detector timestream is dominated by stationary noise with a power spectral density (PSD) of the form:
\begin{eqnarray}
P(f)= \sigma_0^2\,\left(\,1\,+\,\left(\,\frac{f+f_0}{f_k}\,\right)^{\alpha}\,\right)
\label{eqn:noisepowermodel}
\end{eqnarray}
Assuming at first order that the data is only made of noise, we compute the average periodogram of each detector time stream, and fit it to the above model. The obtained PSD is then inverted and inverse Fourier transformed to obtain the inverse noise autocorrelation which is later smoothed with a gaussian window $\mathcal{N}(0,\lambda/2)$, and cut at the \textit{a priori} specified half bandwidth $\lambda$. The smoothing avoids any ringing effects, and the resulting autocorrelation represents the first row of the corresponding Toeplitz block of the inverse noise covariance matrix. In general, even if efforts are made to use as good an approximation as possible, the weights do not have to reflect the actual inverse noise covariance. In such cases the estimated map will be still unbiased but not optimal in the sense of ensuring the lowest possible uncertainty. Nevertheless, for simplicity, we refer to this approach hereafter as a maximum likelihood map-making; and,
\item a variant of the method which allows for arbitrary templates but assumes the weights to be diagonal. This method includes all the predefined templates provided in the software as described in~\ref{sect:appA}. We call this method in the following a template map-making.
\end{enumerate}

\subsubsection{MAPPRAISER solvers}
Linear systems can be solved using either direct or iterative methods. Direct methods typically compute some decomposition of the system matrix and consequently are time consuming scaling in general as a cube of the size of the system. Their advantage is that they give a (numerically) precise solution. Iterative methods form successive approximations that converge to the exact solution, and are particularly well-adapted to solving large sparse systems. They require only efficient routines for a product of the system matrix and an arbitrary vector, and are very efficient in terms of any extra storage. This first property is particularly relevant in the map-making case where the inverse of the system matrix cannot be typically explicitly computed due to both memory and computational cycle issues. Instead, the form of the system matrix allows for an efficient computation of a product of the matrix by a vector. Indeed, this can be implemented~\cite{Cantalupo_2010} by performing the operations from right to left. In the case of the maximum likelihood method, this involves a calculation of the matrix-vector product in the form,
\begin{eqnarray}
\mathbf{P}^\mathrm{T}\,\mathbf{W}\,\mathbf{P}\,\mathbf{x},
\end{eqnarray}
which can be done by performing three subsequent operations solely  on vectors: (1) unpointing; (2) a block-diagonal Toeplitz product; (3) pointing, all of which are implemented in the MIDAPACK library described earlier.
In the case of the template map-making the respective product reads,
\begin{eqnarray}
\mathbf{P}^\mathrm{T}\,\mathbf{F_T}\,\mathbf{P}\,\mathbf{x},
\end{eqnarray}
and again requires three operations performed one after another with the Toeplitz product above replaced now by an application of the deprojection/weighting operator, $\mathbf{F_T}$. This in turn requires a precomputation of the inverse kernel and then merely a series of template pointing and depointing operations all available in MIDAPACK.

MAPPRAISER provides two iterative solvers for the map-making equation. Both are based on the Conjugate Gradient approach. These are Krylov projection methods \cite{golub13} suitable for symmetric (Hermitian) positive definite matrices adapted to the specificity of the map-making problem. We discuss them below.

We note that other iterative solvers could be and have been applied in this context, e.g.~\cite{Dore_2001}, and some alternatives are actively investigated~\cite{Huffenberger_2018, Huffenberger_2021}.
\\

\noindent\textbf{Preconditioned conjugate gradient solver}\\
The first one is the Preconditioned Conjugate Gradient (PCG), which is used to solve Symmetric Positive Definite (SPD) systems.  This solver has already been extensively used in the literature to solve the map-making problem ~\cite{Natoli_2001, Yvon_2005, Cantalupo_2010, Szydlarski_2014}. We refer the reader to these papers for the technicalities of the method. Below we illustrate the algorithm of the PCG solver, where $\mathbf{M}$ is representing a preconditioner. The objective of applying a preconditioner is to reduce the condition number of the system matrix. In MAPPRAISER we implemented two kinds of preconditioners. These are the Block Diagonal preconditioner ($BD$) and the two level preconditioner ($2lvl$). The block-diagonal preconditioner defines the current standard in the field. It is easy to compute and in many cases performs already very well. The two-level preconditioner has been proposed and demonstrated for the map-making problem in~\cite{Szydlarski_2014} and extended to component separation in~\cite{Papez_2020}. It is more involved and requires a precomputation. To the best of our knowledge the MAPPRAISER implementation of this preconditioner is the very first and only existing implementation in actual map-making software. 
\begin{algorithm}
\hrule
\vskip 5pt
	\DontPrintSemicolon
Compute ${\mathbf{r}_0 = \mathbf{b} - \mathbf{A} \mathbf{x}_0}$, $\mathbf{z}_0 = \mathbf{M} \mathbf{r}_0$ and ${\mathbf{p}_0 = \mathbf{z}_0}$,\; 
\For {$k = 0, 1, \ldots, k_{max}$} {
	$\alpha_k = \langle\mathbf{r}_k, \mathbf{z}_k\rangle / \langle\mathbf{A} \mathbf{p}_k, \mathbf{p}_k\rangle$\;
	$\mathbf{x}_{k+1} = \mathbf{x}_k + \alpha_k \mathbf{p}_k$\;
	$\mathbf{r}_{k+1} = \mathbf{r}_k - \alpha_k \mathbf{A} \mathbf{p}_k$\;
	\lIf{$\Vert \mathbf{r}_{k+1} \Vert_2 < \varepsilon \, \Vert b \Vert_2$}{
	  stop}
	$\mathbf{z}_{k+1} = \mathbf{M} \mathbf{r}_{k+1}$\;
	$\beta_k = \langle\mathbf{r}_{k+1}, \mathbf{z}_{k+1}\rangle/\langle\mathbf{r}_k,\mathbf{z}_k\rangle$\;
	$\mathbf{p}_{k+1} = \mathbf{z}_{k+1} + \beta_k \mathbf{p}_k$\;
}
\vskip 5pt
\hrule
\vskip 5pt
\hskip -20pt \caption{The PCG algorithm for the system $\mathbf{A}\mathbf{x} = \mathbf{b}$ with a preconditioner $\mathbf{M}$ \cite{saad_2003}.}
\label{algo:pcg}
\end{algorithm}

\noindent\textbf{Enlarged conjugate gradient solver}\\
The second iterative solver is the Enlarged Conjugate Gradient (ECG) \cite{grigori_2016}. In this approach the matrix $\mathbf{A}$ is partitioned into $\mathcal{N}$ subdomains. Then the unknown vector $\mathbf{x}$ is split into $t$ vectors, $\mathbf{X}^{1 \leq i \leq t}$, such that we can still retrieve the original vector by summing them: $\mathbf{x} = \sum_{i=1}^\mathrm{T}\mathbf{X}^{(i)}$. The parameter $t$ is called the enlarging factor, and the method is in practice not sensitive to the choice of the splitting scheme \cite{Grigori_Tissot2019}. Following these considerations, the method can be derived in a similar way to the standard CG solver, replacing the residuals and search directions by $\mathcal{N}\times t$ matrices, and the optimal step which is a scalar in CG becomes a $t \times t$ matrix. The algorithm is given as Algorithm~\ref{algo:ecg}. Two different variants are implemented: Orthomin and Orthodir. They are equivalent in well behaved cases, however, the Orthodir method is more robust than Orthomin in cases where $\mathbf{Z}_k^\mathrm{T}\mathbf{AZ}_k$ is (nearly) singular. However, the Orthodir method is also more expensive than Orthomin as it requires twice as much memory and flops~\cite{Grigori_Tissot2019}. The square root of $\mathbf{Z}_k^\mathrm{T}\mathbf{AZ}_k$ is computed using a Cholesky factorization as it is faster than alternatives. The Cholesky decomposition can be preceded by a QR factorization of $\mathbf{Z}_k$ to avoid break downs of the Orthomin method in singular cases~\cite{LL_2014}. This is not implemented in the present software release however. 
\begin{algorithm}
\hrule
\vskip 5pt
	\DontPrintSemicolon
Compute ${\mathbf{r}_0 := \mathbf{b} - \mathbf{A} \mathbf{x}_0}$, and split it into a matrix with $t$ columns, $\mathbf{R}_0^e$, s.t. ${\mathbf{r}_0 = \sum_{i=1}^\mathrm{T}  \mathbf{R}_0^{e(i)}}$,\;
${\mathbf{Z}_1 = \mathbf{MR}_0^e}$,\; 
\For{$k = 1, 2, \dots, k_{max}$}{
    $\mathbf{P}_k = \mathbf{Z}_k\left(\mathbf{Z}_k^\mathrm{T}\mathbf{AZ}_k\right)^{-1/2}$\;
	$\boldsymbol{\alpha}_k = \mathbf{P}_k^T\mathbf{R}_{k-1} \quad \triangleright \, t \times t$ matrix\;
	$\mathbf{X}_{k} = \mathbf{X}_{k-1} + \mathbf{P}_k\boldsymbol{\alpha}_k$\;
	$\mathbf{R}_{k} = \mathbf{R}_{k-1} - \mathbf{AP}_k\boldsymbol{\alpha}_k$\;
    \lIf{$\Vert \sum_{i=1}^\mathrm{T}\mathbf{R}_k^{(i)}\Vert_2 < \varepsilon \, \Vert b \Vert_2$}{
	  stop}
    Orthomin:\;
	${\mathbf{Z}_{k+1}} = (\mathbf{I}-\mathbf{P}_k\mathbf{P}_k^T\mathbf{A})\mathbf{MR}_k$\;
	Orthodir:\;
	${\mathbf{Z}_{k+1}} = (\mathbf{I}-\mathbf{P}_k\mathbf{P}_k^T\mathbf{A}-\mathbf{P}_{k-1}\mathbf{P}_{k-1}^T\mathbf{A})\mathbf{MAP}_k$\;
}
${\mathbf{x}_k = \sum_{i=1}^\mathrm{T}\mathbf{X}_k^{(i)}}$.\; 
\vskip 5pt
\hrule
\vskip 5pt
\hskip -20pt \caption{The ECG algorithm for the system $\mathbf{A}\mathbf{x}=\mathbf{b}$ with a preconditioner $\mathbf{M}$ \cite{Grigori_Tissot2019}.}
\label{algo:ecg}
\end{algorithm}

The rate of convergence of the ECG is given by the following result demonstrated in \cite{Grigori_Tissot2019}: if $\mathbf{x}_k$ is the approximate solution given by the Enlarged Conjugate Gradient with an enlarging factor $t$ at step $k$, and $\mathbf{x}_*$ the true solution then we have
\begin{eqnarray}
\|\mathbf{x}_k\,-\,\mathbf{x}_*\|_{\mathbf{A}}^2 \, \leq \, C \, \bigg(\frac{\sqrt[]{\kappa_t}-1}{\sqrt[]{\kappa_t}+1}\bigg)^{2k},
\label{eqn:RateConvECG}
\end{eqnarray}
where $\kappa_t=\lambda_n/\lambda_t$, the ratio of the largest eigenvalue of $\mathbf{A}$, and its $t$ smallest eigenvalue, and $C$ a constant independent of $k$. As such the solver behaves as a deflated-CG \cite{Saad_2000}, and converges faster than CG. However, the rate of convergence is not a sufficient metric to assess the performance of the method. In practice, a more relevant figure of merit is instead time to solution.
In the case of ECG the expanded map space leads to an increase of the number of operations required on each iteration of the solver by the enlarging factor $t$, therefore a trade-off needs to be found between the gain in the rate of convergence and the linear increase in the algorithm's complexity. In the scaling experiments we conducted which results are given in Sect.~\ref{sect:solvers-tests}, we adopt a naive implementation, where the product of the matrices $\mathbf{A}$ and $\mathbf{X}$ is computed by looping on the columns of the enlarged map, and performing the standard CG $\mathbf{A}\,\mathbf{x}$ operation for each column. This baseline implementation of the solver demonstrates the improvement in the number of iterations with increasing the enlarging factor as theoretically predicted. The gain in terms of the number of iterations is however sublinear and does not compensate for the increased cost of each iteration. Consequently, the overall run time increases. We expect this to change with an improved approach to performing the product of the system matrix and expanded map. One particularly promising avenue is to capitalize on GPU-accelerators. This is left for future work. As is, the main interest of the ECG solver is in the fact that it does not require sophisticated preconditioning and is nevertheless very efficient in the cases when the system matrix includes a number of small eigenvalues.

\subsubsection{Preconditioners for PCG}

The objective of preconditioning is to improve the convergence and stability of the iterative solvers.
The preconditioner, $\mathbf{M},$ can be thought of as an approximate to the inverse of the system matrix, $\mathbf{A}$. In preconditioned solvers, instead of the original problem we solve a modified system with the system matrix replaced by a product of the system matrix  and the preconditioner, i.e., $\mathbf{M}\,\mathbf{A}$. Unlike the actual inverse of the system matrix the preconditioner should be easy to compute, and  the preconditioned system matrix, $\mathbf{M}\,\mathbf{A}$ should be better conditioned and its eigenspectrum more clustered. Both of these properties make the system more stable and speed-up the convergence. The MAPPRAISER package offers two preconditioning options. \\

\noindent \textbf{Block-Jacobi preconditioner:}\\
The first preconditioner is the standard Block-Jacobi preconditioner defined as,
\begin{eqnarray}
\mathbf{M}_{BD}\,=\,(\mathbf{P^\mathrm{T}}\,\mathrm{diag\,}\mathbf{N}^{-1}\,\mathbf{P})^{-1},
\end{eqnarray}
where the inner most matrix denotes a diagonal matrix given by the diagonal of the inverse noise covariance matrix. This preconditioner is block-diagonal with square diagonal blocks sizes of which are determined by the number of non-zero elements per row of the pointing matrix. The construction of this preconditioner is done in two steps. First, for each pixel we sum over the (non-zero) elements of all the rows of the pointing matrix corresponding to this pixel while weighting them by the diagonal noise weights. On the second step, we invert each of the blocks using a direct method. On this second step, we also evaluate a condition number of each block, and use it as a criterion to excise pixels with ill-conditioned blocks from the map-making.\\

\noindent \textbf{Two level preconditioner:}\\
The first level of the two-level preconditioners considered in MAPPRAISER is based on the deflation technique. Its purpose is to suppress (deflate) an unwanted subspace of a matrix, so as a result it is contained in its null space. In the context of preconditioning this is ideally the subspace which contains all the smallest eigenvalues of the system matrix. The second level then consists of adding an extra contribution to the deflated matrix in order to shift the eigenvalues of the deflation subspace to unity.
There are different variants of the two-level preconditioners which have been shown to be efficient in different applications. Here, we follow the proposal of \cite{Szydlarski_2014} and focus on the so called ``Adapted Deflation Variant 1'' (A-DEF1) option in ~\cite{Tang_2009}. The preconditioner is defined as follows,
\begin{eqnarray}
\mathbf{M}_{2lvl}
& = & \mathbf{M}_{BD}\,(\mathbf{I}\,  - \, \mathbf{A}\,\mathbf{Q})\,  + \, \mathbf{Q},
\label{eqn:2lvlDef}
\\
\mathbf{Q} & = & \mathbf{Z}\,(\mathbf{Z}^\mathrm{T}\,\mathbf{A}\,\mathbf{Z})^{-1} \mathbf{Z}^\mathrm{T},
\end{eqnarray}
where $\mathbf{Z}$ is a deflation subspace matrix. Its columns span a subspace to be deflated. This is a tall and skinny matrix with a limited number of columns each of which are pixel-domain vectors. The number of columns of $\mathbf{Z}$ defines the volume of the deflation subspace. $\mathbf{A}$ stands for our system matrix.
The first term in Eq.~\eqref{eqn:2lvlDef} defines the first level of the preconditioner. Indeed, we have that $(\mathbf{I}\,  - \, \mathbf{A}\,\mathbf{Q})\,\mathbf{A}\,\mathbf{Z} \, = \, 0$. The second term is the second level correction.

A key element determining the performance of the two level preconditioner is a construction of the deflation subspace matrix, $\mathbf{Z}$. The MAPPRAISER implementation of the PCG solver with the two level preconditioner accepts any user provided matrix $\mathbf{Z}$. In addition, MAPPRAISER includes functionality which allows for calculating the deflation matrix based on two different approaches. We describe those in the following. 

We note that the two-level preconditioner is in principle a dense, square matrix with the dimension determined by the number of the observed sky pixels. For this reason we never explicitly compute it as in many applications this would be prohibitive both from the point of view of the required computations and the computer memory. Instead, 
for any available deflation matrix, $\mathbf{Z}$, we only precompute matrix $\mathbf{Z'} \, \equiv\, \mathbf{A}\, \mathbf{Z}$, the so-called coarse matrix, $\mathbf{E} \equiv  \mathbf{Z}^\mathrm{T}\,\mathbf{A}\, \mathbf{Z}\, = \, \mathbf{Z}^\mathrm{T}\,\mathbf{Z'}$, and its inverse. Computing $\mathbf{Z'}$ is computationally demanding as we have to apply the system matrix to as many vectors as the number of columns of the matrix $\mathbf{Z}$. As each such product is also a dominant computational cost of each iteration of the PCG solver, this computation is as costly as performing as many iterations as the volume of the deflation subspace. As we discuss later this cost can be however effectively hidden in some of the procedures aiming at the calculation of the deflation matrix, $\mathbf{Z}$. This is indeed the case in one of the specific constructions implemented in MAPPRAISER. The matrix $\mathbf{E}$ is then computed by calculating $\mathbf{Z}^\mathrm{T}\,\mathbf{Z'}$, which entails negligible cost. 
The coarse matrix is in most of the cases a small, dense matrix. We therefore can compute its inversion with help of  standard, highly efficient, dense linear algebra routines. If it is known a priori that the columns of $\mathbf{Z}$ are all linearly independent, and therefore the coarse operator is positive definite this could be done very efficiently using the Cholesky decomposition. Otherwise, one has to use a pseudo-inverse, which relies on the calculation of the full  eigenspectrum decomposition of the coarse matrix and is therefore appropriately more costly. In our implementation we opted for the second approach as the calculation of the inverse is in any case subdominant. We use the \emph{dgelss} LAPACK routine in the MKL implementation for this purpose. All these operations have to be performed only once at the beginning of the solution. In addition we need to precompute the block-Jacobi preconditioner following the procedure outlined earlier at negligible cost.

The preconditioner has to be applied once at every step of the PCG iterations. This involves a computation of a product of  the preconditioner and some arbitrary pixel-domain vector. For this purpose we represent the preconditioner in Eq.~\eqref{eqn:2lvlDef} as,
\begin{eqnarray}
\mathbf{M}_{2lvl}
& = & \mathbf{M}_{BD}\,  - \, \mathbf{M}_{BD}\, \mathbf{Z'}\,\mathbf{E}^{-1} \mathbf{Z}^\mathrm{T}\,+\, \mathbf{Z}\,\mathbf{E}^{-1} \mathbf{Z}^\mathrm{T}.
\label{eqn:2lvlDef0}
\end{eqnarray}
We apply the preconditioner by multiplying all factors of each of its terms sequentially, from right to left to a vector capitalizing on precomputed and stored in memory objects. All these objects are either pixel-domain or deflation-subspace objects and the required operations are respective matrix-vector products which require at most $\mathcal{O}(\mathrm{\mathcal{N}_{pix}\,dim}\,\mathbf{Z})$ floating point operations. Unless $\mathrm{dim}\,\mathbf{Z}$ is very large this extra cost is usually subdominant to the cost of applying the system matrix to a vector.  
Each column of the matrix $\mathbf{Z}$ is distributed between the MPI processes in the same way as any other map. This saves the memory but requires one extra global communication call of the \texttt{allreduce} type per iteration. The size of the reduced data is given by the dimension of the deflation space and the time overhead therefore negligible.
Consequently, the total cost of the application of the preconditioner is subdominant and the application of the system matrix to a vector remains the dominant cost of a single iteration.  Therefore, the time of a single iteration is essentially the same for the two-level and block-Jacobi preconditioners.
\\
MAPPRAISER provides two approaches to calculating the deflation subspace matrix, $\mathbf{Z}$. These are,
\begin{itemize}
    \item \textbf{2lvl a priori construction}\\
This construction is based on the structure of the data set, which is assumed to be divided into a number of time intervals (such as noise stationary intervals corresponding to the blocks of the noise covariance, $\mathbf{N}$). This construction is applicable only if total intensity is part of the data model. In this case we calculate the number of observations of pixel $p_i$ during the $j$-th interval, denoted as $s_i^j$. $s_i$ is then the total number of observations of pixel $p_i$. The $(i,j)$ element of the deflation matrix, $\mathbf{Z}$, which corresponds to the total intensity signal is given by the fractional number of observations of the pixel $p_i$ in the $j$-th interval, i.e., $\mathbf{Z}^{tint}_{ij} = s_i^j/s_i$. 
Consequently every such row of $\mathbf{Z}^{tint}$ represents a partition of the unity, since $\sum_j \mathbf{Z}^{tint}(p, j) = 1$. The rows corresponding to $Q$ and $U$ Stokes parameters are set to $0$ ~\cite{Szydlarski_2014}.\\
 
 \item \noindent\textbf{2lvl a posteriori construction}\\
In this construction we attempt to estimate the relevant eigenpairs of the system matrix, $\mathbf{A}$, We denote these as $(\theta, \mathbf{y})$ standing for an eigenvalue and the corresponding eigenvector, respectively. We use the Lanczos algorithm~\cite{golub13, saad_2003}---given explicitely below in algorithm~\ref{algo:lanczos}---to calculate an $m$-by-$m$ tridiagonal matrix, $\mathbf{T}_m$ and an $\mathcal{N}_{pix}$-by-$m$ column orthonormal matrix $\mathbf{V}_m$ such as,
\begin{eqnarray}
\mathbf{A} \mathbf{V}_m & = & \mathbf{V}_m\mathbf{T}_m.
\label{eqn:lanczosDecomp}
\end{eqnarray}
The Lanczos algorithm is an iterative procedure and $m$ defines the number of the iterations. It is bound to be smaller than the size of the system matrix. In our case, $m \ll \mathcal{N}_{pix}$.
The algorithm is applicable to any symmetric positive-definitive (SPD) matrix and requires only matrix-vector products.  Here, these are implemented using the same routines as used by the PCG algorithm. Given the structure and the limited size of the matrix $\mathbf{T}_m$, its eigen-structure can be straightforwardly calculated. Given Eq.~\eqref{eqn:lanczosDecomp}, if $(\theta, \mathbf{x})$ is an eigenpair of $\mathbf{T}$ then $(\theta, \mathbf{V}_m\,\mathbf{x})$ is a corresponding eigenpair of $\mathbf{A}$. Therefore $\mathbf{V}_m\,\mathbf{x}$ define the columns of the deflation matrix, $\mathbf{Z}$. We take all the precomputed vectors, given that the cost of applying the two level preconditioner is subdominant.

We note that the Lanczos algorithm computes internally products of the system matrix and columns of the matrix $\mathbf{V}_m$, which are usually discarded. We store those, i.e., the matrix $\mathbf{A}\,\mathbf{V}_m$, together with the matrix $\mathbf{V}_m$ and use them to compute directly the matrices $\mathbf{A}\,\mathbf{Z}$ and $\mathbf{Z}$. This saves significant precomputation time needed to construct the preconditioner as discussed earlier.

This procedure is a special case of a more general procedure proposed in~\cite{Szydlarski_2014} and employed in~\cite{Puglisi_2018}, which uses the so-called GMRES algorithm, which is more numerically stable and applicable to a general matrix. We find the Lanczos procedure to be more computationally efficient and sufficiently robust in our applications. The main difference between the two approaches is that our procedure is applied to the system matrix and not to the system matrix preconditioned by the block-Jacobi preconditioner as in~\cite{Szydlarski_2014, Puglisi_2018}.

\begin{algorithm}
\hrule
\vskip 5pt
	\DontPrintSemicolon
Compute ${\mathbf{r}_0 \,=\, \mathbf{b} - \mathbf{A}\,\mathbf{x}_0}$, $\beta_1 \,=\,||\mathbf{r}_0||_2$, and $\mathbf{v}_1\,=\,1/\beta_1\mathbf{r}_0$\; 
\For{$k = 1, 2, \dots, m$}{
    $\mathbf{w}_k\,=\,\mathbf{A}\,\mathbf{v}_k-\beta_k\,\mathbf{v}_{k-1}$ (if $k=1$ set $\beta_1\,\mathbf{v}_0 \equiv \mathbf{0}$)\;
    store $\mathbf{A}\,\mathbf{v}_k$\;
    $\alpha_k \,=\, \langle \mathbf{w}_k,\,\mathbf{v}_k \rangle$\;
    $\mathbf{w}_k \,:=\, \mathbf{w}_k - \alpha_k\,\mathbf{v}_k$\;
    $\beta_{k+1} = ||\mathbf{w}_k||_2$\;
    \lIf{$\beta_{k+1} < \varepsilon$}{$m := k$ and break loop\;}
    $\mathbf{v}_{k+1} \,=\, 1/\beta_{k+1}\,\mathbf{w}_{k}$\;
}
return $\mathbf{T}_m \,=\, \text{tridiag}(\beta_i,\,\alpha_i,\,\beta_{i+1})$, $\mathbf{V}_m \,=\,\left[\,\mathbf{v}_1,\,\dots,\,\mathbf{v}_m\right]$, and $\mathbf{A}\,\mathbf{V}_m$.\; 
\vskip 5pt
\hrule
\vskip 5pt
\hskip -20pt \caption{The Lanczos algorithm~\cite{saad_2003} for the matrix $\mathbf{A}\,=\,\mathbf{P}^\mathrm{T}\,\mathbf{W}^{-1}\,\mathbf{P}$. The threshold, $\varepsilon$, is set to machine precision, and was never reached in our applications.}
\label{algo:lanczos}
\end{algorithm}
\end{itemize}
We compare advantages of different solvers in detail in the following sections. To quantify the quality of the current iterate, $\mathbf{x}_i$, we first define a residual, which for a linear system, $\mathbf{A}\,\mathbf{x}\,=\,\mathbf{b}$, is given by,
\begin{equation}
    \mathbf{r}_i \, = \, \|\mathbf{b} \, - \, \mathbf{A}\,\mathbf{x}_i\|.
    \label{eqn:CGresidual}
\end{equation}
If $\mathbf{x}_0 = 0$, as it is always the case in our test runs, then $\mathbf{r}_0 = \mathbf{b}$.
For the convergence we require that the relative residual, $\|\mathbf{r}_i\|/\|\mathbf{b}\| \, = \, \|\mathbf{r}_i\|/\|\mathbf{r}_0\|$, is smaller than some positive threshold value, $\epsilon$. A typical value of $\epsilon$ used in the map-making applications, and therefore also adopted here, is $\epsilon = 10^{-6}$. A key performance metric is consequently a number of the PCG iterations needed to convergence. We denote it, hereafter, by $n_{iter}$ with a superscript added to highlight the preconditioner choice.

As a rule of thumb, we note that each iteration of the PCG algorithm is as costly as applying the system matrix to a single column of the deflation matrix, $\mathbf{Z}$, and as costly as a single iteration of the Lanczos procedure.
Solving the map-making problem using the block-Jacobi preconditioner requires $n_{iter}^{BD}$ products of the system matrix times a vector. For two-level preconditioners, this amounts to $n_{iter}^{2lvl}\,+ \, \mathrm{dim}\,\mathbf{Z}$ where the second term is due to the Lanczos iterations for the a posteriori preconditioner and the application of the system matrix to the deflation matrix in the case of the a priori preconditioner.
All the other costs are then typically subdominant. 
If the same system or similar systems (see~\cite{Puglisi_2018}) need to be solved multiple times the respective costs are: $n_{solves}\,n_{iter}^{BD}$ and $n_{solves}\,n_{iter}^{2lvl}\,+ \, \mathrm{dim}\,\mathbf{Z}$. 
Obviously the key parameter is the number of iterations required to reach desired precision for different types of the preconditioners and different dimensions of the deflation space. 
It can be argued on theoretical grounds that $n_{iter}^{BD} \ge n_{iter}^{2lvl}$ independently of the deflation matrix~\cite{Szydlarski_2014}. This is because the two-level preconditioner cannot make worse the condition number of the preconditioned system matrix. The gain can be however very significant if $\mathbf{Z}$ encompasses the smallest eigenvalues of the system matrix. In such cases, the precomputation time can be offset very quickly if multiple, similar map-making problems have to be solved, leading to substantial overall net gain for the full computation. In fact, whenever the number of iterations per solve is known, we can estimate the minimal number of the map-making solves required to reach this regime.
We investigate this in detail in the next section.

We also note that the deflation matrix computed with the Lanczos algorithm encodes useful information about the system matrix and therefore the error covariance of the estimated map. This information can be useful on the subsequent steps of the analysis.

\section{Applications}
\label{sect:applications}

In this section, we present some of the applications that we have applied our framework to, and demonstrate the computational and scaling performance of the software. 
We emphasize that it is not our goal here to discuss the scientific impact of the obtained results. This is left for future work.

We consider two cases: one with partial sky coverage as expected from ground-based experiments and another with a full sky coverage as expected from a satellite mission.

We run the simulations on two different HPC platforms: 
\begin{itemize}
    \item \textbf{Cori}: This is a Cray XC40 machine based at NERSC\footnote{\scriptsize National Energy Research Scientific Computing Center: \url{https://www.nersc.gov}}. It can reach a peak performance of about 30 petaflops. It has two partitions, the first one is made of 2,388 Intel Xeon ``Haswell'' processor nodes, and the second one is made of 9,688 Intel Xeon Phi ``Knight's Landing'' (KNL) nodes. In our work we use the ``Haswell'' partition. Each Haswell node has two sockets, each of them populated with a 2.3 GHz 16-core Haswell processor, therefore allowing the use of 32 cores per node. Each core supports 2 hyper-threads, and has two 256-bit-wide vector units, allowing a theoretical peak of 36.8 Gflops/core. In terms of available memory, each node has a 128 GB DDR4 2133 MHz memory. 
    \item \textbf{Joliot-Curie}: This is a Bull Sequana machine based at TGCC\footnote{\scriptsize Très Grand Centre de calcul du CEA (The French Alternative Energies and Atomic Energy Commission Computing Center): \url{http://www-hpc.cea.fr/fr/complexe/tgcc.htm}}. The supercomputer features four partitions, however in our work we use a single one; the ``SKL'' partition, which is made of 1,656 nodes, each populated with 2.7 GHz Intel Skylake 8168 bi-processors, each processor hosting 24 cores, for a total of 79,488 cores and a total peak performance of 6.86 Pflops. Each node has a 192 GB DDR4 memory.
\end{itemize}

\subsection{\textbf{Case of a ground-based experiment}}
\subsubsection{Simulations description}
We simulate CMB observations using the TOAST\footnote{\label{foot:Toast}\scriptsize \url{https://github.com/hpc4cmb/toast}} framework. In this section we describe the results obtained with TOAST simulations of a typical ground-based experiment with specifications roughly following those of the Simons Observatory (SO)~\cite{SO_instru, SO_goals}. 
We simulate two sets of observations each comprised of data collected by roughly $6,000$ detectors of an SO-like, Small Aperture Telescope (SAT) operating at 150 GHz, 
\begin{itemize}
    \item \textbf{S1:} corresponds to the observation of a $20 \times 10 \deg^2$ sky patch. The observations consist of 33 constant elevations scans (CESs), each between 45 min and 1 h long, totaling to about 24 hours of observation. The sampling rate is set to $132$ Hz, yielding a total of $\mathcal{O}(10^{11})$ time samples, i.e., $\sim$10 M time samples per detector. The time-domain objects required for this run have a total size of about $\sim 5$~TB worth of memory. During the data staging process from TOAST to MAPPRAISER, the different detector data are concatenated at the level of each MPI process, thereby further constraining the minimum memory requirements as the high-water mark reaches around twice the total size at about $\sim 10$~TB. We use this simulation to evaluate the scaling performance of the solvers.
    \item \textbf{S2:} corresponds to the observation of a smaller sky patch of $10 \times 10 \deg^2$. The observations consist of 95 CESs, spanning 5 consecutive days, each about 15 min long, for a total of 24 hours of observation. The sampling rate is set to $37$ Hz, yielding a total of $\mathcal{O}(10^{10})$ time samples, i.e., $\sim$1 M time samples per detector. The memory requirement for this run is about 10 times lower than the previous case at $\sim 1$~TB. We use this simulation to study the convergence of the two-level preconditioners and to validate the templates marginalization procedure.
\end{itemize}
The scanning speed is $1 \deg/$s in both simulations, and the half wave plate rotation frequency is set to $2$ Hz. The input map of the simulations is pixelized in HEALPix\footnote{\scriptsize \url{http://healpix.sourceforge.net}} format \cite{Gorski_2005, Zonca2019} with a resolution corresponding to nside = 512, and includes a CMB realization from best-fit Planck cosmology~\cite{Planck_2018_cosmo} and a Gaussian lensing potential. The resulting sky patch from \textbf{S1} counts around $\sim 250,000$ pixels, while the one from \textbf{S2} counts about $\sim 140,000$ pixels. In addition, we simulate instrumental 1/f noise and atmosphere. The instrumental noise has a characteristic knee frequency of 50 mHz ($f_k$ in Eq.~\eqref{eqn:noisepowermodel}), a slope of $\sim -1$ ($\alpha$ in Eq.~\eqref{eqn:noisepowermodel}), and a detector NET of 400 $\mu$K$\sqrt{s}$. The native atmospheric simulation built in the TOAST framework is based on a 3d turbulence model~\cite{Errard_2015} sensitive to parameters such as wind properties, weather conditions, and turbulence characteristics, calibrated on real data from experiments in Chile such as POLARBEAR~\cite{PB_SPIE} and ACT~\cite{ACT}. A 3d volume is generated and propagated according to the wind direction and speed as it is scanned by the detectors, thereby generating correlations between time samples as well as cross-correlations across the focal plane between detectors.

\subsubsection{Strong scaling tests and collective communication performance}

\begin{figure}[h!]
\centering
\includegraphics[width=0.48\textwidth]{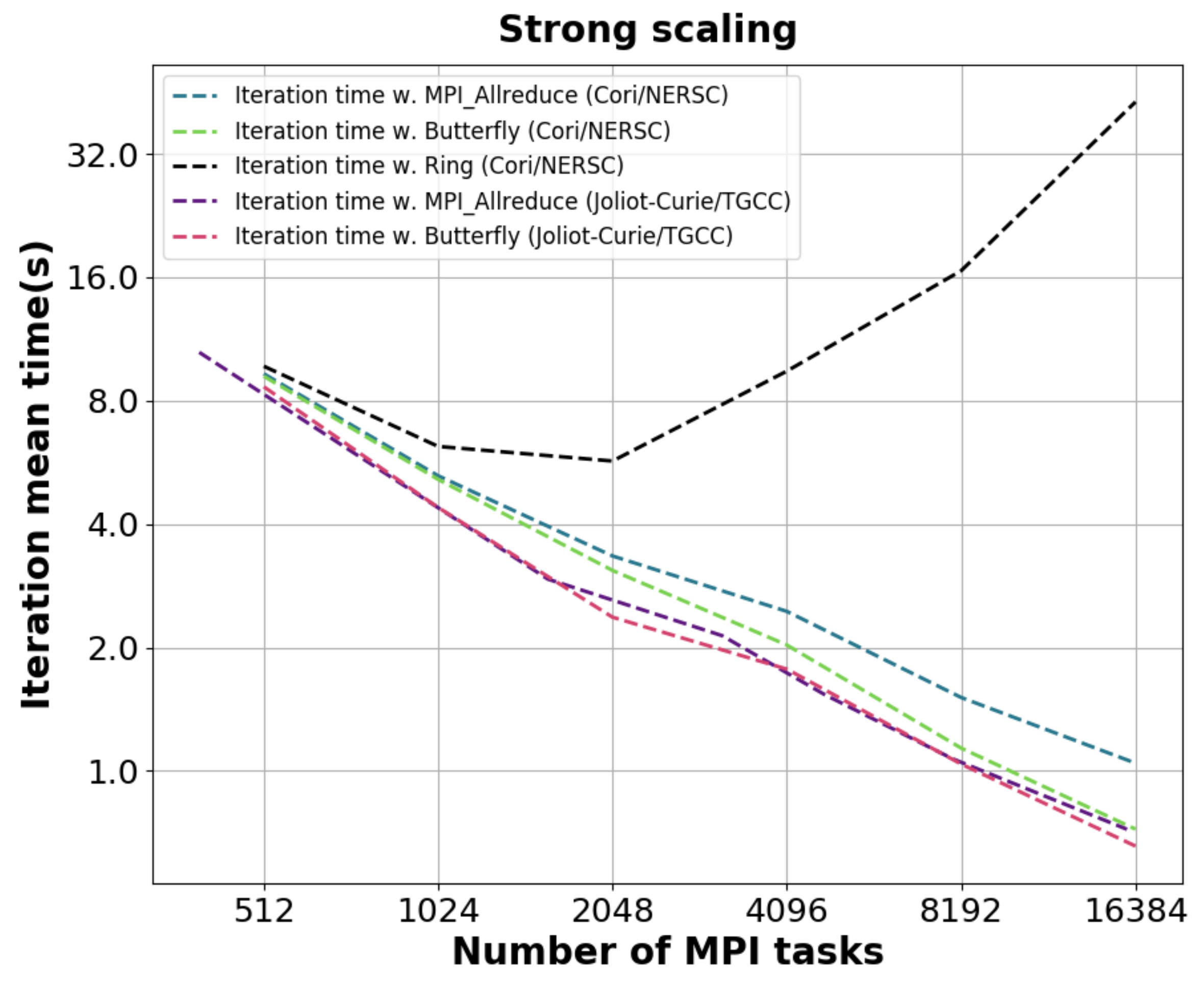}
\caption{\label{iteration_scaling} 
Strong scaling of the mean time per iteration comparing 5 sets of runs: (1) runs executed on Cori at NERSC using the \texttt{MPI\_Allreduce} communication scheme; (2) runs executed on Cori at NERSC using the Butterfly communication scheme; (3) runs executed on Cori at NERSC using the Ring communication scheme; (4) runs executed on Joliot-Curie at TGCC using the \texttt{MPI\_Allreduce} communication scheme; (5) runs executed on Joliot-Curie at TGCC using the Butterfly communication scheme.}
\end{figure}

\begin{figure}[h!]
\centering
\includegraphics[width=0.48\textwidth]{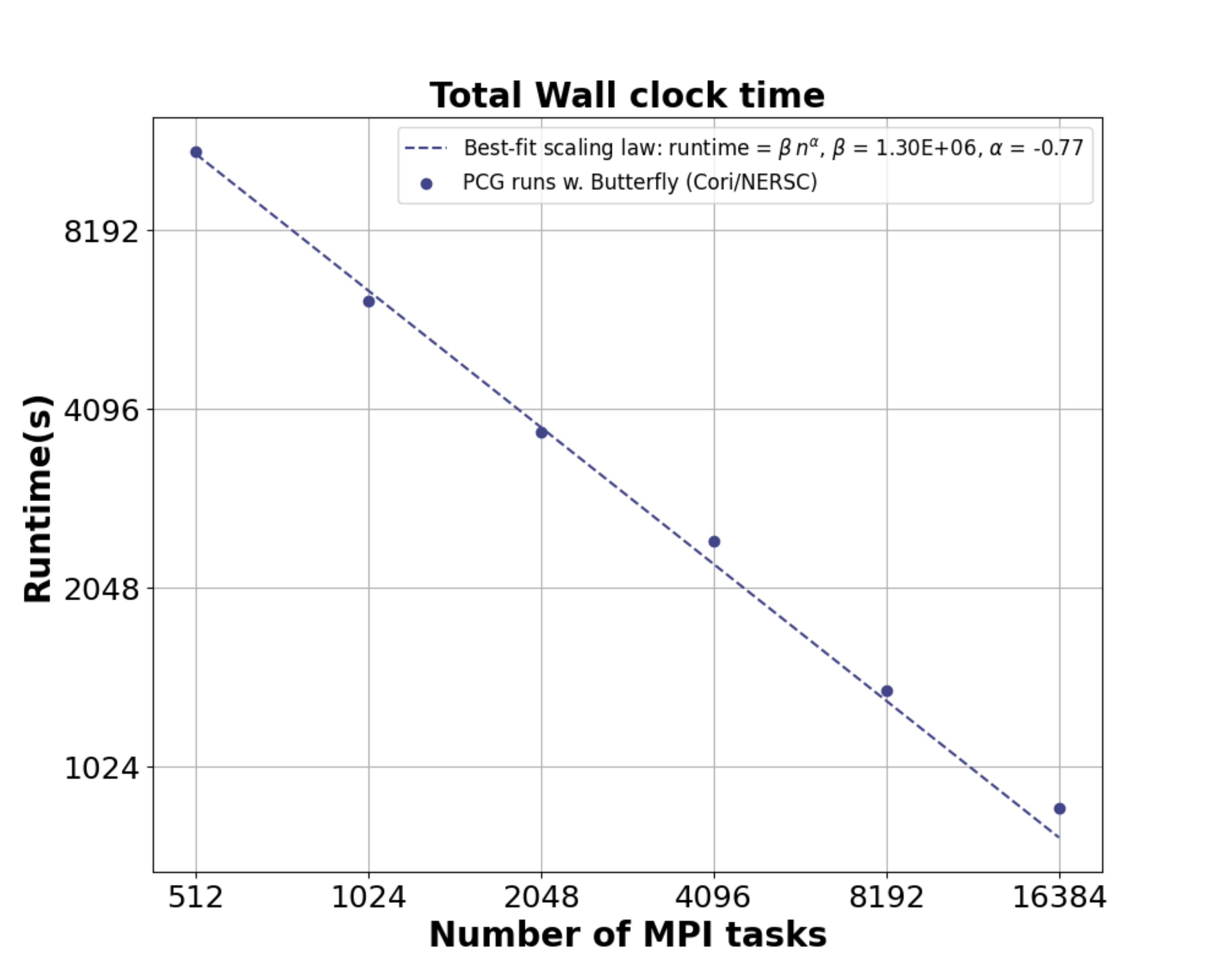}
\caption{\label{Wtime} 
Total wall clock time of the runs executed on Cori at NERSC using the Butterfly communication scheme.}
\end{figure}

We perform several strong scaling tests, using the \textbf{S1} simulation on Cori and Joliot-Curie to demonstrate the ability of the code to run on a large number of cores and handle the data reduction of a large data set. Given that the Skylake processors (Joliot-Curie) are more powerful than the Haswells (Cori) we expect to see longer runtimes with the second machine for the same set of the sky reconstruction parameters. In the runs presented here we assume the maximum-likelihood map-making approach with a Toeplitz-structured noise covariance and a half-bandwith set to 2,048 samples. This corresponds to a time-domain correlation length of $\sim 15$ s, given the adopted sampling rate. The solver converges in around $\sim 1,200$ iterations for the case considered. Given that the two machines had different versions of TOAST installed, the noise simulations were not exactly the same resulting in slight changes in the total number of iterations to convergence between the two machines. However, these changes do not affect the computational performance metric adopted here, which is the mean time per iteration. We choose this metric because it depends only on the half-bandwidth and the total size of the problem which are exactly the same for all the tests performed on both machines. The results are shown in figure \ref{iteration_scaling}. For reference, the total wall clock time is shown in figure \ref{Wtime}, for the \texttt{Butterfly} case executed on Cori, along with the best-fit scaling power law.

\begin{figure}[h!]
\centering
\includegraphics[width=0.48\textwidth]{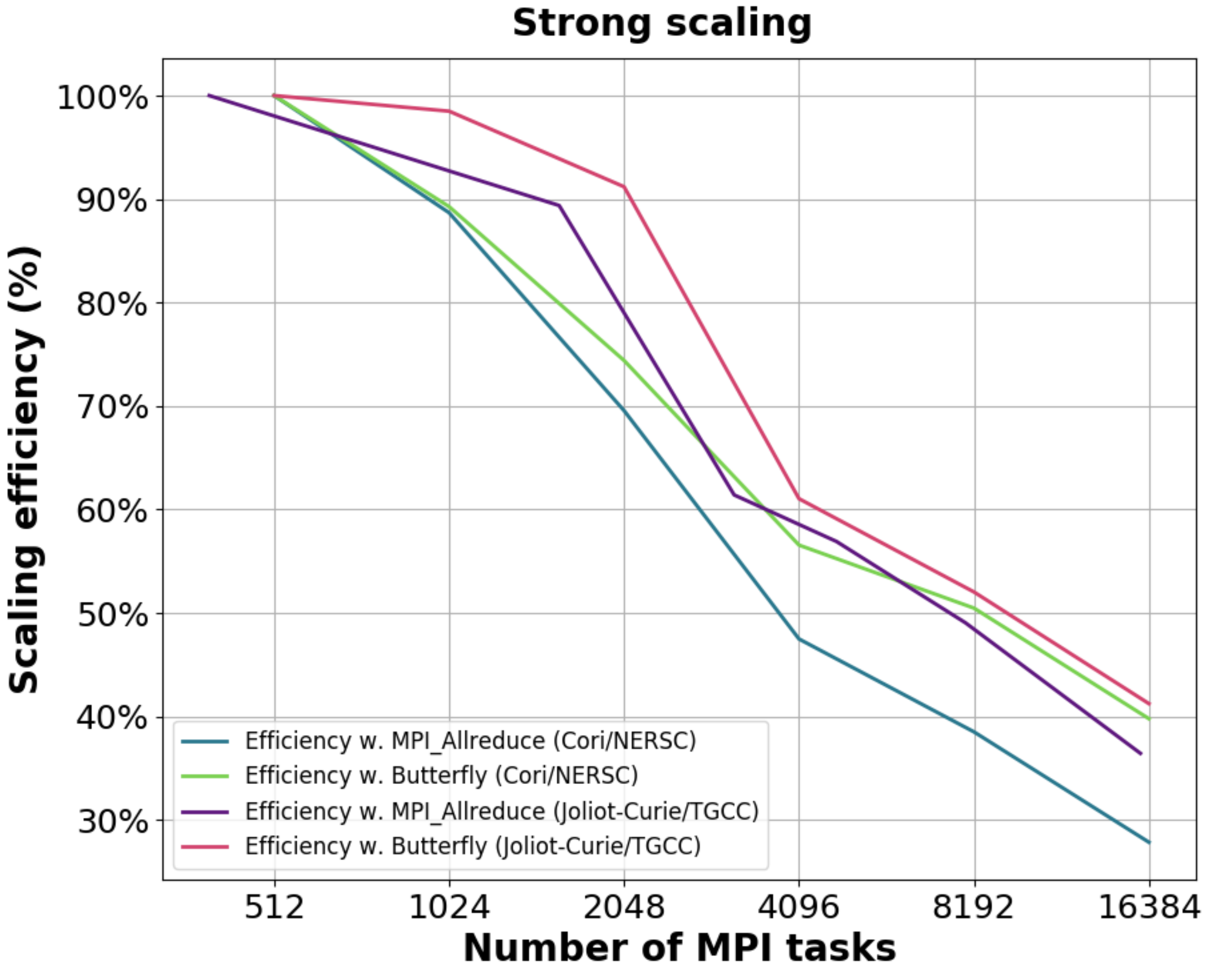}
\caption{\label{scaling_efficiency} 
Strong scaling efficiency comparing 4 sets of runs: (1) runs executed at the Cori machine at NERSC using the \texttt{MPI\_Allreduce} communication scheme; (2) runs executed on Cori  at NERSC using the Butterfly communication scheme; (3) runs executed on Joliot-Curie at TGCC using the \texttt{MPI\_Allreduce} communication scheme; (4)  runs executed on Joliot-Curie at TGCC using the Butterfly communication scheme.}
\end{figure}

The \texttt{Ring} runs do not scale properly as we increase the number of MPI tasks handling the problem, and completely diverge when using a large number of processes, therefore we exclude it from any further analysis. The customized \texttt{Butterfly} scheme exhibits better mean times per iteration compared with the standard \texttt{MPI\_Allreduce} on the Cori machine, particularly as we increase the number of MPI processes. On the Joliot-Curie machine, the \texttt{Butterfly} scheme shows similar or slightly better performance than the standard \texttt{MPI\_Allreduce} depending on the number of MPI tasks. To get a more quantitative view of the scaling performance, we measure the scaling efficiency given by
\begin{equation}
    \varepsilon = \frac{\delta t_1}{N\times \delta t_N},
\end{equation}
where $\delta t_1$ is the reference mean time per iteration to which the scaling efficiency is compared, theoretically this should correspond to the iteration time in a serial run (with a single process), and $\delta t_N$ is the mean time per iteration for a run with $N$ MPI processes. In practice, the size of the problem does not fit into a single node and thus we cannot perform serial runs, hence we set the reference to be the run with the lowest number of MPI processes, and $N$ to the factor by which we increase the number of processes with respect to the reference. In these tests, this corresponds to 396 MPI processes for the \texttt{MPI\_Allreduce} tests on Joliot-Curie and to 512 MPI processes for the remaining cases. The change of reference point should not affect our conclusions: given the proximity between the two references one can reasonably assume that the scaling between the two is close to $100\%$. Results are shown in figure \ref{scaling_efficiency}.

The \texttt{Butterfly} scheme performs better than  \texttt{MPI\_Allreduce} by a few percents on both machines. The scaling efficiency at $\sim 16$ K processes reaches around $40\%$ for the \texttt{Butterfly} scheme, while it is below $30\%$ for \texttt{MPI\_Allreduce} on Cori. Nevertheless, in the considered cases and for the number of MPI processes used, the communication overhead is small as compared to the computation time, and the current implementation of the \texttt{Butterfly} scheme is restricted to numbers of processes which are powers of 2. While the algorithm can be straightforwardly extended, this requires more work, and is left for future extensions of the software. Instead, given the satisfactory performance of the standard \texttt{MPI\_Allreduce} scheme, we select it as the default option for the rest of this work. 

We point out that both schemes, \texttt{MPI\_Allreduce} and \texttt{Butterfly}, show good scaling for these types of applications on both machines, demonstrating good overall portability of the software.

\subsubsection{Solvers and preconditioners performance}
\label{sect:solvers-tests}
\noindent \textbf{Enlarged CG:}

\begin{figure}[h]
\centering
\includegraphics[width=0.48\textwidth]{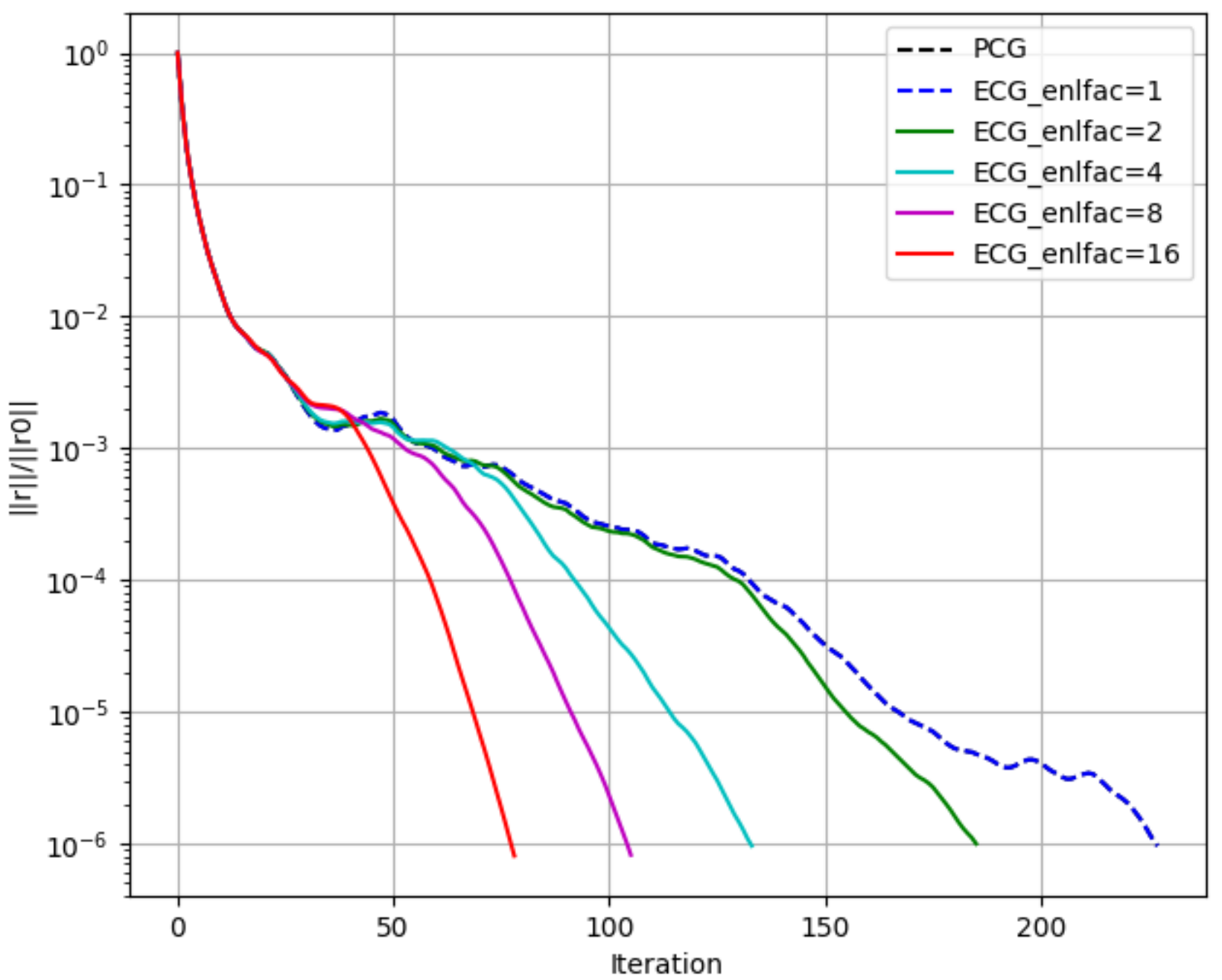}
\caption{\label{ECGtest} 
Dependence of ECG iterations on the enlarging factor, using 1 hour of the SAT simulated data.}
\end{figure}

In figure \ref{ECGtest}, we show a selection of tests of the ECG solver for different enlarging factors. We use the Orthomin method, as the case studied is well behaved and does not warrant the use of the more involved Orthodir method. The data used here constitute a 1 h subset of the \textbf{S1} simulation. This corresponds to a single constant elevation scan. We also exclude the atmospheric noise from the input, and instead artificially increase the instrumental noise $f_{\text{knee}}$ to $1$ Hz. We assume a half bandwidth, $\lambda = 2^{18}$, corresponding to a correlation length of about 33 minutes. This configuration results in a degraded conditioning of the system matrix showcasing the ability of the ECG solver to deal with such cases. We execute multiple ECG runs varying the enlarging factor from 1 to 16, and compare it to a standard PCG run. The comparison between the PCG convergence and the ECG one with enlarging factor equal to unity provides the basis for the validation of the latter: the theoretical expectation is that the two curves should coincide as it is indeed the case in figure \ref{ECGtest}. In addition to this, we observe that the number of iterations required to reach convergence (set in the present case by a tolerance parameter of $10^{-6}$) is monotonically decreasing with the enlarging factor, in particular we gain by about a factor 3 with an enlarging factor of 16 compared to standard PCG. This behavior is also theoretically expected from Eq.~\eqref{eqn:RateConvECG}. However as previously mentioned this does not translate automatically to a net gain in the run time. For that purpose, one should ensure that the implemented algebraic routines used by the solver are sufficiently optimized to compensate for the increase in the cost of each iteration. This is not the case for the naive implementation we used in this test. Our results suggest that this solver could be indeed an interesting, precomputation-free alternative for the map-making problem, in particular given the advent of new computer architectures, motivating future work directly targeting the optimization of the underlying operations.
\\\\
\noindent \textbf{PCG with two-level preconditioners:}

\begin{figure}[h]
\centering
\includegraphics[width=0.48\textwidth]{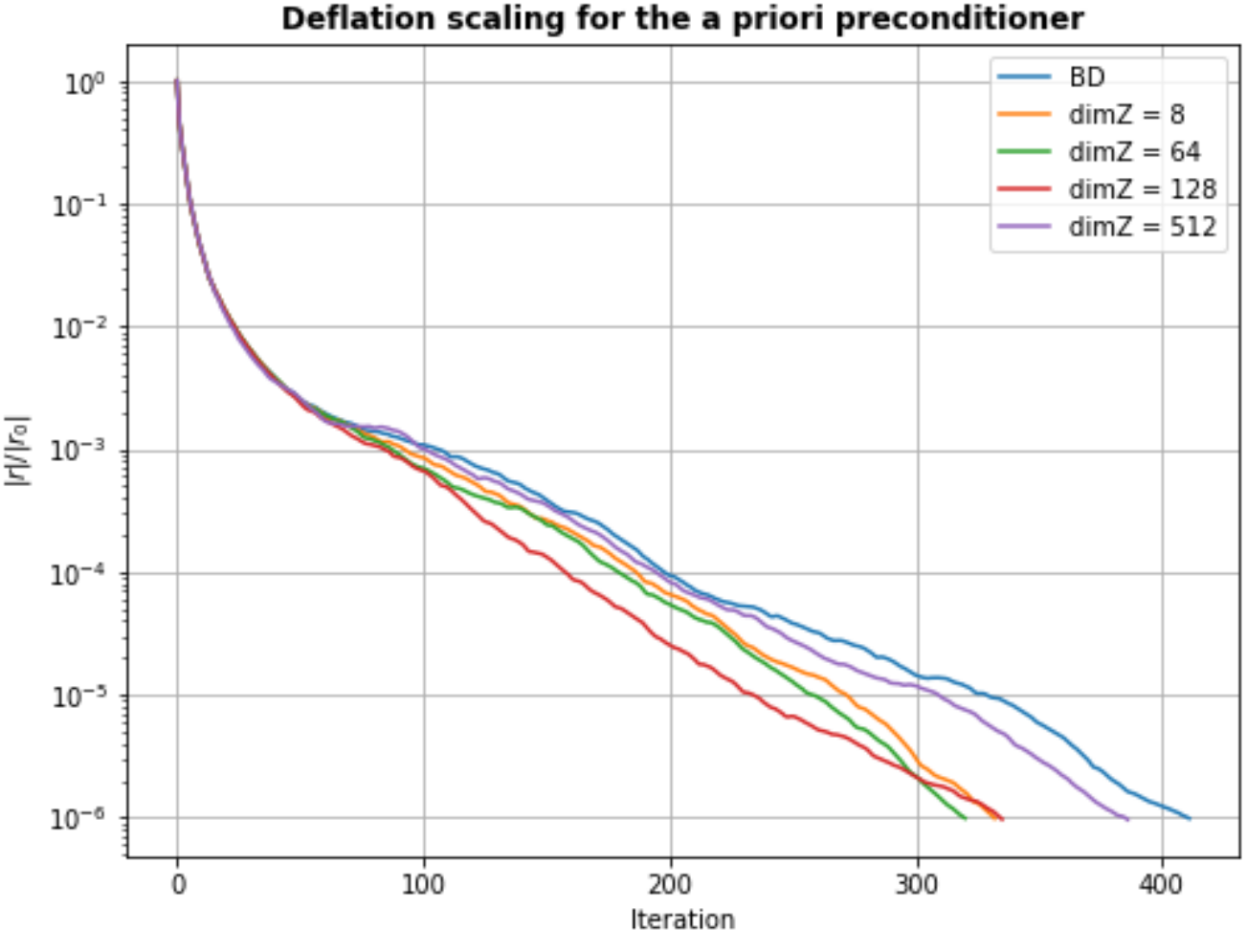}
\caption{\label{convergence_it_apriori} 
Convergence in terms of the number of iterations of the BD preconditioned CG, compared with the \textit{a priori} two-level preconditioned CG, for different dimensions of the deflation subspace.}
\end{figure}

\begin{figure}[!h]
\centering
\includegraphics[width=0.48\textwidth]{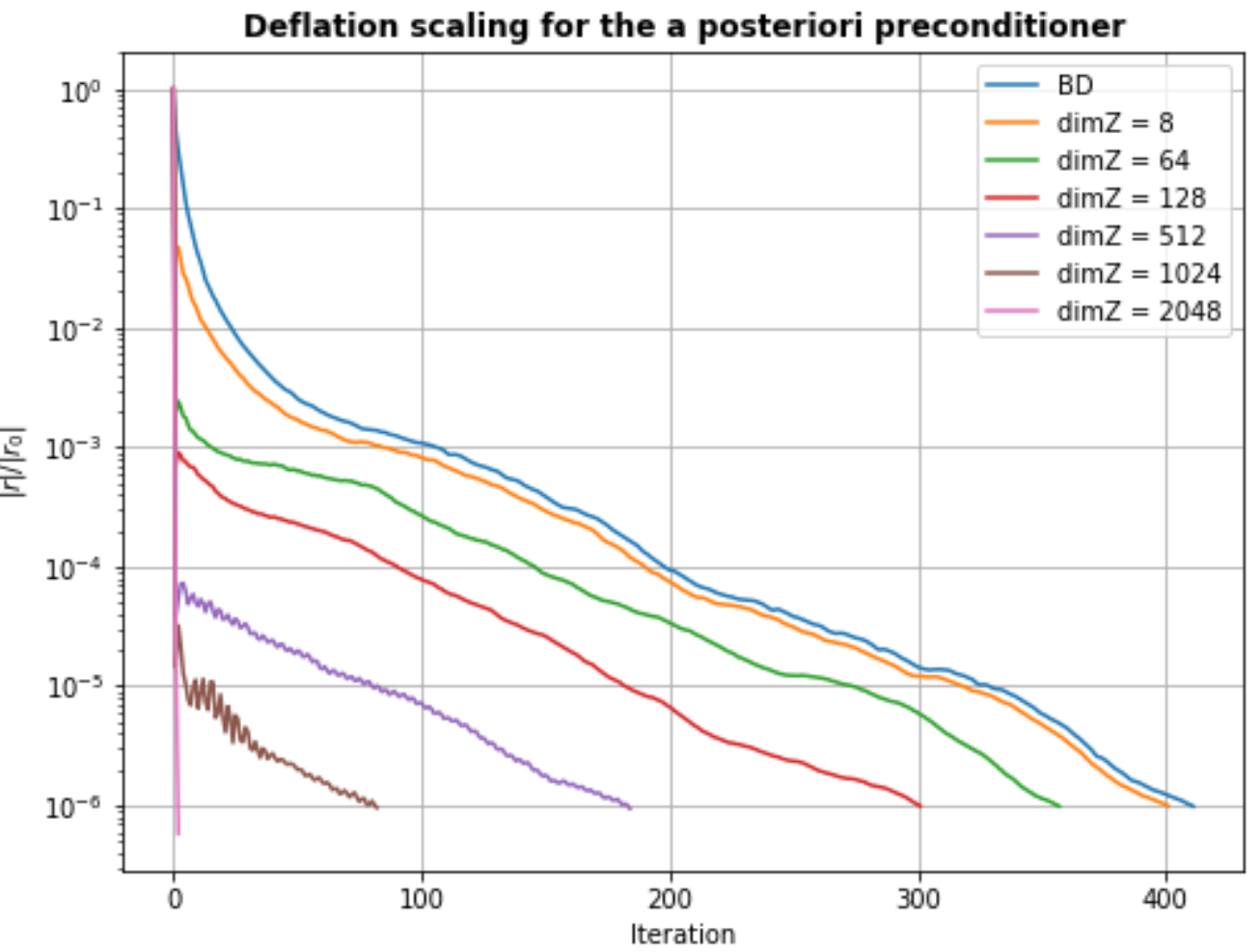}
\caption{\label{convergence_it_lanczos} 
Convergence in terms of the number of iterations of the BD preconditioned CG, compared with the \textit{a posteriori} two-level preconditioned CG, for different dimensions of the deflation subspace.}
\end{figure}

\begin{figure*}[!h]
\centering
\includegraphics[width=1.0\textwidth]{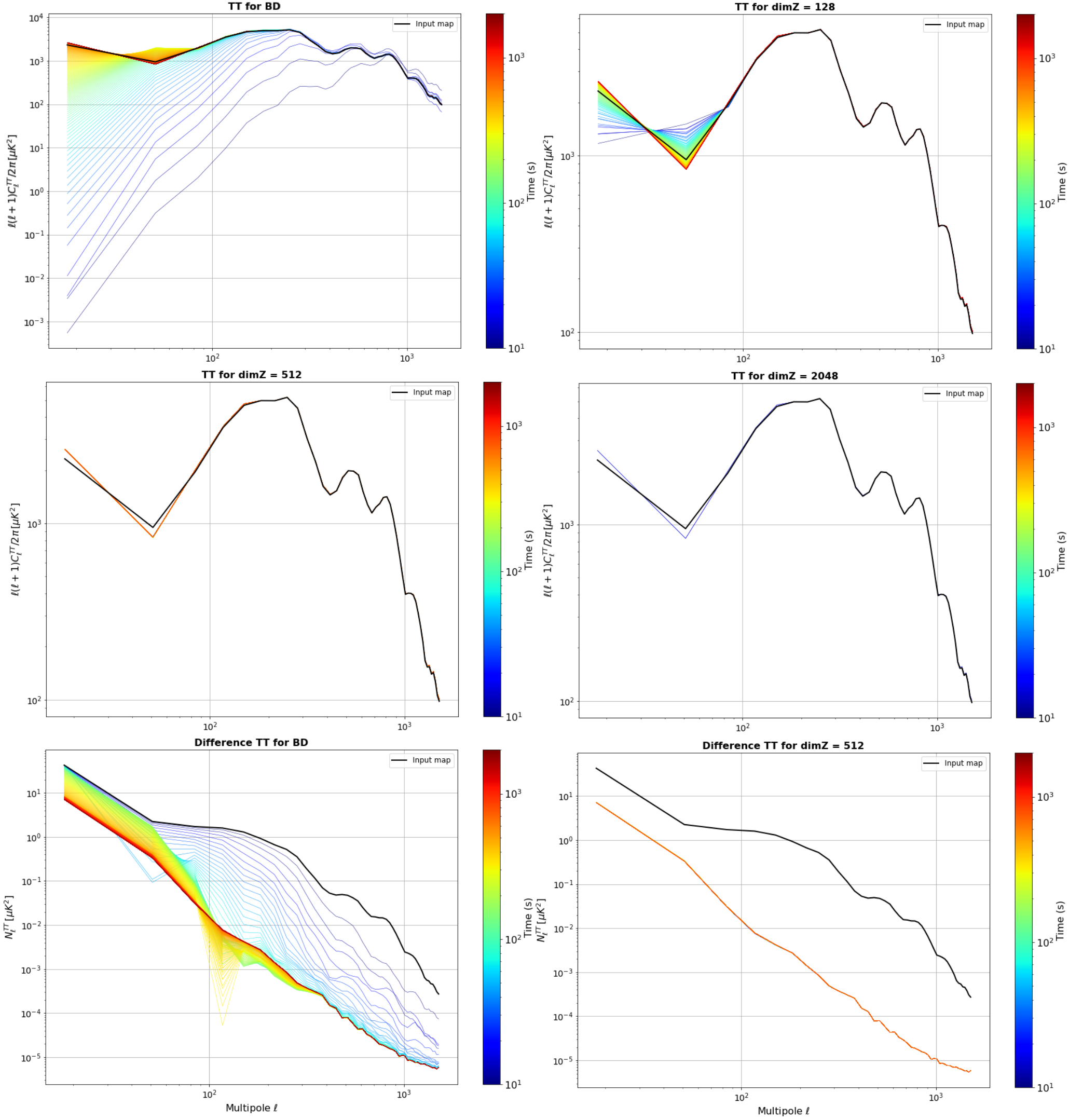}
\caption{\label{PS_TT} 
\textit{(Top and middle rows):} Temperature angular power spectra of the maps at different stages of CG convergence for various preconditioning settings. \textit{(Bottom row):} Temperature angular power spectra of the difference between the input map and the reconstructed maps at different stages of CG convergence.}
\end{figure*}

\begin{figure*}[!h]
\centering
\includegraphics[width=1.0\textwidth]{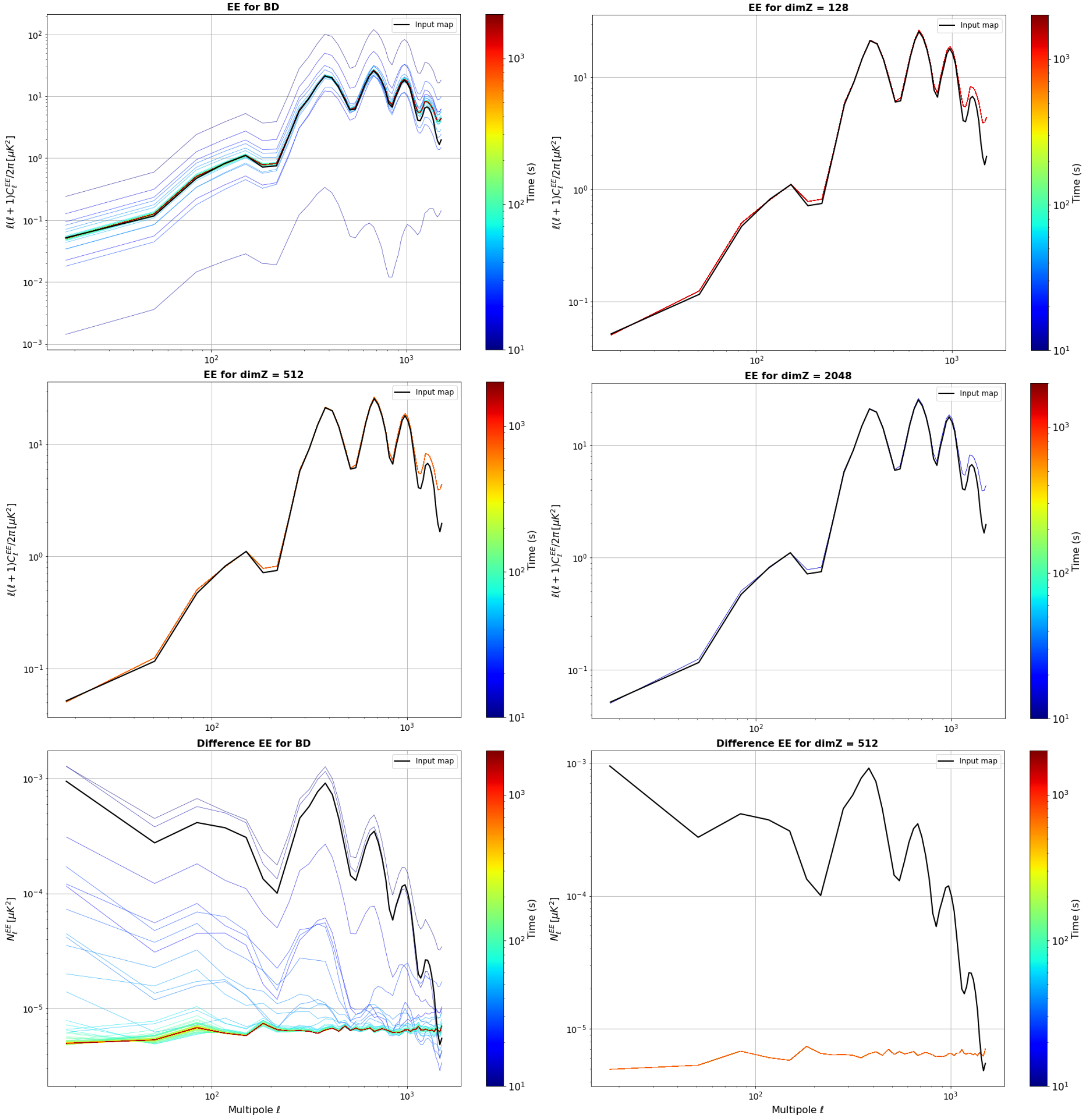}
\caption{\label{PS_EE} 
\textit{(Top and middle rows):} EE angular power spectra of the maps at different stages of CG convergence for various preconditioning settings. \textit{(Bottom row):} EE angular power spectra of the difference between the input map and the reconstructed maps at different stages of CG convergence.}
\end{figure*}

In the present subsection, we compare the performance of the three preconditioners offered by MAPPRAISER. For this purpose we run an adapted version of the \textbf{S2} simulation on Cori, where we cut down the number of detectors by a factor 14, i.e., to 430 detectors, to run on a small number of processes (64 MPI tasks) and we scale down the noise to emphasize the sky signal and study the angular convergence as well. The convergence properties should not be affected by such changes as they don't depend on the noise normalization in general. We use the block-diagonal (BD) preconditioner, the \textit{a priori} and \textit{a posteriori} two-level preconditioners varying the size of the deflation space, and then compare the performance of the three preconditioners in terms of the number of iterations and wall clock time necessary to converge (figures \ref{convergence_it_apriori}, \ref{convergence_it_lanczos} and \ref{wallclock_convergence}). In addition, we also show the TT and EE angular power spectra of the reconstructed maps at all iterations to assess the convergence of the different angular scales (figures \ref{PS_TT} and \ref{PS_EE}). The BB spectra are noise dominated in this regime and therefore not considered here. The power spectra are estimated using the NaMaster library\footnote{\scriptsize \url{https://github.com/LSSTDESC/NaMaster}}~\cite{Alonso_2019}, correcting for the mask, however not for the noise bias. The \emph{a priori} preconditioner tests summarized in figure \ref{convergence_it_apriori} show only marginal gain on the number of iterations to convergence, demonstrating that, at least in the case studied, there is no significant benefit in using the \emph{a priori} preconditioner over the standard BD preconditioner. In fact, the \emph{a priori} preconditioner is included mostly for historical reasons.  Indeed, the main early interest in this preconditioner was driven by the possibility of saving on the precomputation time needed to construct $\mathbf{Z}$~\cite{Szydlarski_2014, Grigori_2012}, which is essentially guessed and therefore computationally cheap. However, constructing the preconditioner still requires a computation of the $\mathbf{A}\,\mathbf{Z}$ matrix.
This is often as costly as the computation of $\mathbf{Z}$ itself. As we discuss in the following in the novel \emph{a posteriori} construction we propose in this work, the cost of the former can be hidden in the calculation of the latter. Consequently, this new class of the \emph{a posteriori} constructions is not only performing better but also requires as much precomputations as the \emph{a priori} technique. Notwithstanding these considerations, the \emph{a priori} construction may still be of interest, in particular, in the cases where there is a limited number of nearly singular eigenvectors spanning a known low-dimensionality subspace. In the following we focus solely on the convergence properties of the \emph{a posteriori} preconditioner.

Figure \ref{convergence_it_lanczos} shows that, for the \emph{a posteriori} preconditioner, the number of iterations required for convergence monotonically decreases with the size of the deflation subspace, $\dim \mathbf{Z}$. For any sufficiently high $\dim \mathbf{Z}$ (here $\gtrsim 128$), the \textit{a posteriori} two-level preconditioner performs consistently better than the \textit{a priori} two-level preconditioner. This is expected since the former uses a more involved procedure to construct the deflation subspace, yielding better estimates of the eigenvectors of the system matrix, once the Lanczos procedure runs for a sufficient number of iterations.

Additionally, the convergence gain as shown in the figure mainly stems from a sharp drop of the residual at the very first CG steps of the solver. This can be understood by examining the convergence of the different angular scales in figures \ref{PS_TT} and \ref{PS_EE}. The colorbar encodes the wall clock time from the start of the solver until convergence is reached, not including precomputation time. The colors are in $\log$ scale and the curves correspond to the power spectra at different iterations. Comparing the various TT and EE plots, we can see that the higher the size of the deflation subspace, the wider the range of the angular scales recovered from the first CG step. As a matter of fact, the precomputation allows the solver to compress information encoded in the system matrix, $\mathbf{A}$, to a small set of modes contained in the deflation matrix, $\mathbf{Z}$. When running the PCG steps, that information allows to recover the smallest angular scales immediately in the first step, while subsequent steps are spent correcting the largest angular scales.
In the extreme cases, $\dim \mathbf{Z} \geq 512$, we can see that the power spectra of the different iterations coincide with each other, and therefore that we practically recover the full range of angular scales of the solution from the first step. Indeed, in the case of $\dim \mathbf{Z} = 2048$, the formal convergence is reached in two steps, meaning that we compress practically all of the information of the system matrix, $\mathbf{A}$, in the relatively low number of modes of the deflation subspace. The \emph{a posteriori} two level preconditioner thus not only improves convergence, it also gives us some insight into the noise covariance structure of the computed maps. However, for $\dim \mathbf{Z} = 512$ or $1024$, while the spectra reach their final shape in at most a couple of steps, e.g., figures~\ref{PS_TT} and~\ref{PS_EE},
the iterations continue for nearly 200 or 100, respectively, steps more in order to attain the formal convergence.
This is related to the fact that the formal residual defined in Eq.~\eqref{eqn:CGresidual} is weighted towards the noisiest pixels often located at the outskirts of the observed sky patch and which matter little in the power spectrum computation. This suggests that if the power spectra are of ultimate interest for the analysis, a more suitable convergence metric could be proposed leading to significant performance gain. The software discussed here is straightforwardly extensible to allow for such tailored, user-defined convergence metrics.

To complement these results, we also show the TT and EE angular power spectra of the difference between the (noiseless) input map and the (noisy) reconstructed maps at each iteration. These are given in the bottom rows of figures \ref{PS_TT}  and \ref{PS_EE} for both the BD case and the $\dim \mathbf{Z} = 512$ case. The obtained spectra are compared to the power spectrum of the input map and can be seen to stabilize on the level corresponding to the expected noise of the recovered maps. This level is several orders of magnitude below the input signal in all the cases but the smallest angular scales for the EE spectrum where, near the pixel scale, the signal becomes comparable to the noise.
These results demonstrate the proper convergence of the maps both in terms of amplitudes and phases of the sky signal and on all angular scales down to the level of the observational precision. We observe that as before, the two-level preconditioner case considered here allows to reconstruct the full range of angular scales essentially in a single step. These observations hold for the other deflation space dimensions as well.

\begin{figure}[!h]
\centering
\includegraphics[width=0.48\textwidth]{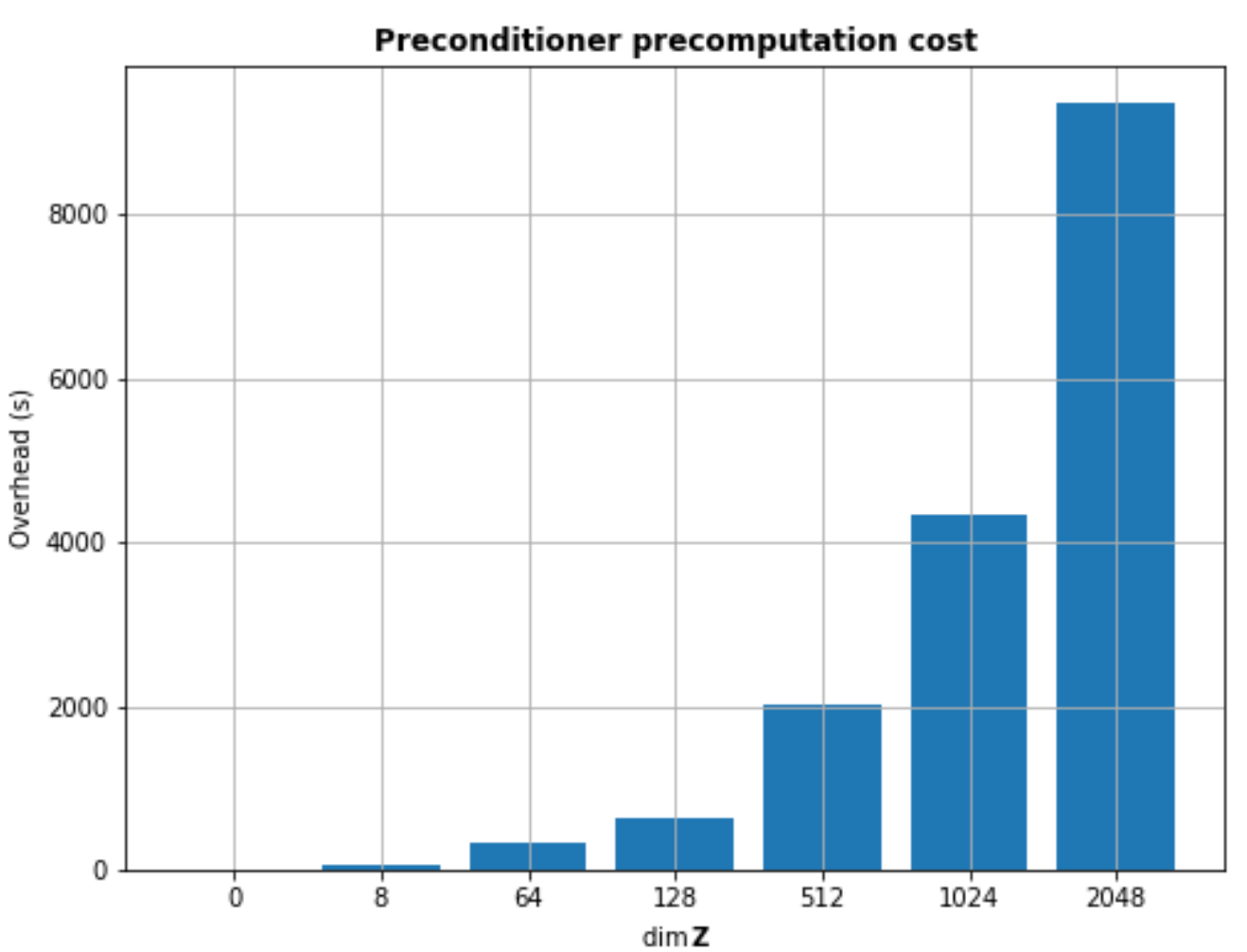}
\caption{\label{precomputation} 
Precomputation cost of the \textit{a posteriori} preconditioner as a function of the dimension of the deflation subspace matrix $\dim \mathbf{Z}$. In practice this dimension is the number of iterations performed by the Lanczos procedure.}
\end{figure}

\begin{figure}[!h]
\centering
\includegraphics[width=0.48\textwidth]{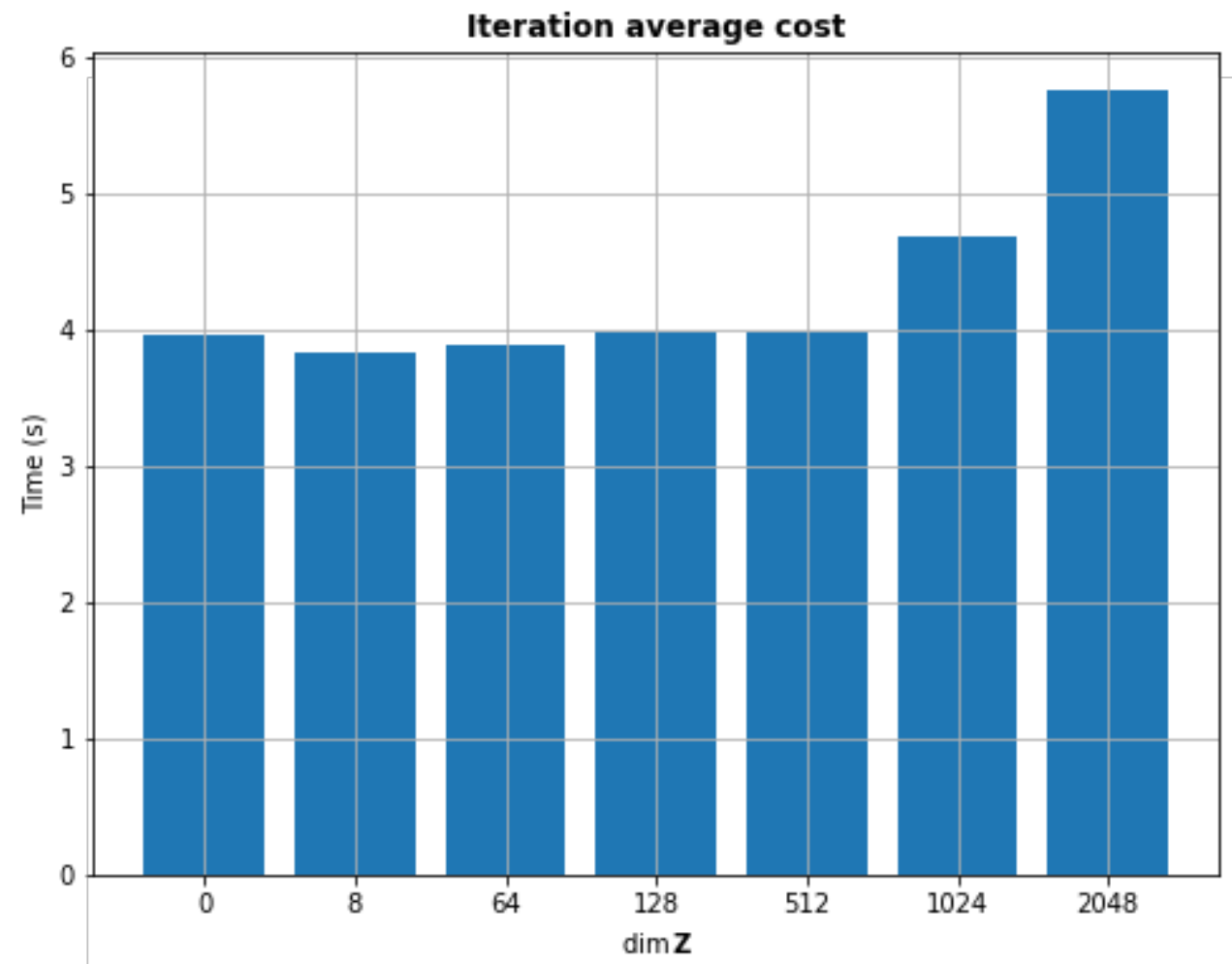}
\caption{\label{iteration_cost} 
Average iteration cost of PCG runs with the \textit{a posteriori} preconditioner as a function of the dimension of the deflation subspace matrix $\dim \mathbf{Z}$.}
\end{figure}

\begin{figure}[!h]
\centering
\includegraphics[width=0.48\textwidth]{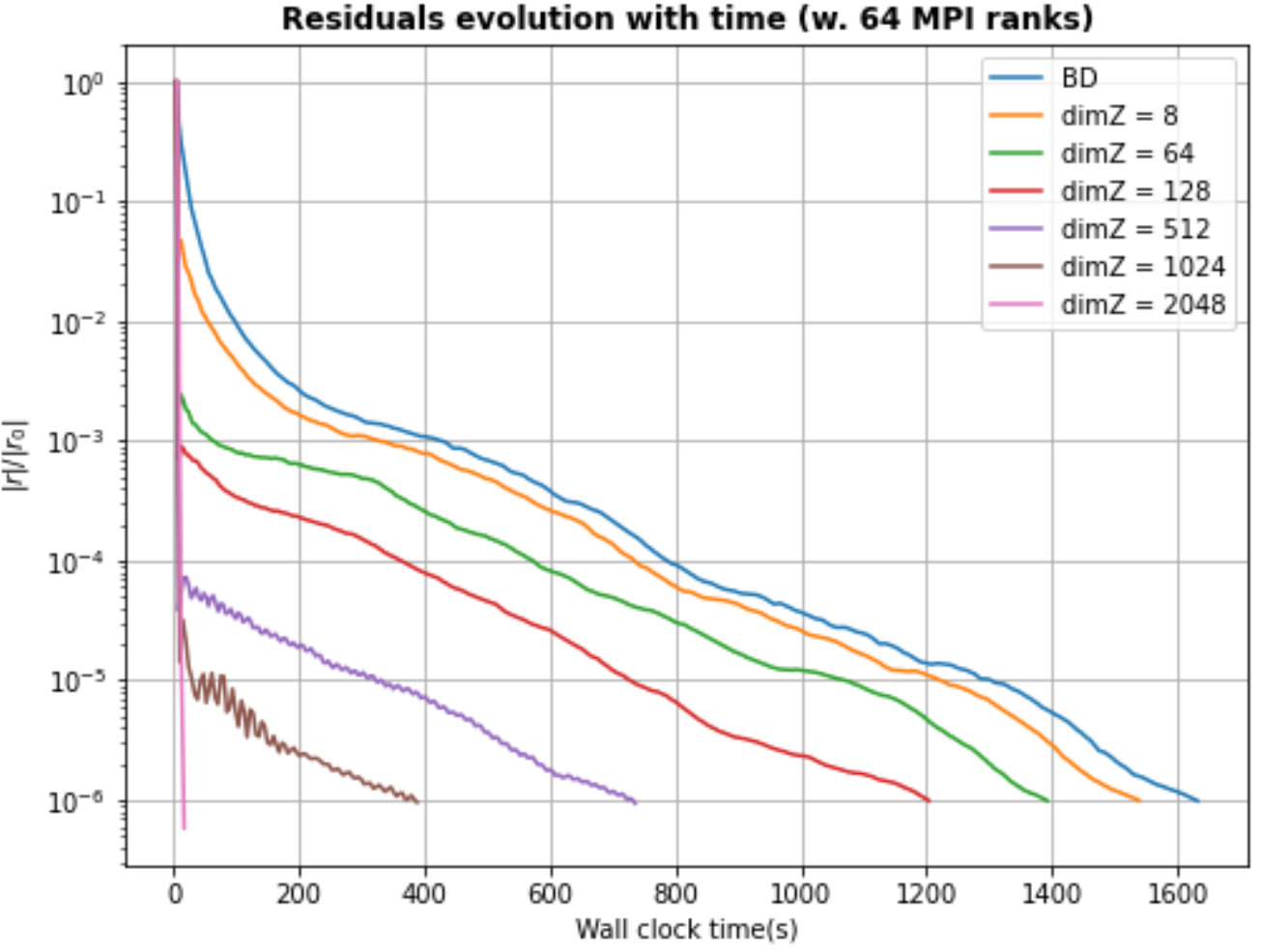}
\caption{\label{wallclock_convergence} 
Convergence in wall clock time of the BD preconditioned CG, compared with the \textit{a posteriori} two-level preconditioned CG, for different dimensions of the deflation subspace. The precomputation time is not included here.}
\end{figure}

The cost of precomputation for different sizes of the deflation subspace constructed with the Lanczos procedure is shown in figure \ref{precomputation}. This cost scales linearly with $\dim \mathbf{Z}$, and can be as costly as a few PCG runs. This makes this method mostly advantageous when we want to solve multiple systems, with roughly the same system matrix but different realizations of the right hand side, as it is the case for example when performing null tests, or Monte Carlo simulations. Then the preconditioner needs to be constructed only once, stored in memory, and all subsequent runs can benefit from the same convergence boost virtually for free. The cost of each iteration does not substantially change in a way that may impact the net gain in runtime as shown in figures \ref{iteration_cost} and \ref{wallclock_convergence}. 

The slight increase in the average iteration cost seen for the cases $\dim \mathbf{Z} = 1024$ or $2048$, is due to the fact that operations start getting non negligible contributions from the $\dim \mathbf{Z}$ map-domain operations. However, for the case considered, this increase does not affect the results in any significant way, but may be of concern for high resolution maps, in the context of LAT observations for instance. In this case, memory space is also a limiting factor and one may be forced to downsample the columns of $\mathbf{Z}$ for the method to work. In such a case, the small scale modes would not be captured by the deflation subspace, but these typically converge very well regardless. This is however beyond the scope of the present work, and may be the object of future investigation.

We can consider some numbers to illustrate the effective gain in runtime enabled by this technique. In the present case studied, the precomputation cost with $\dim \mathbf{Z} = 2048$ is roughly $\sim 5.5$ PCG runs with the standard block-diagonal preconditioner, and the solving time is roughly $10$ s. Solving 100 similar systems for example, would then cost around $\sim 9,500 + 10\times100 = 10,500$ s, while it would cost around $1,650 \times 100 = 165,000$ s running with the block-diagonal preconditioner. We gain about a factor $\sim 16$ solving 100 similar systems. Note that the gain is only going to increase as we increase the number of similar systems to solve, hence alleviating the cost of map-making in Monte Carlo runs for example, as we become only limited by the cost of the simulations.

\subsubsection{Validation of templates marginalization}
In this subsection, we validate the templates marginalization procedure at the map level. For this purpose, we select the input signals in the \textbf{S2} simulation such that we include only the cosmological CMB signal and, each time, one specific time-domain systematic signal modelled by a given template for which we know the exact parameters. We then reconstruct the sky maps assuming 3 different noise models: 
\begin{itemize}
    \item A white noise model (Binned mapper).
    \item An incorrect template model: we assume the correct template class but with the wrong parameters.
    \item A correct template model: we assume both the correct template class and the correct parameters.
\end{itemize}
The binned mapper allows us to see the full impact of the systematic signal on the map if not taken into account in the reconstruction. Using an incorrect template model, allows us to qualitatively assess the impact of model errors, and the last configuration using the correct template model allows us to validate the procedure by verifying that we recover a correct estimate of the CMB map.\\

\noindent \textbf{Scan synchronous signal subtraction:}

\begin{figure}[!h]
\includegraphics[width=0.5\textwidth]{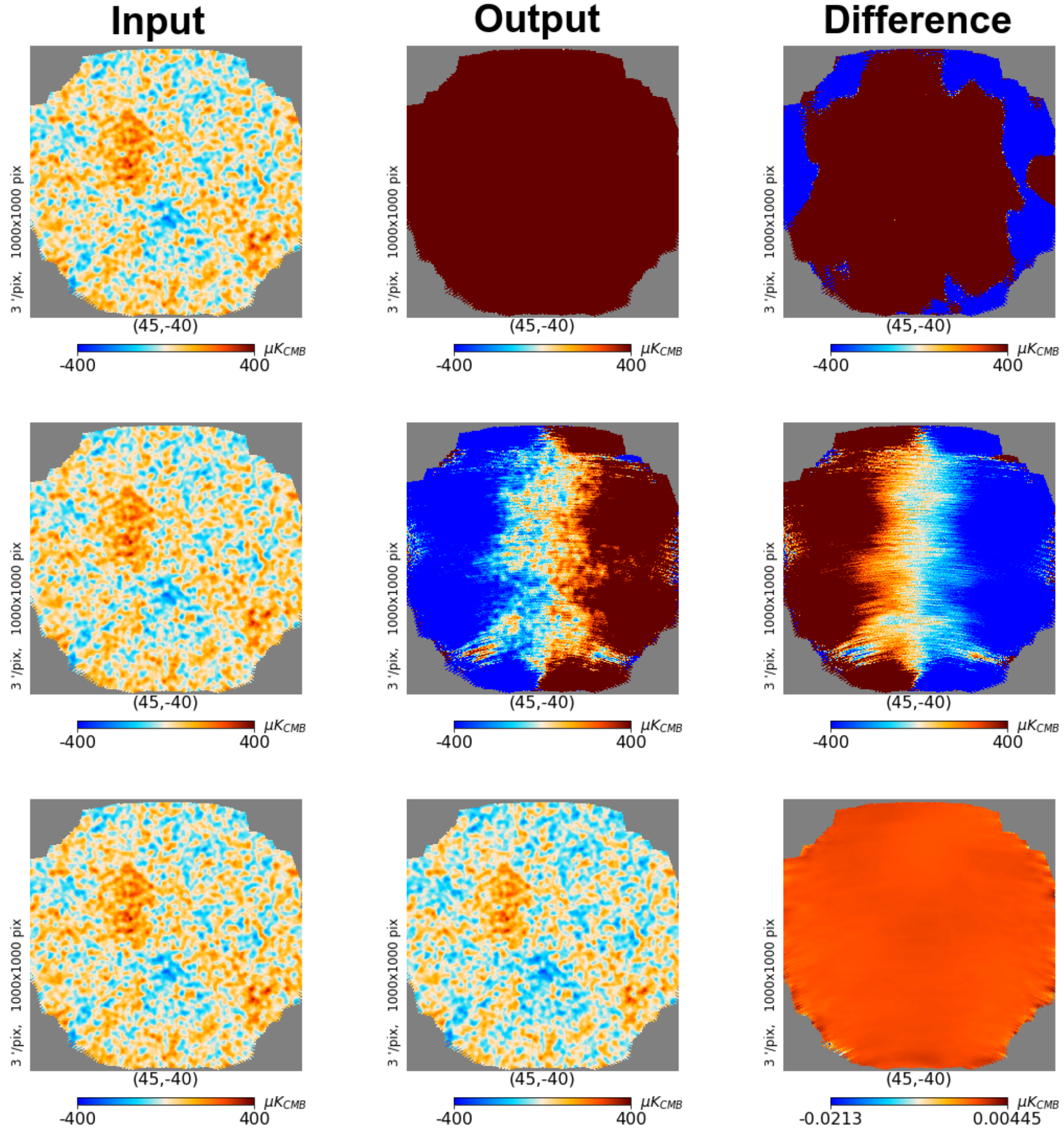}
\caption{\label{SSStest} 
Temperature maps of CMB+SSS mapped with respectively, a binned map-maker, a 10 azimuth bins, and 30 azimuth bins SSS templates map-makers (top to bottom). From left to right, we show respectively the input, the output, and the difference maps.}
\end{figure}

We inject a ground-pickup signal using the scan synchronous signal template which takes the form of a step function in azimuth $g(\varphi)$; that is the ground map is binned in azimuth, with some fixed azimuth step $\Delta \varphi$, and in any given constant elevation scan, 
$$\forall \varphi \in\left[\varphi_{min}+i\Delta\varphi, \varphi_{min}+(i+1)\Delta\varphi\right[,\,g(\varphi)=g_i=\text{const.}$$
Therefore at any given instant $t$, if the telescope's boresight is pointing towards the $i$-th bin in azimuth, we can simply write: $SSS(t)=g_i$. We use a linear ground map of the form $g_i = 1+0.01i$ (K), with an azimuth bin $\Delta \varphi = 0.33\deg$. For the map reconstruction, given that the azimuth range of the constant elevation scans is of $10$ degrees, the correct template model accounts for 30 bins in azimuth for each CES, and we choose to account for only 10 bins in azimuth for each CES in the incorrect model run. The reconstructed temperature maps are shown in figure \ref{SSStest}. 

The total offset of the map cannot be recovered, since it is fully degenerate with the ground template. Therefore, it is removed from the difference maps shown in the right panel of the figure. In the binned map, the coherent ground structure completely dominates the sky signal as one would expect, and we essentially see the projection of the offset of the ground map. The middle row maps show an improvement, as the ground signal is partially filtered, and the ground offset removed, however since we are assuming larger azimuth bins, we still get power from the sub-bin fluctuations leaking to our reconstructed sky signal, and given that the corresponding amplitudes fluctuations is of order $10$ mK, the ground leakage still dominates over the CMB signal. In the bottom row, we recover the correct CMB map effectively filtering the ground signal. The difference map in the bottom right panel, shows a low amplitude (below $\sim 0.02\, \mu$K) smooth residual structure, which essentially comes from degeneracies between the sky and the ground template. These degenerate modes correspond to the projections of the azimuth bins of the ground template on the sky, which, as the sky rotates, leave stripes which are constant in right ascension (such as spherical harmonics with $m=0$), with unconstrained relative offsets. Some of these degeneracies can be broken when data from multiple detectors, with shifted and overlapping azimuth bins, is combined, or with the sky pixels shared between adjacent bins. Nonetheless, these constraints are typically weak, leaving a number of ill-constrained modes in declination which characteristics depend on the pixelization, the size of the azimuthal bins and the scanning strategy~\cite{Poletti_2017, Naess_2014}. 

With these results we can validate the SSS filtering operators and the corresponding templates marginalization procedure.\\

\noindent \textbf{HWP synchronous signal subtraction:}

\begin{figure}[!t]
\includegraphics[width=0.5\textwidth]{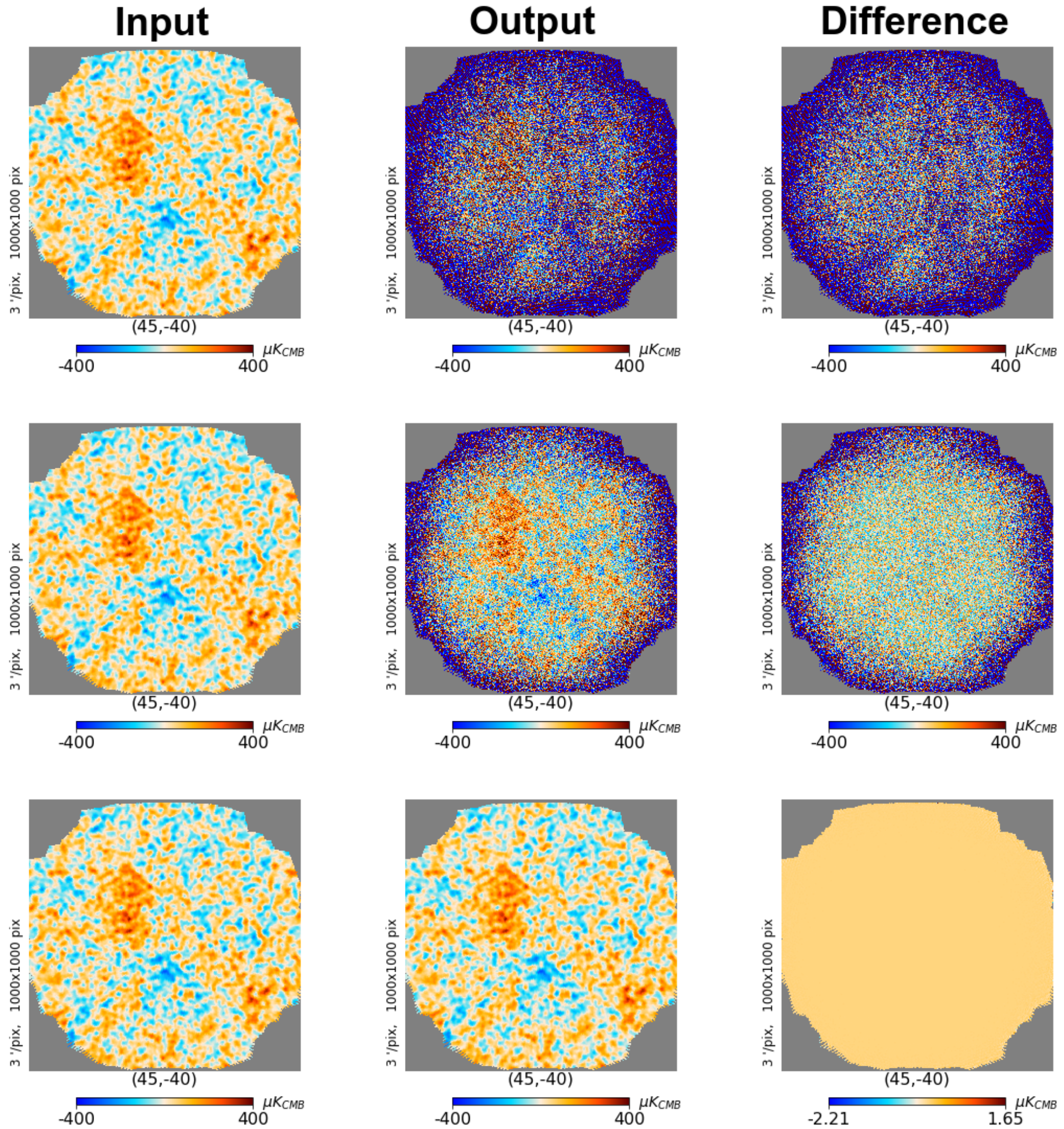}
\caption{\label{hwpss_t} 
Temperature maps of CMB+HWPSS mapped with respectively, a binned map-maker, a fourth, and sixth order HWPSS templates map-makers (top to bottom). From left to right, we show respectively the input, the output, and the difference maps.}
\end{figure}

\begin{figure}[!t]
\includegraphics[width=0.5\textwidth]{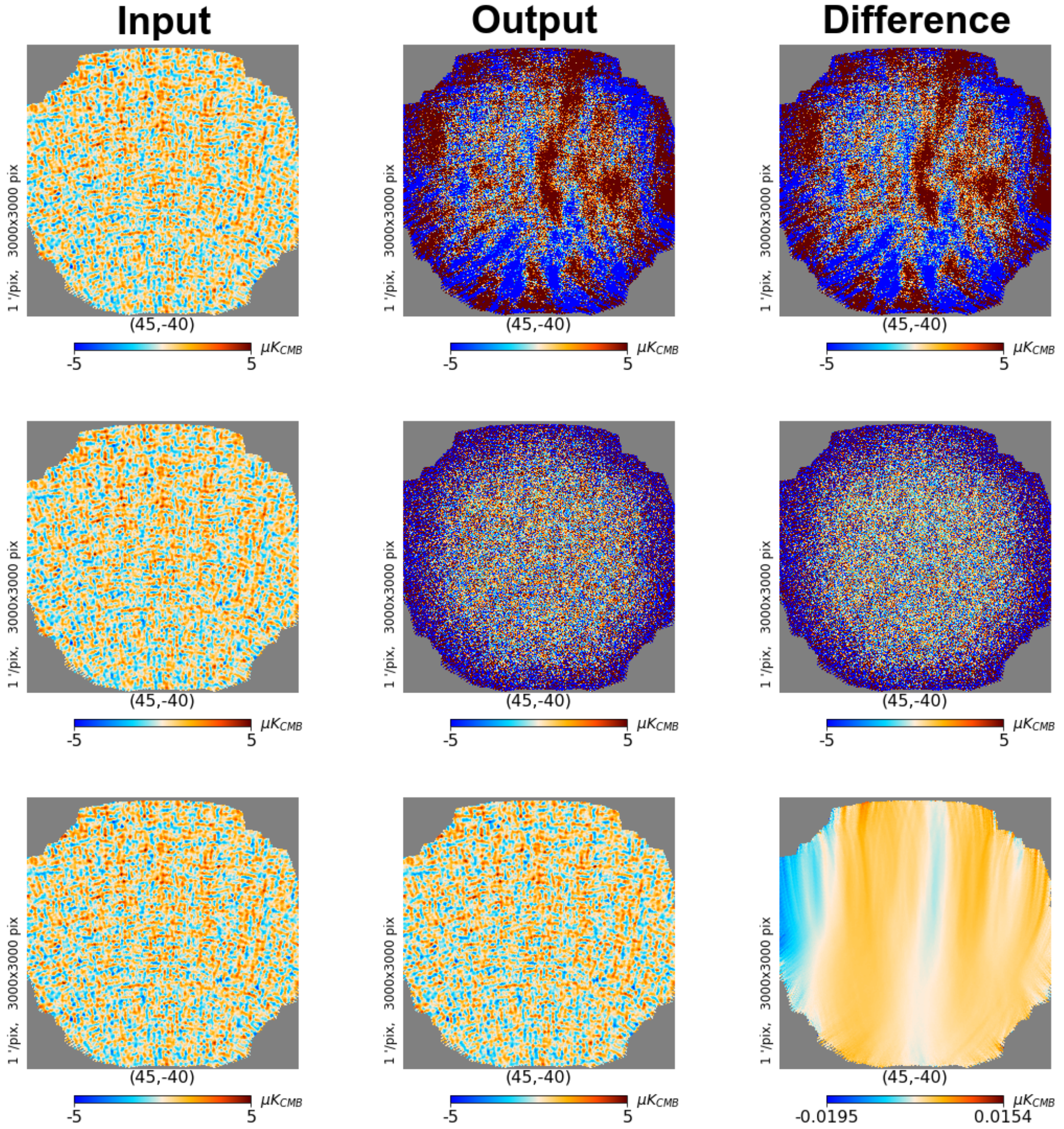}
\caption{\label{hwpss_q} 
Q polarization maps of CMB+HWPSS mapped with respectively, a binned map-maker, a fourth, and sixth order HWPSS templates map-makers (top to bottom). From left to right, we show respectively the input, the output, and the difference maps.}
\end{figure}

We inject a HWP synchronous signal modelled by a set of harmonics of the HWP rotation frenquency, coupled with a linear drift term~\cite{Maxipol_2008, Kusaka_2014, Takakura_2017}. Therefore, for each CES and each detector, it takes the following form:
\begin{equation}
    \mathcal{H}(t)=\sum_n A_n(1+\varepsilon_nt)\cos(n\phi_t)+B_n(1+\varepsilon_nt)\sin(n\phi_t),
\end{equation}
where $\phi_t$ is the HWP angle at time $t$. The harmonics amplitudes $A_n$ and $B_n$ are drawn from a normal distribution $\mathcal{N}(\mu,\sigma)$, with $\mu=100$ mK, and a dispersion $\sigma/\mu$ of $1\%$. The linear drift coefficient $\varepsilon_n$ is set such that we get a $1\%$ drift every $1000$ s ($\sim 16$ min). We inject a total of six harmonics. For the reconstruction, we consider a HWPSS template model for which the baseline length (i.e., the time interval over which the harmonics amplitudes are considered to be constant) is $10$ s. We consider two template configurations, one which takes into account only four harmonics of the HWP rotation frequency, $f_\text{HWP}$, and another one which handles all of the six harmonics. The reconstructed temperature and Q polarization maps are shown, respectively, in figures \ref{hwpss_t} and \ref{hwpss_q}.

The binned maps, in the top rows, show the direct projection of the contaminated TOD into the map, and given the amplitude of the HWPSS, the CMB signal is completely dominated by the spurious signal. In the middle row of figure \ref{hwpss_t}, the filtering of the first four harmonics reduces the noise level of the temperature map at large scales, as we start to see the sky signal emerging, however the remaining $5f_\text{HWP}$ and $6f_\text{HWP}$ harmonics still leak as small scale noise given that these components reside in the high frequency tail of the signal. For polarization, the signal is modulated at $4f_\text{HWP}$ and the filtering of the first four harmonics removes the large scale stripy structure we see in the top row of figure \ref{hwpss_q}, caused by spurious signals varying more slowly than the polarization modulator, but the leakage from the $5f_\text{HWP}$ and $6f_\text{HWP}$ harmonics is still sufficently strong for the CMB signal to remain subdominant. This is due to two aspects: (1) the $5f_\text{HWP}$ and $6f_\text{HWP}$ harmonics are closer to the center of the science band in the case of polarization than in temperature because of the HWP modulation, and (2) the polarization signal is fainter in amplitude than temperature. In the bottom rows, we recover the correct input CMB signal, and the HWPSS is effectively filtered thereby validating the HWPSS filtering operators and the full templates marginalization procedure. The residual structure seen in the difference map is below $\sim 0.02\, \mu$K in amplitude, and is due to the linear drift term that is not perfectly modelled in the map reconstruction, since the harmonics amplitudes are considered to be constant over each 10 s interval.

\subsection{\textbf{Case of a satellite mission.}}
The purpose of this section is to demonstrate our software in the context of full sky, satellite-like observations. Our study case is based roughly on a satellite mission, LiteBIRD~\cite{Hazumi_2020}. The presented results complement those discussed in the previous sections, which considered partial-sky observations. In addition, we employ here idealized simulations free of the presence of systematic effects and focus the discussion on statistical uncertainties of maps derived in different map-making set-ups.
\subsubsection{Simulations description}

We use TOAST to simulate full sky observations assuming roughly specifications of the LiteBIRD mission~\cite{Sugai_2020}. We simulate one year of observations at multiple frequency bands corresponding to those of LiteBIRD. The input sky includes CMB and galactic signal assuming the d1s1 foreground model from PySM \cite{PySM}. We set nside to 256, i.e., the total number of pixels in the maps is $786,432$. We do not include beams in the simulations. The scan strategy from the second Lagrange Point (L2) is described by the rotation of the satellite around the spin axis, with a period of 20 min, and the spin angle---the angle between the spin axis and the boresight direction---is $\beta = 50\,\deg$. In addition, the spin axis also precesses around the Sun-Earth axis with a period of $\sim 200$ min and a precession angle $\alpha = 45\,\deg$. Finally, the yearly revolution of the Sun-Earth axis around the Sun is also modeled.

We also simulate 1/f instrumental noise, with characteristic knee frequencies of about 20 mHz ($f_k$ in Eq.~\eqref{eqn:noisepowermodel}), and a slope of $\sim -1$ ($\alpha$ in Eq.~\eqref{eqn:noisepowermodel}). A new realization from the detector noise PSDs is drawn every six hours. The sampling rate is $\sim20$ Hz, which means that each detector data amounts to about $\sim 600$~M time samples. Depending on the instrument and the frequency channel, we need to process data from tens ($\sim20$) to hundreds ($\sim500$) detectors, which implies that the sizes of the data sets generated in the simulations range from $\mathcal{O}(10^{10})$ to $\mathcal{O}(10^{11})$ time samples, with the minimum required memory from $\mathcal{O}(1)$~TB to $\mathcal{O}(10)$~TB. In the following, we present results only for three selected channels with center frequencies of $\sim 40$, $\sim 140,$ and $\sim 235$ GHz and assuming $50, 360$ and $250$ detectors per channel, respectively. The half-wave plate rotation frequency is taken to be different for each of the channels, ranging from $\sim40$ to $\sim 60$ rpm, as are the white noise levels, which are taken to be $\sim20$, $\sim3$, and $\sim5$ $\mu$K$\sqrt{\mathrm{s}}$, respectively for the low, medium and high frequency channels considered here. 
All runs are performed on Cori at NERSC.
 
\subsubsection{Full-sky maximum likelihood maps}
\begin{figure}[!t]
\includegraphics[width=0.48\textwidth]{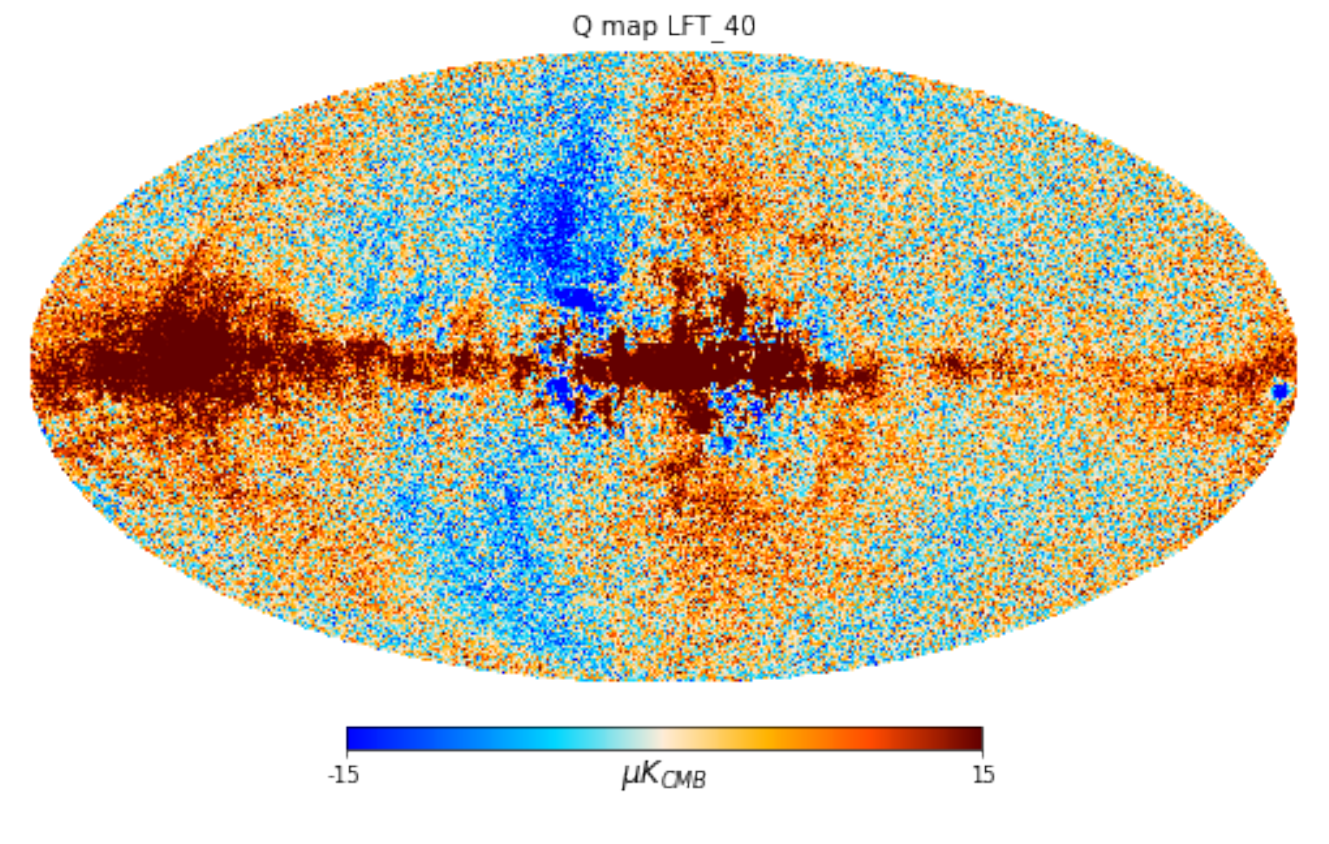}
\includegraphics[width=0.48\textwidth]{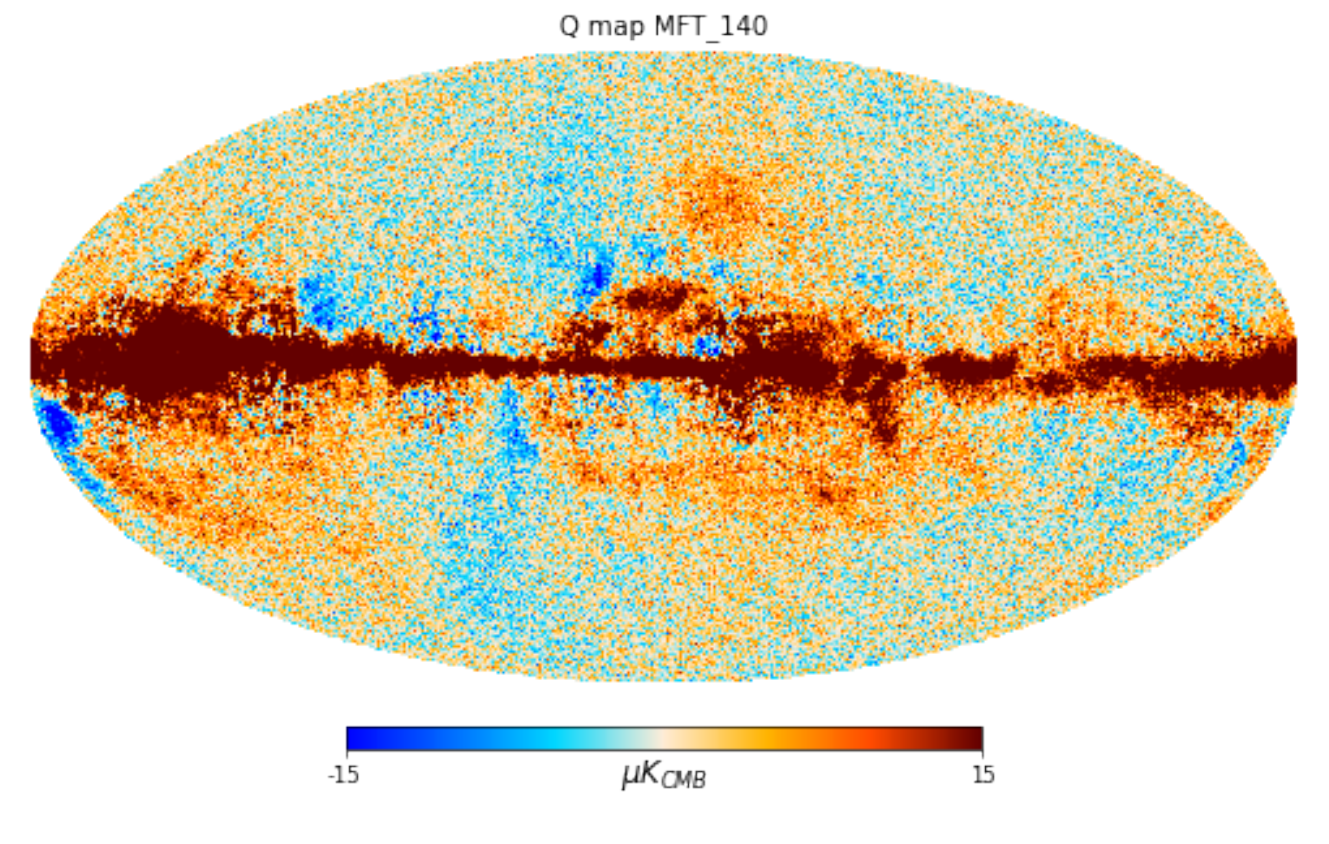}
\includegraphics[width=0.48\textwidth]{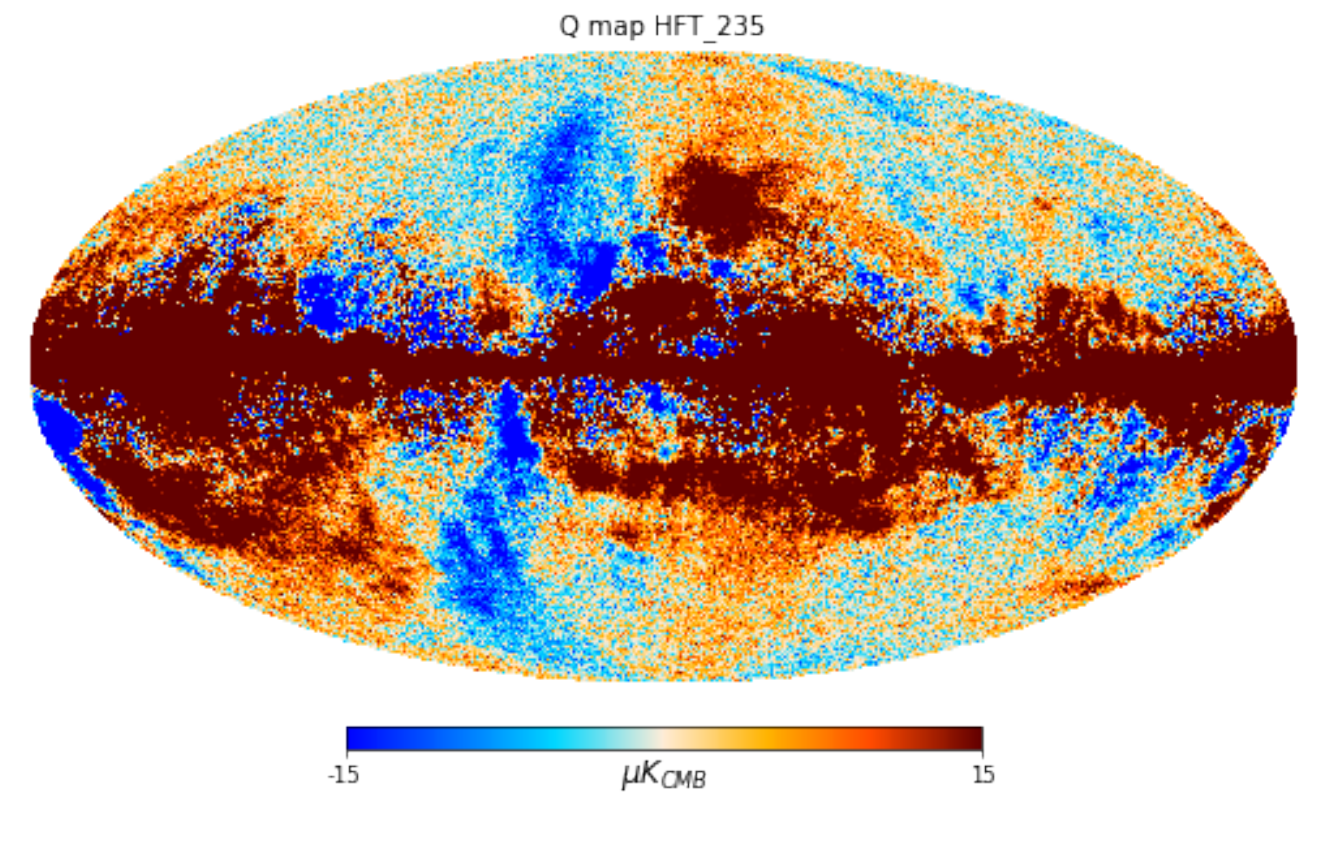}
\caption{\label{MP_LB} 
Full-sky maximum likelihood maps from 1 year of simulated observations of the LiteBIRD space mission. From \emph{top} to \emph{bottom} we show the Q polarization maps obtained in the $40$, $140$, and $235$ GHz frequency bands, respectively.
}
\end{figure}

In the simulated data described above, we expect no spatial noise correlations across the focal plane since the instrumental noise is drawn independently for each detector. This is a simplification of the actual case where the minor detector-detector correlation can, and will be, present due to systematic effects, such as read-out induced cross-talk or common thermal modes. In this case, we can get very close to the true maximum likelihood maps by using the banded block-Toeplitz noise model in the map reconstruction. In figure \ref{MP_LB}, we show the Q polarization full-sky maps derived from the data simulated respectively at 40, 140, and 235 GHz frequency band and processed with a map-maker assuming a Toeplitz-structured noise covariance with a half bandwidth of $\lambda=8,192$ time samples, corresponding to a correlation length of about $\sim 430$ s. Given the expected proximity of the resulting maps to the optimal solution, we hereafter refer to these as maximum likelihood when comparing them to other maps.

\begin{figure}[!ht]
\includegraphics[width=0.48\textwidth]{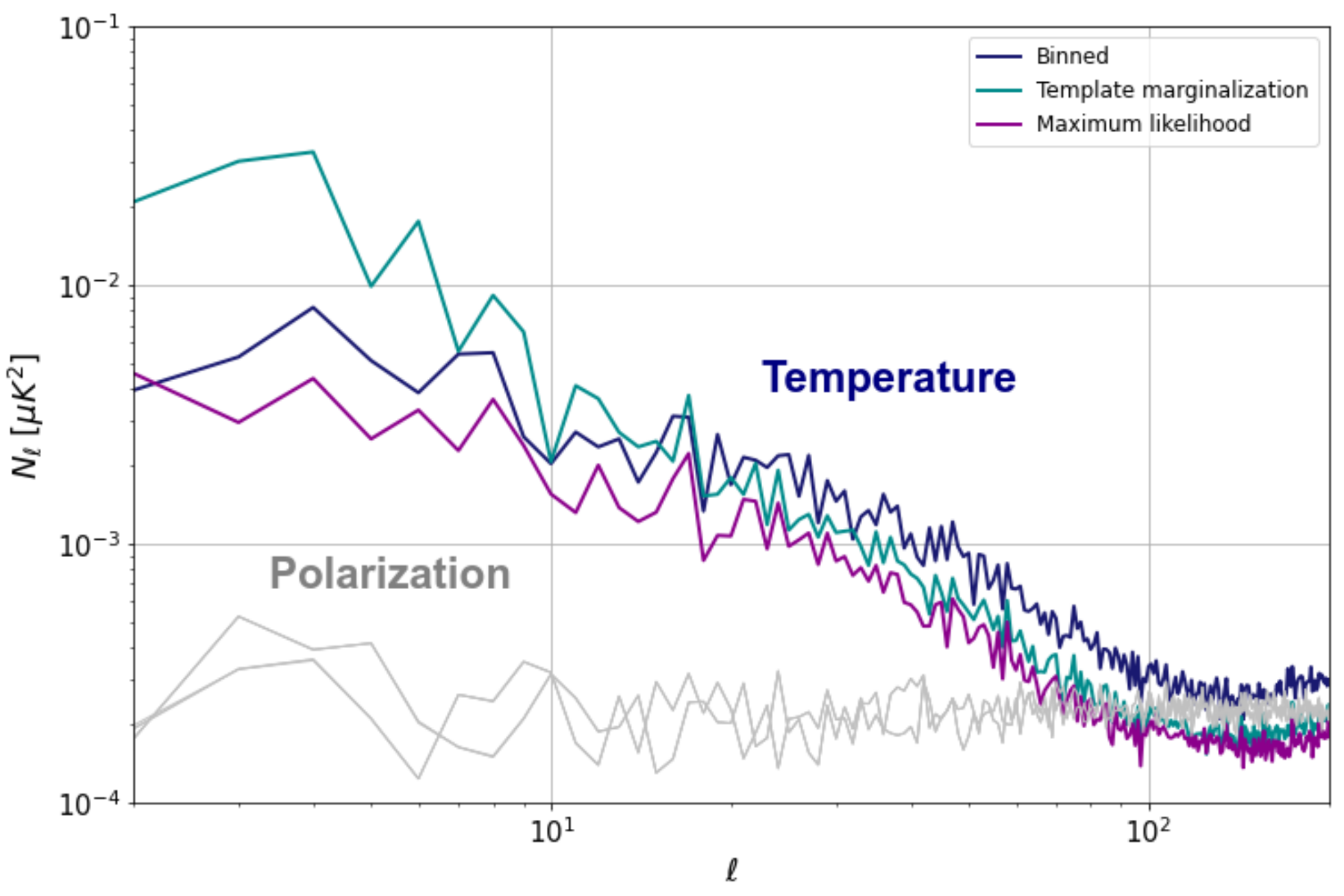}
\caption{\label{noise_power} 
Comparison of the noise power spectra obtained via the different map-making methods available in MAPPRAISER. The baseline for the Template marginalization run is set to $120$s. The maps are computed from the simulated data of the $40$ GHz channel.}
\end{figure}

\subsubsection{Full-sky templates marginalization maps}

In the templates marginalization approach, we use polynomial templates to mitigate the 1/f noise due to the instrument. We fix the polynomial order to 0 but vary the baseline length from 20 to 3,600 s in order to demonstrate its impact on the statistical uncertainties of the derived maps. The procedure is fully equivalent to standard destriping, allowing us to compare our findings with other results in the literature. In all cases the derived maps  are essentially indistinguishable by eye, between themselves and from the ML maps, for all Stokes parameters given their high signal-to-noise ratio. To make the differences more evident, we consider here instead difference maps of the output (noisy) and input (noiseless) maps. Given the unbiased character of all the maps discussed here, the difference maps reflect the noise present in the maps derived from the simulated data allowing us to quantify and compare their statistical uncertainties. The results of these tests are discussed in the following section, where we present the results obtained using data of the 40 GHz channel, but similar observations can be made about the other channels as well. 

\subsection{Noise spectra}

We use the case of the maximum likelihood maps as a reference to which we compare the noise structure in the map domain obtained with other mapping methods. We pay most of the attention to temperature maps, as the noise is essentially white in polarization in all studied cases.

Figure \ref{noise_power} shows the noise power spectra of the temperature and polarization maps computed from the difference between the output and input maps of the 40 GHz channel. The underlying maps are obtained with multiple map-making methods: maximum likelihood with the parameters mentioned above, templates marginalization including only polynomial templates with a baseline length fixed to 120 s, and binned map-making. The EE and BB noise spectra are both shown in grey and are consistent with the polarization white noise level expected from detector NETs. They are essentially flat, due to the HWP modulating the signal away from the low frequency 1/f noise, hence we do not see any difference in polarization between the different mapping approaches in this case. However, as mentioned previously, the simple configuration considered in the input simulation is not realistic, and one should expect a more complex noise structure to arise for the polarization signals when introducing systematics. The TT noise power spectra however show differences between different mapping approaches. In particular, the maximum likelihood estimator yields the least noisy maps as expected. The templates marginalization is closer to the maximum likelihood solution only at small scales, while the noise is boosted at large scales yielding worst estimates than the binned map-maker at $\ell \lesssim 20$.

\begin{figure}[!t]
\includegraphics[width=0.48\textwidth]{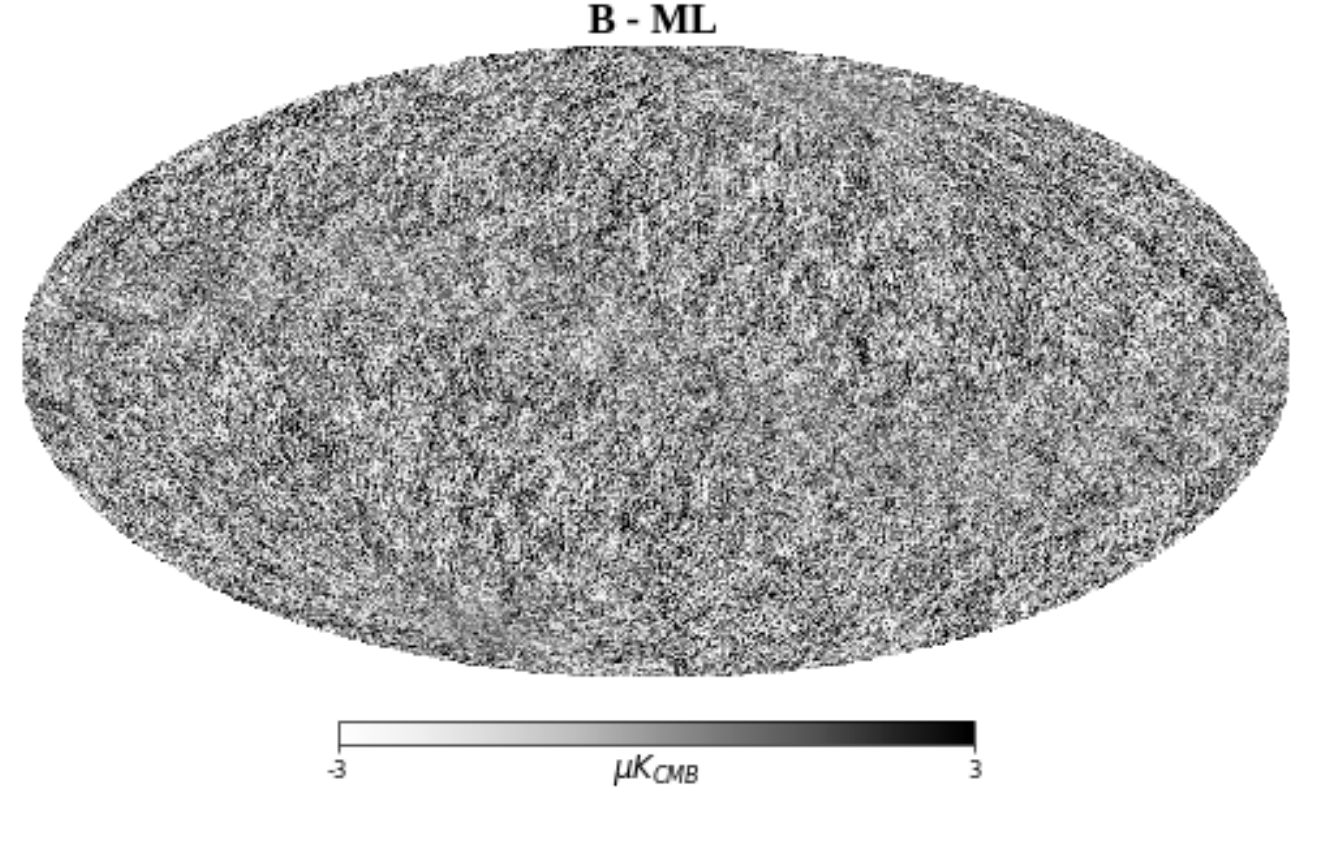}
\includegraphics[width=0.48\textwidth]{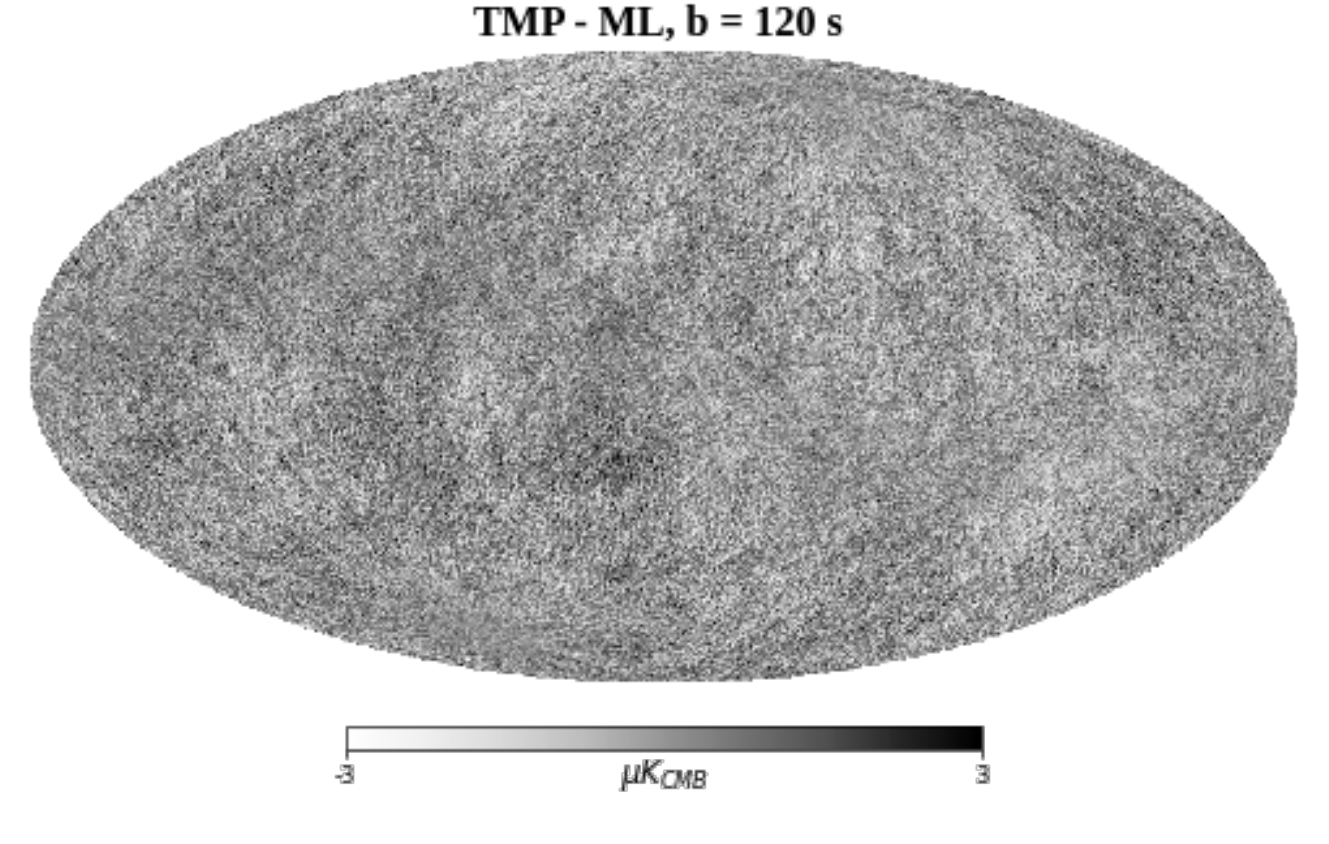}
\includegraphics[width=0.48\textwidth]{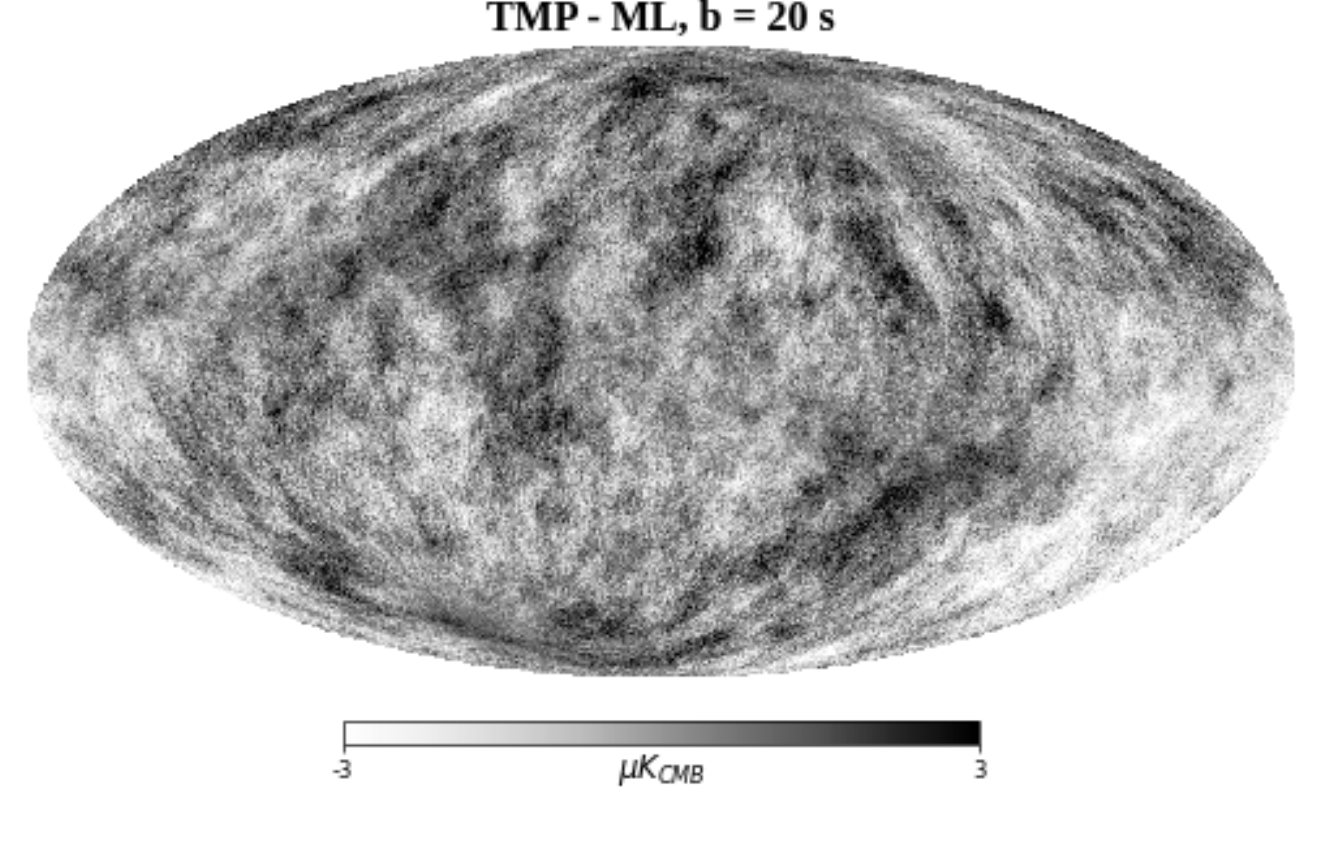}
\caption{\label{noise_diff} 
The upper map shows the difference between the binned temperature noise map and the maximum likelihood one. The middle and bottom maps show the difference between the templates marginalization temperature noise maps, respectively, for a baseline of $120$~s and $20$~s, and the maximum likelihood one. The maps are computed from simulated data of the $40$ GHz channel.}
\end{figure}

\begin{figure*}[!t]
\centering \includegraphics[width=1.0\textwidth]{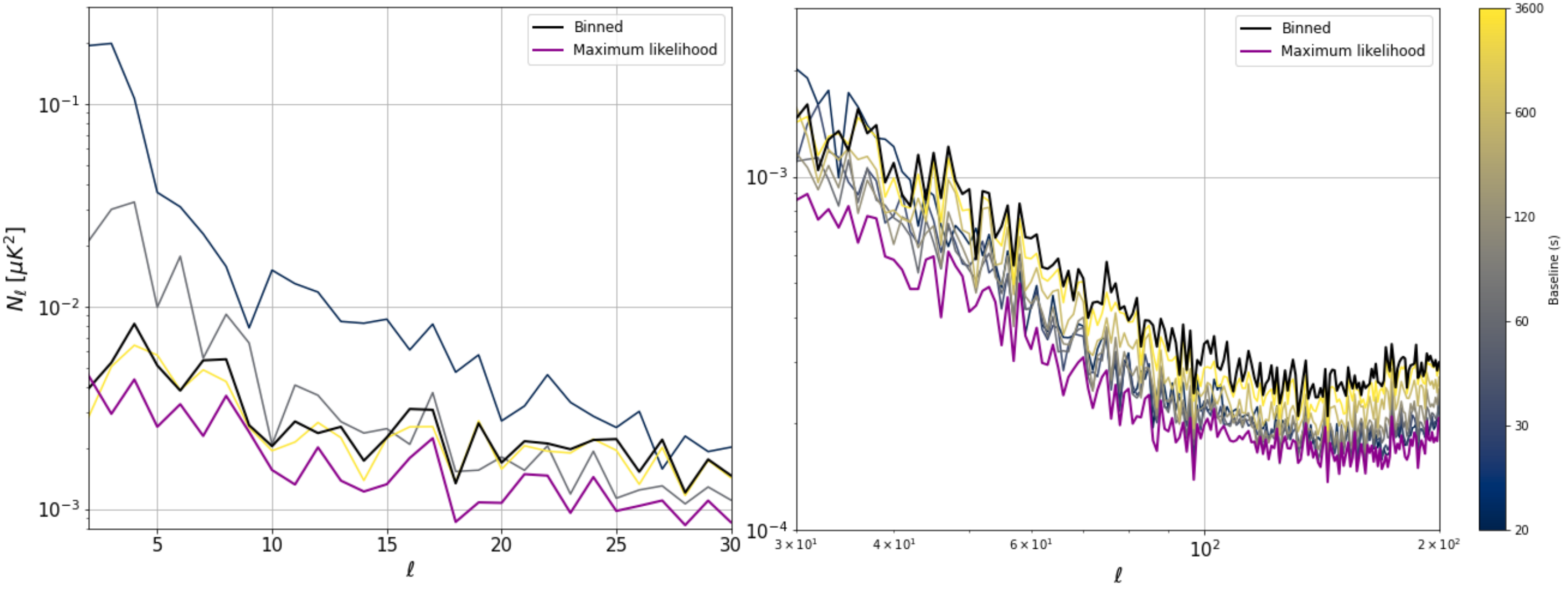}
\caption{\label{TMP_baselines} 
The temperature noise power spectra obtained via the templates marginalization approach for different baselines. The \emph{left} figure shows the $\ell \in \left[2, 30\right]$ range, and the \emph{right} figure shows the $\ell \in [30, 200]$ range. We also include the noise power spectra from the binned and maximum likelihood maps for comparison. The maps are computed from the simulated data of the $40$ GHz channel.}
\end{figure*}

Figure \ref{noise_diff} shows the difference between the binned noise map and the maximum likelihood noise map, as well as the differences between noise maps obtained with templates marginalization assuming baselines of 20~s or 120~s, and the noise maps obtained with the maximum likelihood estimator. Given that the binned mapper does not correct for the noise correlations, we see stripes in the difference maps. However, these features are not as strong as the usual stripe patterns seen in the binned Planck maps, for example Figure 6 of \cite{Keihanen_2010}. This is due to both a lack of sufficient noise power at low frequencies, given the smaller knee frequencies and milder slope of the detector noise assumed here and reflected in the simulation, and to the more optimal cross-linking of the LiteBIRD-like scanning strategy~\cite{Hazumi_2020, LB_science}. Indeed we see in figure \ref{noise_power}, that the TT noise power spectra of the binned case and the maximum likelihood case at the largest scales are roughly of the same order of magnitude, with the major difference being a highly anisotropic noise properties in the former case manifested by the stripes. These stripes are suppressed in the case of the maximum likelihood map-making by down weighting the long temporal modes and in the case of the template marginalization by their explicit filtering. This comes at the price of adding more degrees of freedom, thereby increasing the overall level of the statistical noise. This effect should be particularly noticeable for short baselines, and is clearly visible by comparing the maps of figure \ref{noise_diff}. Processing the maps with short baselines for the polynomial filters boosts the large scale features in the noise map (bottom map), although the map obtained with an intermediate baseline length of 120 s (middle map) shows less stripy patterns than the binned map (top map). 

These effects can be seen more clearly in the power spectra. Figure \ref{TMP_baselines} shows the TT noise power spectra for multiple baselines compared with the binned and maximum likelihood map-makers. The right panel shows the TT noise power spectra at multipoles $\ell \in [30, 200]$. As the baseline shortens towards 20 s, the lowest value tested in these runs, the noise level gets closer to the maximum likelihood estimate, hence the increase of the statistical noise at these scales does not dominate over the correction of the systematic errors induced by the noise correlations. Conversely, as we increase the baseline lengths, we get closer to the binned map estimate, which is expected since the offsets span durations close to the length of the stationary intervals defined in the simulation, and the map-maker does the exact same processing as a binned map-maker to any modes shorter than the offsets baselines, thus not properly handling the 1/f noise. 

The left panel shows the TT noise power spectra at lower multipoles, $\ell \in [2, 30]$. For clarity, only three baselines (20, 60 and 3,600 s) are selected. In this regime, the statistical noise dominates as the number of sky modes becomes low and the maps processed with filter baselines shorter than typically $\sim 1/f_k$ become noisier than the binned maps at these scales. These findings are similar to observations made in \cite{Kurki_Suonio_2009, Keihanen_2010} in applications of destripers to Planck simulations as well as for ground-based observations \cite{Sutton_2009}. In order to be able to process the maps with short baselines ($\lesssim 1/f_k$), one needs to include more information to bring these noise levels down. This could be done by introducing priors on the templates amplitudes, computed from the noise PSDs, as is done for instance in the MADAM or Descart destripers~\cite{Keihanen_2010, Sutton_2009}, this also typically speeds up the convergence of the map-maker. 

\section{Conclusion and future work}
In this paper we provide a detailed description of the MAPPRAISER software framework, which provides tools for computing faithful estimates of sky maps from the time-domain data collected by CMB experiments, in the presence of spurious signals which contribute linearly to the data. These tools benefit from a massively parallel implementation which is also easily extensible and portable. The framework allows the user to choose between different linear solvers and preconditioning techniques to maximize the computational efficiency according to the specific cases that need to be processed. In addition, one can also choose between different map-making methods according to the case at hand. The framework is flexible and extensible and can deal with cases of generalized map-making problems as those arising in the presence of frequency dependent systematic effects~\cite{Verges_2021}. The architecture is modular such as to allow other solvers, preconditioners, map-making approaches, or different templates for handling time-domain systematics to be easily integrated in the framework. 

\sloppy
We demonstrate all these features on simulated data from the TOAST software framework, using ground based and satellite experiments specifications. We first test the scalability and portability of the code by running our map-making code on multiple platforms using different collective communication schemes and show that the code exhibits good scaling up to our tests limit consisting of 16,384 MPI processes demonstrating its ability to handle large data sets. We then show a proof of concept of how convergence can be improved with the ECG solver, which to our knowledge, is the first application of the method in the context of CMB analysis. We conduct an in-depth study of the convergence properties of the two-level preconditioners and demonstrate a significant speedup in the context of multiple solves such as required for null tests or Monte Carlo simulations. We validate the templates marginalization procedure in the case of ground based experiments using two specific templates for ground pickup and half-wave plate synchronous signal. Finally, we also demonstrate the ability of the code to handle data from a full year of simulated observations from a satellite experiment such as LiteBIRD, and explore the noise properties in the recovered maps with the different map-making methods available.

Future work can focus on further optimization of the operations by capitalizing on GPU based architectures such as to exploit the potential of the ECG solver, or to enable efficient implementation of time-domain operations using a noise model which takes into account spatial correlations between detectors induced, for instance, by atmospheric emission, common thermal modes in the focal plane or other systematic effects such as detector crosstalk. Such GPU accelerated operations will also allow efficient implementations of spherical harmonic transforms useful for Wiener filtering or beam deconvolution map-making. The present software framework can also be used as a forecasting tool to assess the impact of different systematics naturally modeled in time-domain and their mitigation using realistic simulations of actual CMB experiments such as Simons Observatory, CMB-S4 and LiteBIRD. Finally, the framework can also be incorporated within broader  non-linear schemes integrating multiple steps of the CMB data analysis such as map-making and component separation into a single one~\cite{Papez_2020}.

\section*{Acknowledgement}
The work was supported by the French National Research Agency (Agence National de Recherche) grant, ANR-B3DCMB, (ANR-17-CE23-0002). We salute members of the B3DCMB collaboration for their multidisciplinary support and insights throughout the implementation of the project. We thank Julian Borill, Reijo Keskitalo, Theodore Kisner for their relevant comments and help with running the simulation software and interfacing it with MAPPRAISER. We also thank Sigurd N\ae ss, Matthew Hasselfield, and all members of the Simons Observatory Pipeline Working Group for their useful comments and discussions. We acknowledge help of Jan Pape\v{z} and Olivier Tissot with the integration of the ECG solver within the framework and the work of Pierre Cargemel and Fr\'ed\'eric Dauvergne on the development of the baseline version of MIDAPACK on which this work has built and improved upon.
We acknowledge the use of the NumPy~\cite{numpy}, Matplotlib~\cite{Matplotlib}, TOAST$^{\ref{foot:Toast}}$, NaMaster~\cite{Alonso_2019}, healpy~\cite{Zonca2019} and HEALPix~\cite{Gorski_2005} software packages.
This research used resources of the National Energy Research Scientific Computing Center (NERSC), a U.S. Department of Energy Office of Science User Facility located at Lawrence Berkeley National Laboratory, operated under Contract No. DE-AC02-05CH11231. 
In France the HPC resources were provided by GENCI–TGCC (allocation A0080411472).

\AtNextBibliography{\small}
\printbibliography

@article{Stompor_2001,
   title={Making maps of the cosmic microwave background: The MAXIMA example},
   volume={65},
   ISSN={1089-4918},
   url={http://dx.doi.org/10.1103/PhysRevD.65.022003},
   DOI={10.1103/physrevd.65.022003},
   number={2},
   journal={Physical Review D},
   publisher={American Physical Society (APS)},
   author={Stompor, R. and Balbi, Amedeo and Borrill, Julian D. and Ferreira, Pedro G. and Hanany, Shaul and Jaffe, Andrew H. and Lee, Adrian T. and Oh, Sang and Rabii, Bahman and Richards, Paul L. and et al.},
   year={2001},
   month=Dec
}

@article{Poletti_2017,
   title={Making maps of cosmic microwave background polarization for B-mode studies: the POLARBEAR example},
   volume={600},
   ISSN={1432-0746},
   url={http://dx.doi.org/10.1051/0004-6361/201629467},
   DOI={10.1051/0004-6361/201629467},
   journal={Astronomy \& Astrophysics},
   publisher={EDP Sciences},
   author={Poletti, D. and Fabbian, Giulio and Le Jeune, Maude and Peloton, Julien and Arnold, Kam and Baccigalupi, Carlo and Barron, Darcy and Beckman, Shawn and Borrill, Julian and Chapman, Scott and et al.},
   year={2017},
   month=Mar,
   pages={A60}
}

@article{Sutton_2009,
    author = {Sutton, D. and Johnson, B. R. and Brown, M. L. and Cabella, P. and Ferreira, P. G. and Smith, K. M.},
    title = "{Map making in small field modulated CMB polarization experiments: approximating the maximum likelihood method}",
    journal = {Monthly Notices of the Royal Astronomical Society},
    volume = {393},
    number = {3},
    pages = {894-910},
    year = {2009},
    month = {02},
    issn = {0035-8711},
    doi = {10.1111/j.1365-2966.2008.14195.x},
    url = {https://doi.org/10.1111/j.1365-2966.2008.14195.x},
    eprint = {https://academic.oup.com/mnras/article-pdf/393/3/894/18569414/mnras0393-0894.pdf},
}

@article{Sutton_2010,
   title={Fast and precise map-making for massively multi-detector CMB experiments},
   volume={407},
   ISSN={0035-8711},
   url={http://dx.doi.org/10.1111/j.1365-2966.2010.16954.x},
   DOI={10.1111/j.1365-2966.2010.16954.x},
   number={3},
   journal={Monthly Notices of the Royal Astronomical Society},
   publisher={Oxford University Press (OUP)},
   author={Sutton, D. and Zuntz, J. A. and Ferreira, P. G. and Brown, M. L. and Eriksen, H. K. and Johnson, B. R. and Kusaka, A. and Naess, S. K. and Wehus, I. K.},
   year={2010},
   month=Aug,
   pages={1387–1402}
}

@ARTICLE{Burigana_1999,
       author = {{Burigana}, C. and {Malaspina}, M. and {Mandolesi}, N. and {Danse}, L. and {Maino}, D. and {Bersanelli}, M. and {Maltoni}, M.},
        title = "{A preliminary study on destriping techniques of PLANCK/LFI measurements versus observational strategy}",
      journal = {arXiv e-prints},
         year = 1999,
        month = jun,
          eid = {astro-ph/9906360},
        pages = {astro-ph/9906360},
archivePrefix = {arXiv},
       eprint = {astro-ph/9906360},
 primaryClass = {astro-ph},
       adsurl = {https://ui.adsabs.harvard.edu/abs/1999astro.ph..6360B}
}

@article{Delabrouille_1998,
	author = {{Delabrouille}, J.},
	title = {Analysis of the accuracy of a destriping method for future cosmic
microwave background mapping with the PLANCK SURVEYOR satellite},
	DOI= "10.1051/aas:1998119",
	url= "https://doi.org/10.1051/aas:1998119",
	journal = {Astronomy \& Astrophysics Supplement Series},
	year = 1998,
	volume = 127,
	number = 3,
	pages = "555-567",
}

@article{Maino_1999,
   title={The Planck-LFI instrument: Analysis of the $1/f$ noise and
implications for the scanning strategy},
   volume={140},
   ISSN={1286-4846},
   url={http://dx.doi.org/10.1051/aas:1999429},
   DOI={10.1051/aas:1999429},
   number={3},
   journal={Astronomy \& Astrophysics Supplement Series},
   publisher={EDP Sciences},
   author={Maino, D. and Burigana, C. and Maltoni, M. and Wandelt, B. D. and Górski, K. M. and Malaspina, M. and Bersanelli, M. and Mandolesi, N. and Banday, A. J. and Hivon, E.},
   year={1999},
   month=Dec,
   pages={383–391}
}

@article{Revenu_2000,
	author = {{Revenu}, B. and {Kim, A.} and {Ansari, R.} and {Couchot, F.} and {Delabrouille, J.} and {Kaplan, J.}},
	title = {Destriping of polarized data in a CMB mission with a circular scanning strategy},
	DOI= "10.1051/aas:2000308",
	url= "https://doi.org/10.1051/aas:2000308",
	journal = {Astronomy \& Astrophysics Supplement Series},
	year = 2000,
	volume = 142,
	number = 3,
	pages = "499-509",
}

@ARTICLE{Keihanen_2004,
       author = {{Keih{\"a}nen}, E. and {Kurki-Suonio}, H. and {Poutanen}, T. and {Maino}, D. and {Burigana}, C.},
        title = "{A maximum likelihood approach to the destriping technique}",
      journal = {Astronomy \& Astrophysics},
         year = 2004,
        month = dec,
       volume = {428},
        pages = {287-298},
          doi = {10.1051/0004-6361:200400060},
archivePrefix = {arXiv},
       eprint = {astro-ph/0304411},
 primaryClass = {astro-ph},
       adsurl = {https://ui.adsabs.harvard.edu/abs/2004A&A...428..287K}
}

@article{Verges_2021,
   title={Framework for analysis of next generation, polarized CMB data sets in the presence of Galactic foregrounds and systematic effects},
   volume={103},
   ISSN={2470-0029},
   url={http://dx.doi.org/10.1103/PhysRevD.103.063507},
   DOI={10.1103/physrevd.103.063507},
   number={6},
   journal={Physical Review D},
   publisher={American Physical Society (APS)},
   author={Vergès, C. and Errard, J. and Stompor, R.},
   year={2021},
   month=Mar
}

@article{Armitage_2004,
   title={Deconvolution map-making for cosmic microwave background observations},
   volume={70},
   ISSN={1550-2368},
   url={http://dx.doi.org/10.1103/PhysRevD.70.123007},
   DOI={10.1103/physrevd.70.123007},
   number={12},
   journal={Physical Review D},
   publisher={American Physical Society (APS)},
   author={Armitage, C. and Wandelt, B.},
   year={2004},
   month=Dec
}

@article{Harrison_2011,
   title={A deconvolution map-making method for experiments with circular scanning strategies},
   volume={532},
   ISSN={1432-0746},
   url={http://dx.doi.org/10.1051/0004-6361/201116986},
   DOI={10.1051/0004-6361/201116986},
   journal={Astronomy \& Astrophysics},
   publisher={EDP Sciences},
   author={Harrison, D. L. and van Leeuwen, F. and Ashdown, M. A. J.},
   year={2011},
   month=Jul,
   pages={A55}
}

@article{Keihanen_2012,
   title={ArtDeco: a beam-deconvolution code for absolute cosmic microwave background measurements},
   volume={548},
   ISSN={1432-0746},
   url={http://dx.doi.org/10.1051/0004-6361/201220183},
   DOI={10.1051/0004-6361/201220183},
   journal={Astronomy \& Astrophysics},
   publisher={EDP Sciences},
   author={Keihänen, E. and Reinecke, M.},
   year={2012},
   month=Nov,
   pages={A110}
}

@article{Szydlarski_2014,
   title={Accelerating the cosmic microwave background map-making procedure through preconditioning},
   volume={572},
   ISSN={1432-0746},
   url={http://dx.doi.org/10.1051/0004-6361/201323210},
   DOI={10.1051/0004-6361/201323210},
   journal={Astronomy \& Astrophysics},
   publisher={EDP Sciences},
   author={Szydlarski, M. and Grigori, L. and Stompor, R.},
   year={2014},
   month=Nov,
   pages={A39}
}

@article{Ferreira_2000,
   title={Simultaneous estimation of noise and signal in cosmic microwave background experiments},
   volume={312},
   ISSN={1365-2966},
   url={http://dx.doi.org/10.1046/j.1365-8711.2000.03108.x},
   DOI={10.1046/j.1365-8711.2000.03108.x},
   number={1},
   journal={Monthly Notices of the Royal Astronomical Society},
   publisher={Oxford University Press (OUP)},
   author={Ferreira, P. G. and Jaffe, A. H.},
   year={2000},
   month=Feb,
   pages={89–102}
}

@article{Wehus_2012,
   title={Bayesian noise estimation for non-ideal cosmic microwave background experiments},
   volume={199},
   ISSN={1538-4365},
   url={http://dx.doi.org/10.1088/0067-0049/199/1/15},
   DOI={10.1088/0067-0049/199/1/15},
   number={1},
   journal={The Astrophysical Journal Supplement Series},
   publisher={American Astronomical Society},
   author={Wehus, I. K. and Næss, S. K. and Eriksen, H. K.},
   year={2012},
   month=Feb,
   pages={15}
}

@article{Maxipol_2008,
author = {Johnson, B. and Collins, Judy and Abroe, M. and Ade, Peter and Bock, J. and Borrill, J. and Boscaleri, A. and Bernardis, P. and Hanany, Shaul and Jaffe, A. and Jones, T. and Lee, A. and Levinson, L. and Matsumura, T. and Rabii, B. and Renbarger, T. and Richards, P. and Smoot, George and Stompor, Radek and Zuntz, and},
year = {2008},
month = {12},
pages = {42},
title = {MAXIPOL: Cosmic Microwave Background Polarimetry Using a Rotating Half-Wave Plate},
volume = {665},
journal = {The Astrophysical Journal},
doi = {10.1086/518105}
}

@article{Kusaka_2014,
   title={Modulation of cosmic microwave background polarization with a warm rapidly rotating half-wave plate on the Atacama B-Mode Search instrument},
   volume={85},
   ISSN={1089-7623},
   url={http://dx.doi.org/10.1063/1.4862058},
   DOI={10.1063/1.4862058},
   number={2},
   journal={Review of Scientific Instruments},
   publisher={AIP Publishing},
   author={Kusaka, A. and Essinger-Hileman, T. and Appel, J. W. and Gallardo, P. and Irwin, K. D. and Jarosik, N. and Nolta, M. R. and Page, L. A. and Parker, L. P. and Raghunathan, S. and et al.},
   year={2014},
   month=Feb,
   pages={024501}
}

@book{num_recipes_3rd,
author = {Press, W. H. and Teukolsky, Saul A. and Vetterling, William T. and Flannery, Brian P.},
title = {Numerical Recipes 3rd Edition: The Art of Scientific Computing},
year = {2007},
isbn = {0521880688},
publisher = {Cambridge University Press},
address = {USA},
edition = {3},
}

@ARTICLE{Prunet_2001,
       author = {{Prunet}, S. and {Ade}, P.~A.~R. and {Bock}, J.~J. and {Bond}, J.~R. and {Borrill}, J. and {Boscaleri}, A. and {Coble}, K. and {Crill}, B.~P. and {de Bernardis}, P. and {De Gasperis}, G. and {De Troia}, G. and {Farese}, P.~C. and {Ferreira}, P.~G. and {Ganga}, K. and {Giacometti}, M. and {Hivon}, E. and {Hristov}, V.~V. and {Iacoangeli}, A. and {Jaffe}, A.~H. and {Lange}, A.~E. and {Martinis}, L. and {Masi}, S. and {Mason}, P. and {Mauskopf}, P.~D. and {Melchiorri}, A. and {Miglio}, L. and {Montroy}, T. and {Netterfield}, C.~B. and {Pascale}, E. and {Piacentini}, F. and {Pogosyan}, D. and {Pongetti}, F. and {Prunet}, S. and {Rao}, S. and {Romeo}, G. and {Ruhl}, J.~E. and {Scaramuzzi}, F. and {Sforna}, D. and {Vittorio}, N.},
        title = "{Noise estimation in CMB time-streams and fast map-making. Application to the BOOMERanG98 data}",
      journal = {arXiv e-prints},
         year = 2001,
        month = jan,
          eid = {astro-ph/0101073},
        pages = {astro-ph/0101073},
archivePrefix = {arXiv},
       eprint = {astro-ph/0101073},
 primaryClass = {astro-ph},
       adsurl = {https://ui.adsabs.harvard.edu/abs/2001astro.ph..1073P}
}

@article{Natoli_2001,
   title={A Map-Making algorithm for the Planck Surveyor},
   volume={372},
   ISSN={1432-0746},
   url={http://dx.doi.org/10.1051/0004-6361:20010393},
   DOI={10.1051/0004-6361:20010393},
   number={1},
   journal={Astronomy \& Astrophysics},
   publisher={EDP Sciences},
   author={Natoli, P. and de Gasperis, G. and Gheller, C. and Vittorio, N.},
   year={2001},
   month=Jun,
   pages={346–356}
}

@article{Cantalupo_2010,
   title={MADmap: A massively parallel maximum likelihood cosmic microwave background map-maker},
   volume={187},
   ISSN={1538-4365},
   url={http://dx.doi.org/10.1088/0067-0049/187/1/212},
   DOI={10.1088/0067-0049/187/1/212},
   number={1},
   journal={The Astrophysical Journal Supplement Series},
   publisher={American Astronomical Society},
   author={Cantalupo, C. M. and Borrill, J. D. and Jaffe, A. H. and Kisner, T. S. and Stompor, R.},
   year={2010},
   month=Mar,
   pages={212–227}
}

@article{Yvon_2005,
   title={Mirage: A new iterative map-making code for CMB experiments},
   volume={436},
   ISSN={1432-0746},
   url={http://dx.doi.org/10.1051/0004-6361:20035920},
   DOI={10.1051/0004-6361:20035920},
   number={2},
   journal={Astronomy \& Astrophysics},
   publisher={EDP Sciences},
   author={Yvon, D. and Mayet, F.},
   year={2005},
   month=May,
   pages={729–739}
}

@article{Papez_2020,
   title={Accelerating linear system solvers for time-domain component separation of cosmic microwave background data},
   volume={638},
   ISSN={1432-0746},
   url={http://dx.doi.org/10.1051/0004-6361/202037687},
   DOI={10.1051/0004-6361/202037687},
   journal={Astronomy \& Astrophysics},
   publisher={EDP Sciences},
   author={Papež, J. and Grigori, L. and Stompor, R.},
   year={2020},
   month=Jun,
   pages={A73}
}

@article{grigori_2016,
  TITLE = {{Enlarged Krylov Subspace Conjugate Gradient Methods for Reducing Communication}},
  AUTHOR = {L. {Grigori} and S. {Moufawad} and F. {Nataf}},
  URL = {https://hal.inria.fr/hal-01357899},
  JOURNAL = {{SIAM Journal on Matrix Analysis and Applications}},
  PUBLISHER = {{Society for Industrial and Applied Mathematics}},
  VOLUME = {37},
  NUMBER = {2},
  PAGES = {744--773},
  YEAR = {2016},
  DOI = {10.1137/140989492},
  HAL_ID = {hal-01357899},
  HAL_VERSION = {v1},
}

@article{Grigori_Tissot2019,
author = {Grigori, L. and Tissot, O.},
title = {Scalable Linear Solvers Based on Enlarged Krylov Subspaces with Dynamic Reduction of Search Directions},
journal = {SIAM Journal on Scientific Computing},
volume = {41},
number = {5},
pages = {C522-C547},
year = {2019},
doi = {10.1137/18M1196285},
URL = { 
        https://doi.org/10.1137/18M1196285
},
eprint = { 
        https://doi.org/10.1137/18M1196285
}
}

@article{Saad_2000,
author = {Saad, Y. and Yeung, M. and Erhel, J. and Guyomarc'h, F.},
title = {A Deflated Version of the Conjugate Gradient Algorithm},
journal = {SIAM Journal on Scientific Computing},
volume = {21},
number = {5},
pages = {1909-1926},
year = {2000},
doi = {10.1137/S1064829598339761},
URL = { 
        https://doi.org/10.1137/S1064829598339761
},
eprint = { 
        https://doi.org/10.1137/S1064829598339761
}
}

@book{saad_2003,
  author = {Saad, Y.},
  doi = {10.1137/1.9780898718003},
  edition = {Second},
  isbn = {978-0-89871-534-7},
  publisher = {SIAM},
  series = {Other Titles in Applied Mathematics},
  title = {{Iterative Methods for Sparse Linear Systems}},
  url = {http://www-users.cs.umn.edu/\~{}saad/IterMethBook\_2ndEd.pdf},
  year = 2003
}

@article{Tang_2009,
author = {Tang, J.M. and Nabben, R and Vuik, C. and Erlangga, Y.A.},
year = {2009},
month = {06},
pages = {},
title = {Comparison of Two-Level Preconditioners Derived from Deflation, Domain Decomposition and Multigrid Methods},
volume = {39},
journal = {Journal of Scientific Computing, 39 (3), 2009},
doi = {10.1007/s10915-009-9272-6}
}

@book{golub13,
  author = {Golub, G. H. and van Loan, C. F.},
  edition = {Fourth},
  publisher = {JHU Press},
  refid = {824733531},
  title = {Matrix Computations},
  url = {http://www.cs.cornell.edu/cv/GVL4/golubandvanloan.htm},
  year = 2013
}

@article{Puglisi_2018,
   title={Iterative map-making with two-level preconditioning for polarized cosmic microwave background data sets},
   volume={618},
   ISSN={1432-0746},
   url={http://dx.doi.org/10.1051/0004-6361/201832710},
   DOI={10.1051/0004-6361/201832710},
   journal={Astronomy \& Astrophysics},
   publisher={EDP Sciences},
   author={Puglisi, G. and Poletti, Davide and Fabbian, Giulio and Baccigalupi, Carlo and Heltai, Luca and Stompor, Radek},
   year={2018},
   month=Oct,
   pages={A62}
}

@ARTICLE{Huffenberger_2021,
       author = {{Chiang}, B.-C. and {Huffenberger}, K. M.},
        title = "{Cooling Improves Cosmic Microwave Background Map-Making When Low-Frequency Noise is Large}",
      journal = {arXiv e-prints},
         year = 2021,
        month = sep,
          eid = {arXiv:2109.11622},
        pages = {arXiv:2109.11622},
archivePrefix = {arXiv},
       eprint = {2109.11622},
 primaryClass = {astro-ph.CO},
       adsurl = {https://ui.adsabs.harvard.edu/abs/2021arXiv210911622C}
}

@ARTICLE{Huffenberger_2018,
       author = {{Huffenberger}, K. M. and {N\ae ss}, S. K.},
        title = "{Cosmic Microwave Background Mapmaking with a Messenger Field}",
      journal = {The Astrophysical Journal},
         year = 2018,
        month = jan,
       volume = {852},
       number = {2},
          eid = {92},
        pages = {92},
          doi = {10.3847/1538-4357/aa9c7d},
archivePrefix = {arXiv},
       eprint = {1705.01893},
 primaryClass = {astro-ph.IM},
       adsurl = {https://ui.adsabs.harvard.edu/abs/2018ApJ...852...92H}
}

@ARTICLE{Dore_2001,
       author = {{Dor{\'e}}, O. and {Teyssier}, R. and {Bouchet}, F.~R. and {Vibert}, D. and {Prunet}, S.},
        title = "{MAPCUMBA: A fast iterative multi-grid map-making algorithm for CMB experiments}",
      journal = {Astronomy \& Astrophysics},
         year = 2001,
        month = jul,
       volume = {374},
        pages = {358-370},
          doi = {10.1051/0004-6361:20010692},
archivePrefix = {arXiv},
       eprint = {astro-ph/0101112},
 primaryClass = {astro-ph},
       adsurl = {https://ui.adsabs.harvard.edu/abs/2001A&A...374..358D}
}

@article{Gorski_2005,
   title={HEALPix: A Framework for High‐Resolution Discretization and Fast Analysis of Data Distributed on the Sphere},
   volume={622},
   ISSN={1538-4357},
   url={http://dx.doi.org/10.1086/427976},
   DOI={10.1086/427976},
   number={2},
   journal={The Astrophysical Journal},
   publisher={American Astronomical Society},
   author={Górski, K. M. and Hivon, E. and Banday, A. J. and Wandelt, B. D. and Hansen, F. K. and Reinecke, M. and Bartelmann, M.},
   year={2005},
   month=Apr,
   pages={759–771}
}

@ARTICLE{PySM,
       author = {{Thorne}, B. and {Dunkley}, J. and {Alonso}, D. and {N{\ae}ss}, S.},
        title = "{The Python Sky Model: software for simulating the Galactic microwave sky}",
      journal = {Monthly Notices of the Royal Astronomical Society},
         year = 2017,
        month = aug,
       volume = {469},
       number = {3},
        pages = {2821-2833},
          doi = {10.1093/mnras/stx949},
archivePrefix = {arXiv},
       eprint = {1608.02841},
 primaryClass = {astro-ph.CO},
       adsurl = {https://ui.adsabs.harvard.edu/abs/2017MNRAS.469.2821T}
}

@article{Keihanen_2010,
   title={Making cosmic microwave background temperature and polarization maps with MADAM},
   volume={510},
   ISSN={1432-0746},
   url={http://dx.doi.org/10.1051/0004-6361/200912813},
   DOI={10.1051/0004-6361/200912813},
   journal={Astronomy \& Astrophysics},
   publisher={EDP Sciences},
   author={Keihänen, E. and Keskitalo, R. and Kurki-Suonio, H. and Poutanen, T. and Sirviö, A.-S.},
   year={2010},
   month=Feb,
   pages={A57}
}

@ARTICLE{Planck_2018_cosmo,
       author = {{Planck Collaboration} and {Aghanim}, N. and {Akrami}, Y. and {Ashdown}, M. and {Aumont}, J. and {Baccigalupi}, C. and {Ballardini}, M. and {Banday}, A.~J. and {Barreiro}, R.~B. and {Bartolo}, N. and {Basak}, S. and {Battye}, R. and {Benabed}, K. and {Bernard}, J. -P. and {Bersanelli}, M. and {Bielewicz}, P. and {Bock}, J.~J. and {Bond}, J.~R. and {Borrill}, J. and {Bouchet}, F.~R. and {Boulanger}, F. and {Bucher}, M. and {Burigana}, C. and {Butler}, R.~C. and {Calabrese}, E. and {Cardoso}, J. -F. and {Carron}, J. and {Challinor}, A. and {Chiang}, H.~C. and {Chluba}, J. and {Colombo}, L.~P.~L. and {Combet}, C. and {Contreras}, D. and {Crill}, B.~P. and {Cuttaia}, F. and {de Bernardis}, P. and {de Zotti}, G. and {Delabrouille}, J. and {Delouis}, J. -M. and {Di Valentino}, E. and {Diego}, J.~M. and {Dor{\'e}}, O. and {Douspis}, M. and {Ducout}, A. and {Dupac}, X. and {Dusini}, S. and {Efstathiou}, G. and {Elsner}, F. and {En{\ss}lin}, T.~A. and {Eriksen}, H.~K. and {Fantaye}, Y. and {Farhang}, M. and {Fergusson}, J. and {Fernandez-Cobos}, R. and {Finelli}, F. and {Forastieri}, F. and {Frailis}, M. and {Fraisse}, A.~A. and {Franceschi}, E. and {Frolov}, A. and {Galeotta}, S. and {Galli}, S. and {Ganga}, K. and {G{\'e}nova-Santos}, R.~T. and {Gerbino}, M. and {Ghosh}, T. and {Gonz{\'a}lez-Nuevo}, J. and {G{\'o}rski}, K.~M. and {Gratton}, S. and {Gruppuso}, A. and {Gudmundsson}, J.~E. and {Hamann}, J. and {Handley}, W. and {Hansen}, F.~K. and {Herranz}, D. and {Hildebrandt}, S.~R. and {Hivon}, E. and {Huang}, Z. and {Jaffe}, A.~H. and {Jones}, W.~C. and {Karakci}, A. and {Keih{\"a}nen}, E. and {Keskitalo}, R. and {Kiiveri}, K. and {Kim}, J. and {Kisner}, T.~S. and {Knox}, L. and {Krachmalnicoff}, N. and {Kunz}, M. and {Kurki-Suonio}, H. and {Lagache}, G. and {Lamarre}, J. -M. and {Lasenby}, A. and {Lattanzi}, M. and {Lawrence}, C.~R. and {Le Jeune}, M. and {Lemos}, P. and {Lesgourgues}, J. and {Levrier}, F. and {Lewis}, A. and {Liguori}, M. and {Lilje}, P.~B. and {Lilley}, M. and {Lindholm}, V. and {L{\'o}pez-Caniego}, M. and {Lubin}, P.~M. and {Ma}, Y. -Z. and {Mac{\'\i}as-P{\'e}rez}, J.~F. and {Maggio}, G. and {Maino}, D. and {Mandolesi}, N. and {Mangilli}, A. and {Marcos-Caballero}, A. and {Maris}, M. and {Martin}, P.~G. and {Martinelli}, M. and {Mart{\'\i}nez-Gonz{\'a}lez}, E. and {Matarrese}, S. and {Mauri}, N. and {McEwen}, J.~D. and {Meinhold}, P.~R. and {Melchiorri}, A. and {Mennella}, A. and {Migliaccio}, M. and {Millea}, M. and {Mitra}, S. and {Miville-Desch{\^e}nes}, M. -A. and {Molinari}, D. and {Montier}, L. and {Morgante}, G. and {Moss}, A. and {Natoli}, P. and {N{\o}rgaard-Nielsen}, H.~U. and {Pagano}, L. and {Paoletti}, D. and {Partridge}, B. and {Patanchon}, G. and {Peiris}, H.~V. and {Perrotta}, F. and {Pettorino}, V. and {Piacentini}, F. and {Polastri}, L. and {Polenta}, G. and {Puget}, J. -L. and {Rachen}, J.~P. and {Reinecke}, M. and {Remazeilles}, M. and {Renzi}, A. and {Rocha}, G. and {Rosset}, C. and {Roudier}, G. and {Rubi{\~n}o-Mart{\'\i}n}, J.~A. and {Ruiz-Granados}, B. and {Salvati}, L. and {Sandri}, M. and {Savelainen}, M. and {Scott}, D. and {Shellard}, E.~P.~S. and {Sirignano}, C. and {Sirri}, G. and {Spencer}, L.~D. and {Sunyaev}, R. and {Suur-Uski}, A. -S. and {Tauber}, J.~A. and {Tavagnacco}, D. and {Tenti}, M. and {Toffolatti}, L. and {Tomasi}, M. and {Trombetti}, T. and {Valenziano}, L. and {Valiviita}, J. and {Van Tent}, B. and {Vibert}, L. and {Vielva}, P. and {Villa}, F. and {Vittorio}, N. and {Wandelt}, B.~D. and {Wehus}, I.~K. and {White}, M. and {White}, S.~D.~M. and {Zacchei}, A. and {Zonca}, A.},
        title = "{Planck 2018 results. VI. Cosmological parameters}",
      journal = {Astronomy \& Astrophysics},
         year = 2020,
        month = sep,
       volume = {641},
          eid = {A6},
        pages = {A6},
          doi = {10.1051/0004-6361/201833910},
archivePrefix = {arXiv},
       eprint = {1807.06209},
 primaryClass = {astro-ph.CO},
       adsurl = {https://ui.adsabs.harvard.edu/abs/2020A&A...641A...6P}
}

@book{pachenko11,
  author = {Pacheco, P.},
  isbn = {9780123742605 0123742609},
  publisher = {Morgan Kaufmann},
  refid = {668986119},
  title = {An Introduction to Parallel Programming},
  year = 2011
}

@INPROCEEDINGS{Grigori_2012,
  author={Grigori, L. and Stompor, R. and Szydlarski, M.},
  booktitle={SC '12: Proceedings of the International Conference on High Performance Computing, Networking, Storage and Analysis}, 
  title={A parallel two-level preconditioner for Cosmic Microwave Background map-making}, 
  year={2012},
  volume={},
  number={},
  pages={1-10},
  doi={10.1109/SC.2012.10}
}

@article{Naess_2014,
   title={The Atacama Cosmology Telescope: CMB polarization at {$200 < \ell < 9000$}},
   volume={2014},
   ISSN={1475-7516},
   url={http://dx.doi.org/10.1088/1475-7516/2014/10/007},
   DOI={10.1088/1475-7516/2014/10/007},
   number={10},
   journal={Journal of Cosmology and Astroparticle Physics},
   publisher={IOP Publishing},
   author={Naess, S. and Hasselfield, M. and McMahon, J. and Niemack, Michael D. and Addison, Graeme E. and Ade, Peter A. R. and Allison, Rupert and Amiri, Mandana and Battaglia, Nick and Beall, James A. and et al.},
   year={2014},
   month=Oct,
   pages={007–007}
}

@article{Takakura_2017,
   title={Performance of a continuously rotating half-wave plate on the POLARBEAR telescope},
   volume={2017},
   ISSN={1475-7516},
   url={http://dx.doi.org/10.1088/1475-7516/2017/05/008},
   DOI={10.1088/1475-7516/2017/05/008},
   number={05},
   journal={Journal of Cosmology and Astroparticle Physics},
   publisher={IOP Publishing},
   author={Takakura, S. and Aguilar, Mario and Akiba, Yoshiki and Arnold, Kam and Baccigalupi, Carlo and Barron, Darcy and Beckman, Shawn and Boettger, David and Borrill, Julian and Chapman, Scott and et al.},
   year={2017},
   month=May,
   pages={008–008}
}

@ARTICLE{LB_science,
       author = {{The LiteBIRD collaboration}},
      journal = {PTEP},
         year = {2021, in prep}
}

@article{Hazumi_2020,
   title={LiteBIRD satellite: JAXA’s new strategic L-class mission for all-sky surveys of cosmic microwave background polarization},
   ISBN={9781510636743},
   url={http://dx.doi.org/10.1117/12.2563050},
   DOI={10.1117/12.2563050},
   journal={Space Telescopes and Instrumentation 2020: Optical, Infrared, and Millimeter Wave},
   publisher={SPIE},
   author={Hazumi, M. and Ade, Peter A. and Adler, Alexandre and Allys, Erwan and Arnold, Kam and Auguste, Didier and Aumont, Jonathan and Aurlien, Ragnhild and Austermann, Jason and Baccigalupi, Carlo and et al.},
   editor={Lystrup, Makenzie and Batalha, Natalie and Tong, Edward C. and Siegler, Nicholas and Perrin, Marshall D.Editors},
   year={2020},
   month=Dec
}

@article{Kurki_Suonio_2009,
   title={Destriping CMB temperature and polarization maps},
   volume={506},
   ISSN={1432-0746},
   url={http://dx.doi.org/10.1051/0004-6361/200912361},
   DOI={10.1051/0004-6361/200912361},
   number={3},
   journal={Astronomy \& Astrophysics},
   publisher={EDP Sciences},
   author={Kurki-Suonio, H. and Keihänen, E. and Keskitalo, R. and Poutanen, T. and Sirviö, A.-S. and Maino, D. and Burigana, C.},
   year={2009},
   month=Aug,
   pages={1511–1539}
}

@article{BBpaper2020,
   title={A Measurement of the Degree-scale CMB B-mode Angular Power Spectrum with POLARBEAR},
   volume={897},
   ISSN={1538-4357},
   url={http://dx.doi.org/10.3847/1538-4357/ab8f24},
   DOI={10.3847/1538-4357/ab8f24},
   number={1},
   journal={The Astrophysical Journal},
   publisher={American Astronomical Society},
   author={Adachi, S. and Aguilar Faúndez, M. A. O. and Arnold, K. and Baccigalupi, C. and Barron, D. and Beck, D. and Beckman, S. and Bianchini, F. and Boettger, D. and et al.},
   year={2020},
   month=Jul,
   pages={55}
}

@article{Planck2018_legacy,
	author = {{Planck Collaboration} and {Aghanim, N.} and {Akrami, Y.} and {Arroja, F.} and {Ashdown, M.} and {Aumont, J.} and {Baccigalupi, C.} and {Ballardini, M.} and {Banday, A. J.} and {Barreiro, R. B.} and {Bartolo, N.} and {Basak, S.} and {Battye, R.} and {Benabed, K.} and {Bernard, J.-P.} and {Bersanelli, M.} and {Bielewicz, P.} and {Bock, J. J.} and {Bond, J. R.} and {Borrill, J.} and {Bouchet, F. R.} and {Boulanger, F.} and {Bucher, M.} and {Burigana, C.} and {Butler, R. C.} and {Calabrese, E.} and {Cardoso, J.-F.} and {Carron, J.} and {Casaponsa, B.} and {Challinor, A.} and {Chiang, H. C.} and {Colombo, L. P. L.} and {Combet, C.} and {Contreras, D.} and {Crill, B. P.} and {Cuttaia, F.} and {de Bernardis, P.} and {de Zotti, G.} and {Delabrouille, J.} and {Delouis, J.-M.} and {D\'esert, F.-X.} and {Di Valentino, E.} and {Dickinson, C.} and {Diego, J. M.} and {Donzelli, S.} and {Dor\'e, O.} and {Douspis, M.} and {Ducout, A.} and {Dupac, X.} and {Efstathiou, G.} and {Elsner, F.} and {En\ss{}lin, T. A.} and {Eriksen, H. K.} and {Falgarone, E.} and {Fantaye, Y.} and {Fergusson, J.} and {Fernandez-Cobos, R.} and {Finelli, F.} and {Forastieri, F.} and {Frailis, M.} and {Franceschi, E.} and {Frolov, A.} and {Galeotta, S.} and {Galli, S.} and {Ganga, K.} and {G\'enova-Santos, R. T.} and {Gerbino, M.} and {Ghosh, T.} and {Gonz\'alez-Nuevo, J.} and {G\'orski, K. M.} and {Gratton, S.} and {Gruppuso, A.} and {Gudmundsson, J. E.} and {Hamann, J.} and {Handley, W.} and {Hansen, F. K.} and {Helou, G.} and {Herranz, D.} and {Hildebrandt, S. R.} and {Hivon, E.} and {Huang, Z.} and {Jaffe, A. H.} and {Jones, W. C.} and {Karakci, A.} and {Keih\"anen, E.} and {Keskitalo, R.} and {Kiiveri, K.} and {Kim, J.} and {Kisner, T. S.} and {Knox, L.} and {Krachmalnicoff, N.} and {Kunz, M.} and {Kurki-Suonio, H.} and {Lagache, G.} and {Lamarre, J.-M.} and {Langer, M.} and {Lasenby, A.} and {Lattanzi, M.} and {Lawrence, C. R.} and {Le Jeune, M.} and {Leahy, J. P.} and {Lesgourgues, J.} and {Levrier, F.} and {Lewis, A.} and {Liguori, M.} and {Lilje, P. B.} and {Lilley, M.} and {Lindholm, V.} and {L\'opez-Caniego, M.} and {Lubin, P. M.} and {Ma, Y.-Z.} and {Mac\'{\i}as-P\'erez, J. F.} and {Maggio, G.} and {Maino, D.} and {Mandolesi, N.} and {Mangilli, A.} and {Marcos-Caballero, A.} and {Maris, M.} and {Martin, P. G.} and {Martinelli, M.} and {Mart\'{\i}nez-Gonz\'alez, E.} and {Matarrese, S.} and {Mauri, N.} and {McEwen, J. D.} and {Meerburg, P. D.} and {Meinhold, P. R.} and {Melchiorri, A.} and {Mennella, A.} and {Migliaccio, M.} and {Millea, M.} and {Mitra, S.} and {Miville-Desch\^enes, M.-A.} and {Molinari, D.} and {Moneti, A.} and {Montier, L.} and {Morgante, G.} and {Moss, A.} and {Mottet, S.} and {M\"unchmeyer, M.} and {Natoli, P.} and {N\o{}rgaard-Nielsen, H. U.} and {Oxborrow, C. A.} and {Pagano, L.} and {Paoletti, D.} and {Partridge, B.} and {Patanchon, G.} and {Pearson, T. J.} and {Peel, M.} and {Peiris, H. V.} and {Perrotta, F.} and {Pettorino, V.} and {Piacentini, F.} and {Polastri, L.} and {Polenta, G.} and {Puget, J.-L.} and {Rachen, J. P.} and {Reinecke, M.} and {Remazeilles, M.} and {Renault, C.} and {Renzi, A.} and {Rocha, G.} and {Rosset, C.} and {Roudier, G.} and {Rubi\~no-Mart\'{\i}n, J. A.} and {Ruiz-Granados, B.} and {Salvati, L.} and {Sandri, M.} and {Savelainen, M.} and {Scott, D.} and {Shellard, E. P. S.} and {Shiraishi, M.} and {Sirignano, C.} and {Sirri, G.} and {Spencer, L. D.} and {Sunyaev, R.} and {Suur-Uski, A.-S.} and {Tauber, J. A.} and {Tavagnacco, D.} and {Tenti, M.} and {Terenzi, L.} and {Toffolatti, L.} and {Tomasi, M.} and {Trombetti, T.} and {Valiviita, J.} and {Van Tent, B.} and {Vibert, L.} and {Vielva, P.} and {Villa, F.} and {Vittorio, N.} and {Wandelt, B. D.} and {Wehus, I. K.} and {White, M.} and {White, S. D. M.} and {Zacchei, A.} and {Zonca, A.}},
	title = {Planck 2018 results - I. Overview and the cosmological legacy of Planck},
	DOI= "10.1051/0004-6361/201833880",
	url= "https://doi.org/10.1051/0004-6361/201833880",
	journal = {A\&A},
	year = 2020,
	volume = 641,
	pages = "A1",
}

@article{Sugai_2020,
   title={Updated Design of the CMB Polarization Experiment Satellite LiteBIRD},
   volume={199},
   ISSN={1573-7357},
   url={http://dx.doi.org/10.1007/s10909-019-02329-w},
   DOI={10.1007/s10909-019-02329-w},
   number={3-4},
   journal={Journal of Low Temperature Physics},
   publisher={Springer Science and Business Media LLC},
   author={Sugai, H. and Ade, P. A. R. and Akiba, Y. and Alonso, D. and Arnold, K. and Aumont, J. and Austermann, J. and Baccigalupi, C. and Banday, A. J. and Banerji, R. and et al.},
   year={2020},
   month=Jan,
   pages={1107–1117}
}

@article{Alonso_2019,
   title={A unified pseudo-{$C_\ell$} framework},
   volume={484},
   ISSN={1365-2966},
   url={http://dx.doi.org/10.1093/mnras/stz093},
   DOI={10.1093/mnras/stz093},
   number={3},
   journal={Monthly Notices of the Royal Astronomical Society},
   publisher={Oxford University Press (OUP)},
   author={Alonso, D. and Sanchez, J. and Slosar, A.},
   year={2019},
   month=Jan,
   pages={4127–4151}
}

@article{Errard_2015,
   title={Modeling atmospheric emission for CMB ground-based observations},
   volume={809},
   ISSN={1538-4357},
   url={http://dx.doi.org/10.1088/0004-637X/809/1/63},
   DOI={10.1088/0004-637x/809/1/63},
   number={1},
   journal={The Astrophysical Journal},
   publisher={American Astronomical Society},
   author={Errard, J. and Ade, P. A. R. and Akiba, Y. and Arnold, K. and Atlas, M. and Baccigalupi, C. and Barron, D. and Boettger, D. and Borrill, J. and Chapman, S. and et al.},
   year={2015},
   month=Aug,
   pages={63}
}

@article{Zonca2019,
  doi = {10.21105/joss.01298},
  url = {https://doi.org/10.21105/joss.01298},
  year = {2019},
  month = mar,
  publisher = {The Open Journal},
  volume = {4},
  number = {35},
  pages = {1298},
  author = {A. Zonca and Leo Singer and Daniel Lenz and Martin Reinecke and Cyrille Rosset and Eric Hivon and Krzysztof Gorski},
  title = {\texttt{healpy}: equal area pixelization and spherical harmonics transforms for data on the sphere in Python},
  journal = {Journal of Open Source Software}
}

@ARTICLE{bunn_etal_1996,
       author = {{Bunn}, E. F. and {Hoffman}, Y. and {Silk}, J.},
        title = "{The Wiener-filtered COBE DMR Data and Predictions for the Tenerife Experiment}",
      journal = {The Astrophysical Journal},
         year = 1996,
        month = jun,
       volume = {464},
        pages = {1},
          doi = {10.1086/177294},
archivePrefix = {arXiv},
       eprint = {astro-ph/9509045},
 primaryClass = {astro-ph},
       adsurl = {https://ui.adsabs.harvard.edu/abs/1996ApJ...464....1B}
}

@INPROCEEDINGS{PB_SPIE,
       author = {{Kermish}, Zigmund D. and {Ade}, Peter and {Anthony}, Aubra and {Arnold}, Kam and {Barron}, Darcy and {Boettger}, David and {Borrill}, Julian and {Chapman}, Scott and {Chinone}, Yuji and {Dobbs}, Matt A. and {Errard}, Josquin and {Fabbian}, Giulio and {Flanigan}, Daniel and {Fuller}, George and {Ghribi}, Adnan and {Grainger}, Will and {Halverson}, Nils and {Hasegawa}, Masaya and {Hattori}, Kaori and {Hazumi}, Masashi and {Holzapfel}, William L. and {Howard}, Jacob and {Hyland}, Peter and {Jaffe}, Andrew and {Keating}, Brian and {Kisner}, Theodore and {Lee}, Adrian T. and {Le Jeune}, Maude and {Linder}, Eric and {Lungu}, Marius and {Matsuda}, Frederick and {Matsumura}, Tomotake and {Meng}, Xiaofan and {Miller}, Nathan J. and {Morii}, Hideki and {Moyerman}, Stephanie and {Myers}, Mike J. and {Nishino}, Haruki and {Paar}, Hans and {Quealy}, Erin and {Reichardt}, Christian L. and {Richards}, Paul L. and {Ross}, Colin and {Shimizu}, Akie and {Shimon}, Meir and {Shimmin}, Chase and {Sholl}, Mike and {Siritanasak}, Praween and {Spieler}, Helmuth and {Stebor}, Nathan and {Steinbach}, Bryan and {Stompor}, Radek and {Suzuki}, Aritoki and {Tomaru}, Takayuki and {Tucker}, Carole and {Zahn}, Oliver},
        title = "{The POLARBEAR experiment}",
    booktitle = {Millimeter, Submillimeter, and Far-Infrared Detectors and Instrumentation for Astronomy VI},
         year = 2012,
       editor = {{Holland}, Wayne S. and {Zmuidzinas}, Jonas},
       series = {Society of Photo-Optical Instrumentation Engineers (SPIE) Conference Series},
       volume = {8452},
        month = sep,
          eid = {84521C},
        pages = {84521C},
          doi = {10.1117/12.926354},
archivePrefix = {arXiv},
       eprint = {1210.7768},
 primaryClass = {astro-ph.IM},
       adsurl = {https://ui.adsabs.harvard.edu/abs/2012SPIE.8452E..1CK}
}

@article{ACT,
   title={The Atacama Cosmology Telescope: The polarization-sensitive ACTPol instrument},
   volume={227},
   ISSN={1538-4365},
   url={http://dx.doi.org/10.3847/1538-4365/227/2/21},
   DOI={10.3847/1538-4365/227/2/21},
   number={2},
   journal={The Astrophysical Journal Supplement Series},
   publisher={American Astronomical Society},
   author={Thornton, R. J. and Ade, P. A. R. and Aiola, S. and Angilè, F. E. and Amiri, M. and Beall, J. A. and Becker, D. T. and Cho, H-M. and Choi, S. K. and Corlies, P. and et al.},
   year={2016},
   month=Dec,
   pages={21}
}

@Article{numpy,
 title         = {Array programming with {NumPy}},
 author        = {Charles R. Harris and K. Jarrod Millman and St{\'{e}}fan J.
                 van der Walt and Ralf Gommers and Pauli Virtanen and David
                 Cournapeau and Eric Wieser and Julian Taylor and Sebastian
                 Berg and Nathaniel J. Smith and Robert Kern and Matti Picus
                 and Stephan Hoyer and Marten H. van Kerkwijk and Matthew
                 Brett and Allan Haldane and Jaime Fern{\'{a}}ndez del
                 R{\'{i}}o and Mark Wiebe and Pearu Peterson and Pierre
                 G{\'{e}}rard-Marchant and Kevin Sheppard and Tyler Reddy and
                 Warren Weckesser and Hameer Abbasi and Christoph Gohlke and
                 Travis E. Oliphant},
 year          = {2020},
 month         = sep,
 journal       = {Nature},
 volume        = {585},
 number        = {7825},
 pages         = {357--362},
 doi           = {10.1038/s41586-020-2649-2},
 publisher     = {Springer Science and Business Media {LLC}},
 url           = {https://doi.org/10.1038/s41586-020-2649-2}
}

@Article{Matplotlib,
  Author    = {Hunter, J. D.},
  Title     = {Matplotlib: A 2D graphics environment},
  Journal   = {Computing in Science \& Engineering},
  Volume    = {9},
  Number    = {3},
  Pages     = {90--95},
  publisher = {IEEE COMPUTER SOC},
  doi       = {10.1109/MCSE.2007.55},
  year      = 2007
}

@inproceedings{SO_instru,
author = {Nicholas Galitzki and Aamir Ali and Kam S. Arnold and Peter C. Ashton and Jason E. Austermann and Carlo Baccigalupi and Taylor Baildon and Darcy Barron and James A. Beall and Shawn Beckman and Sarah Marie M. Bruno and Sean Bryan and Paolo G. Calisse and Grace E. Chesmore and Yuji Chinone and Steve K. Choi and Gabriele Coppi and Kevin D. Crowley and Kevin T. Crowley and Ari Cukierman and Mark J. Devlin and Simon Dicker and Bradley Dober and Shannon M. Duff and Jo Dunkley and Giulio Fabbian and Patricio A. Gallardo and Martina Gerbino and Neil Goeckner-Wald and Joseph E. Golec and Jon E. Gudmundsson and Erin E. Healy and Shawn Henderson and Charles A. Hill and Gene C. Hilton and Shuay-Pwu Patty Ho and Logan A. Howe and Johannes Hubmayr and Oliver Jeong and Brian Keating and Brian J. Koopman and Kenji Kiuchi and Akito Kusaka and Jacob Lashner and Adrian T. Lee and Yaqiong Li and Michele Limon and Marius Lungu and Frederick Matsuda and Philip D. Mauskopf and Andrew J. May and Nialh McCallum and Jeff McMahon and Federico Nati and Michael D. Niemack and John L. Orlowski-Scherer and Stephen C. Parshley and Lucio Piccirillo and Mayuri Sathyanarayana Rao and Christopher Raum and Maria Salatino and Joseph S. Seibert and Carlos Sierra and Max Silva-Feaver and Sara M. Simon and Suzanne T. Staggs and Jason R. Stevens and Aritoki Suzuki and Grant Teply and Robert Thornton and Calvin Tsai and Joel N. Ullom and Eve M. Vavagiakis and Michael R. Vissers and Benjamin Westbrook and Edward J. Wollack and Zhilei Xu and Ningfeng Zhu},
title = {{The Simons Observatory: instrument overview}},
volume = {10708},
booktitle = {Millimeter, Submillimeter, and Far-Infrared Detectors and Instrumentation for Astronomy IX},
editor = {Jonas Zmuidzinas and Jian-Rong Gao},
organization = {International Society for Optics and Photonics},
publisher = {SPIE},
pages = {1 -- 13},
year = {2018},
doi = {10.1117/12.2312985},
URL = {https://doi.org/10.1117/12.2312985}
}

@article{SO_goals,
	doi = {10.1088/1475-7516/2019/02/056},
	url = {https://doi.org/10.1088/1475-7516/2019/02/056},
	year = 2019,
	month = {feb},
	publisher = {{IOP} Publishing},
	volume = {2019},
	number = {02},
	pages = {056--056},
	author = {Peter Ade and James Aguirre and Zeeshan Ahmed and Simone Aiola and Aamir Ali and David Alonso and Marcelo A. Alvarez and Kam Arnold and Peter Ashton and Jason Austermann and Humna Awan and Carlo Baccigalupi and Taylor Baildon and Darcy Barron and Nick Battaglia and Richard Battye and Eric Baxter and Andrew Bazarko and James A. Beall and Rachel Bean and Dominic Beck and Shawn Beckman and Benjamin Beringue and Federico Bianchini and Steven Boada and David Boettger and J. Richard Bond and Julian Borrill and Michael L. Brown and Sarah Marie Bruno and Sean Bryan and Erminia Calabrese and Victoria Calafut and Paolo Calisse and Julien Carron and Anthony Challinor and Grace Chesmore and Yuji Chinone and Jens Chluba and Hsiao-Mei Sherry Cho and Steve Choi and Gabriele Coppi and Nicholas F. Cothard and Kevin Coughlin and Devin Crichton and Kevin D. Crowley and Kevin T. Crowley and Ari Cukierman and John M. D{\textquotesingle}Ewart and Rolando Dünner and Tijmen de Haan and Mark Devlin and Simon Dicker and Joy Didier and Matt Dobbs and Bradley Dober and Cody J. Duell and Shannon Duff and Adri Duivenvoorden and Jo Dunkley and John Dusatko and Josquin Errard and Giulio Fabbian and Stephen Feeney and Simone Ferraro and Pedro Flux{\`{a}} and Katherine Freese and Josef C. Frisch and Andrei Frolov and George Fuller and Brittany Fuzia and Nicholas Galitzki and Patricio A. Gallardo and Jose Tomas Galvez Ghersi and Jiansong Gao and Eric Gawiser and Martina Gerbino and Vera Gluscevic and Neil Goeckner-Wald and Joseph Golec and Sam Gordon and Megan Gralla and Daniel Green and Arpi Grigorian and John Groh and Chris Groppi and Yilun Guan and Jon E. Gudmundsson and Dongwon Han and Peter Hargrave and Masaya Hasegawa and Matthew Hasselfield and Makoto Hattori and Victor Haynes and Masashi Hazumi and Yizhou He and Erin Healy and Shawn W. Henderson and Carlos Hervias-Caimapo and Charles A. Hill and J. Colin Hill and Gene Hilton and Matt Hilton and Adam D. Hincks and Gary Hinshaw and Ren{\'{e}}e Hlo{\v{z}}ek and Shirley Ho and Shuay-Pwu Patty Ho and Logan Howe and Zhiqi Huang and Johannes Hubmayr and Kevin Huffenberger and John P. Hughes and Anna Ijjas and Margaret Ikape and Kent Irwin and Andrew H. Jaffe and Bhuvnesh Jain and Oliver Jeong and Daisuke Kaneko and Ethan D. Karpel and Nobuhiko Katayama and Brian Keating and Sarah S. Kernasovskiy and Reijo Keskitalo and Theodore Kisner and Kenji Kiuchi and Jeff Klein and Kenda Knowles and Brian Koopman and Arthur Kosowsky and Nicoletta Krachmalnicoff and Stephen E. Kuenstner and Chao-Lin Kuo and Akito Kusaka and Jacob Lashner and Adrian Lee and Eunseong Lee and David Leon and Jason S.-Y. Leung and Antony Lewis and Yaqiong Li and Zack Li and Michele Limon and Eric Linder and Carlos Lopez-Caraballo and Thibaut Louis and Lindsay Lowry and Marius Lungu and Mathew Madhavacheril and Daisy Mak and Felipe Maldonado and Hamdi Mani and Ben Mates and Frederick Matsuda and Loïc Maurin and Phil Mauskopf and Andrew May and Nialh McCallum and Chris McKenney and Jeff McMahon and P. Daniel Meerburg and Joel Meyers and Amber Miller and Mark Mirmelstein and Kavilan Moodley and Moritz Munchmeyer and Charles Munson and Sigurd Naess and Federico Nati and Martin Navaroli and Laura Newburgh and Ho Nam Nguyen and Michael Niemack and Haruki Nishino and John Orlowski-Scherer and Lyman Page and Bruce Partridge and Julien Peloton and Francesca Perrotta and Lucio Piccirillo and Giampaolo Pisano and Davide Poletti and Roberto Puddu and Giuseppe Puglisi and Chris Raum and Christian L. Reichardt and Mathieu Remazeilles and Yoel Rephaeli and Dominik Riechers and Felipe Rojas and Anirban Roy and Sharon Sadeh and Yuki Sakurai and Maria Salatino and Mayuri Sathyanarayana Rao and Emmanuel Schaan and Marcel Schmittfull and Neelima Sehgal and Joseph Seibert and Uros Seljak and Blake Sherwin and Meir Shimon and Carlos Sierra and Jonathan Sievers and Precious Sikhosana and Maximiliano Silva-Feaver and Sara M. Simon and Adrian Sinclair and Praween Siritanasak and Kendrick Smith and Stephen R. Smith and David Spergel and Suzanne T. Staggs and George Stein and Jason R. Stevens and Radek Stompor and Aritoki Suzuki and Osamu Tajima and Satoru Takakura and Grant Teply and Daniel B. Thomas and Ben Thorne and Robert Thornton and Hy Trac and Calvin Tsai and Carole Tucker and Joel Ullom and Sunny Vagnozzi and Alexander van Engelen and Jeff Van Lanen and Daniel D. Van Winkle and Eve M. Vavagiakis and Clara Verg{\`{e}}s and Michael Vissers and Kasey Wagoner and Samantha Walker and Jon Ward and Ben Westbrook and Nathan Whitehorn and Jason Williams and Joel Williams and Edward J. Wollack and Zhilei Xu and Byeonghee Yu and Cyndia Yu and Fernando Zago and Hezi Zhang and Ningfeng Zhu and},
	title = {The Simons Observatory: science goals and forecasts},
	journal = {Journal of Cosmology and Astroparticle Physics}
}

@article{Planck18_HFI_dataproc,
	author = {{Planck Collaboration} and {Aghanim, N.} and {Akrami, Y.} and {Ashdown, M.} and {Aumont, J.} and {Baccigalupi, C.} and {Ballardini, M.} and {Banday, A. J.} and {Barreiro, R. B.} and {Bartolo, N.} and {Basak, S.} and {Benabed, K.} and {Bernard, J.-P.} and {Bersanelli, M.} and {Bielewicz, P.} and {Bond, J. R.} and {Borrill, J.} and {Bouchet, F. R.} and {Boulanger, F.} and {Bucher, M.} and {Burigana, C.} and {Calabrese, E.} and {Cardoso, J.-F.} and {Carron, J.} and {Challinor, A.} and {Chiang, H. C.} and {Colombo, L. P. L.} and {Combet, C.} and {Couchot, F.} and {Crill, B. P.} and {Cuttaia, F.} and {de Bernardis, P.} and {de Rosa, A.} and {de Zotti, G.} and {Delabrouille, J.} and {Delouis, J.-M.} and {Di Valentino, E.} and {Diego, J. M.} and {Dor\'e, O.} and {Douspis, M.} and {Ducout, A.} and {Dupac, X.} and {Efstathiou, G.} and {Elsner, F.} and {En\ss{}lin, T. A.} and {Eriksen, H. K.} and {Falgarone, E.} and {Fantaye, Y.} and {Finelli, F.} and {Frailis, M.} and {Fraisse, A. A.} and {Franceschi, E.} and {Frolov, A.} and {Galeotta, S.} and {Galli, S.} and {Ganga, K.} and {G\'enova-Santos, R. T.} and {Gerbino, M.} and {Ghosh, T.} and {Gonz\'alez-Nuevo, J.} and {G\'orski, K. M.} and {Gratton, S.} and {Gruppuso, A.} and {Gudmundsson, J. E.} and {Handley, W.} and {Hansen, F. K.} and {Henrot-Versill\'e, S.} and {Herranz, D.} and {Hivon, E.} and {Huang, Z.} and {Jaffe, A. H.} and {Jones, W. C.} and {Karakci, A.} and {Keih\"anen, E.} and {Keskitalo, R.} and {Kiiveri, K.} and {Kim, J.} and {Kisner, T. S.} and {Krachmalnicoff, N.} and {Kunz, M.} and {Kurki-Suonio, H.} and {Lagache, G.} and {Lamarre, J.-M.} and {Lasenby, A.} and {Lattanzi, M.} and {Lawrence, C. R.} and {Levrier, F.} and {Liguori, M.} and {Lilje, P. B.} and {Lindholm, V.} and {L\'opez-Caniego, M.} and {Ma, Y.-Z.} and {Mac\'{\i}as-P\'erez, J. F.} and {Maggio, G.} and {Maino, D.} and {Mandolesi, N.} and {Mangilli, A.} and {Martin, P. G.} and {Mart\'{\i}nez-Gonz\'alez, E.} and {Matarrese, S.} and {Mauri, N.} and {McEwen, J. D.} and {Melchiorri, A.} and {Mennella, A.} and {Migliaccio, M.} and {Miville-Desch\^enes, M.-A.} and {Molinari, D.} and {Moneti, A.} and {Montier, L.} and {Morgante, G.} and {Moss, A.} and {Mottet, S.} and {Natoli, P.} and {Pagano, L.} and {Paoletti, D.} and {Partridge, B.} and {Patanchon, G.} and {Patrizii, L.} and {Perdereau, O.} and {Perrotta, F.} and {Pettorino, V.} and {Piacentini, F.} and {Puget, J.-L.} and {Rachen, J. P.} and {Reinecke, M.} and {Remazeilles, M.} and {Renzi, A.} and {Rocha, G.} and {Roudier, G.} and {Salvati, L.} and {Sandri, M.} and {Savelainen, M.} and {Scott, D.} and {Sirignano, C.} and {Sirri, G.} and {Spencer, L. D.} and {Sunyaev, R.} and {Suur-Uski, A.-S.} and {Tauber, J. A.} and {Tavagnacco, D.} and {Tenti, M.} and {Toffolatti, L.} and {Tomasi, M.} and {Tristram, M.} and {Trombetti, T.} and {Valiviita, J.} and {Vansyngel, F.} and {Van Tent, B.} and {Vibert, L.} and {Vielva, P.} and {Villa, F.} and {Vittorio, N.} and {Wandelt, B. D.} and {Wehus, I. K.} and {Zonca, A.}},
	title = {Planck 2018 results - III. High Frequency Instrument data processing and frequency maps},
	DOI= "10.1051/0004-6361/201832909",
	url= "https://doi.org/10.1051/0004-6361/201832909",
	journal = {A\&A},
	year = 2020,
	volume = 641,
	pages = "A3",
}

@article{Planck18_LFI_dataproc,
	author = {{Planck Collaboration} and {Akrami, Y.} and {Arg\"ueso, F.} and {Ashdown, M.} and {Aumont, J.} and {Baccigalupi, C.} and {Ballardini, M.} and {Banday, A. J.} and {Barreiro, R. B.} and {Bartolo, N.} and {Basak, S.} and {Benabed, K.} and {Bernard, J.-P.} and {Bersanelli, M.} and {Bielewicz, P.} and {Bonavera, L.} and {Bond, J. R.} and {Borrill, J.} and {Bouchet, F. R.} and {Boulanger, F.} and {Bucher, M.} and {Burigana, C.} and {Butler, R. C.} and {Calabrese, E.} and {Cardoso, J.-F.} and {Colombo, L. P. L.} and {Crill, B. P.} and {Cuttaia, F.} and {de Bernardis, P.} and {de Rosa, A.} and {de Zotti, G.} and {Delabrouille, J.} and {Di Valentino, E.} and {Dickinson, C.} and {Diego, J. M.} and {Donzelli, S.} and {Ducout, A.} and {Dupac, X.} and {Efstathiou, G.} and {Elsner, F.} and {En\ss{}lin, T. A.} and {Eriksen, H. K.} and {Fantaye, Y.} and {Finelli, F.} and {Frailis, M.} and {Franceschi, E.} and {Frolov, A.} and {Galeotta, S.} and {Galli, S.} and {Ganga, K.} and {G\'enova-Santos, R. T.} and {Gerbino, M.} and {Ghosh, T.} and {Gonz\'alez-Nuevo, J.} and {G\'orski, K. M.} and {Gratton, S.} and {Gruppuso, A.} and {Gudmundsson, J. E.} and {Handley, W.} and {Hansen, F. K.} and {Herranz, D.} and {Hivon, E.} and {Huang, Z.} and {Jaffe, A. H.} and {Jones, W. C.} and {Karakci, A.} and {Keih\"anen, E.} and {Keskitalo, R.} and {Kiiveri, K.} and {Kim, J.} and {Kisner, T. S.} and {Krachmalnicoff, N.} and {Kunz, M.} and {Kurki-Suonio, H.} and {Lamarre, J.-M.} and {Lasenby, A.} and {Lattanzi, M.} and {Lawrence, C. R.} and {Leahy, J. P.} and {Levrier, F.} and {Liguori, M.} and {Lilje, P. B.} and {Lindholm, V.} and {L\'opez-Caniego, M.} and {Ma, Y.-Z.} and {Mac\'{\i}as-P\'erez, J. F.} and {Maggio, G.} and {Maino, D.} and {Mandolesi, N.} and {Mangilli, A.} and {Maris, M.} and {Martin, P. G.} and {Mart\'{\i}nez-Gonz\'alez, E.} and {Matarrese, S.} and {Mauri, N.} and {McEwen, J. D.} and {Meinhold, P. R.} and {Melchiorri, A.} and {Mennella, A.} and {Migliaccio, M.} and {Molinari, D.} and {Montier, L.} and {Morgante, G.} and {Moss, A.} and {Natoli, P.} and {Pagano, L.} and {Paoletti, D.} and {Partridge, B.} and {Patanchon, G.} and {Patrizii, L.} and {Peel, M.} and {Perrotta, F.} and {Pettorino, V.} and {Piacentini, F.} and {Polenta, G.} and {Puget, J.-L.} and {Rachen, J. P.} and {Racine, B.} and {Reinecke, M.} and {Remazeilles, M.} and {Renzi, A.} and {Rocha, G.} and {Roudier, G.} and {Rubi\~no-Mart\'{\i}n, J. A.} and {Salvati, L.} and {Sandri, M.} and {Savelainen, M.} and {Scott, D.} and {Seljebotn, D. S.} and {Sirignano, C.} and {Sirri, G.} and {Spencer, L. D.} and {Suur-Uski, A.-S.} and {Tauber, J. A.} and {Tavagnacco, D.} and {Tenti, M.} and {Terenzi, L.} and {Toffolatti, L.} and {Tomasi, M.} and {Trombetti, T.} and {Valiviita, J.} and {Vansyngel, F.} and {Van Tent, B.} and {Vielva, P.} and {Villa, F.} and {Vittorio, N.} and {Wandelt, B. D.} and {Watson, R.} and {Wehus, I. K.} and {Zacchei, A.} and {Zonca, A.}},
	title = {Planck 2018 results - II. Low Frequency Instrument data processing},
	DOI= "10.1051/0004-6361/201833293",
	url= "https://doi.org/10.1051/0004-6361/201833293",
	journal = {A\&A},
	year = 2020,
	volume = 641,
	pages = "A2",
}

@article{NPIPE_dataproc,
	author = {{Planck Collaboration} and {Akrami, Y.} and {Andersen, K. J.} and {Ashdown, M.} and {Baccigalupi, C.} and {Ballardini, M.} and {Banday, A. J.} and {Barreiro, R. B.} and {Bartolo, N.} and {Basak, S.} and {Benabed, K.} and {Bernard, J.-P.} and {Bersanelli, M.} and {Bielewicz, P.} and {Bond, J. R.} and {Borrill, J.} and {Burigana, C.} and {Butler, R. C.} and {Calabrese, E.} and {Casaponsa, B.} and {Chiang, H. C.} and {Colombo, L. P. L.} and {Combet, C.} and {Crill, B. P.} and {Cuttaia, F.} and {de Bernardis, P.} and {de Rosa, A.} and {de Zotti, G.} and {Delabrouille, J.} and {Di Valentino, E.} and {Diego, J. M.} and {Dor\'e, O.} and {Douspis, M.} and {Dupac, X.} and {Eriksen, H. K.} and {Fernandez-Cobos, R.} and {Finelli, F.} and {Frailis, M.} and {Fraisse, A. A.} and {Franceschi, E.} and {Frolov, A.} and {Galeotta, S.} and {Galli, S.} and {Ganga, K.} and {Gerbino, M.} and {Ghosh, T.} and {Gonz\'alez-Nuevo, J.} and {G\'orski, K. M.} and {Gruppuso, A.} and {Gudmundsson, J. E.} and {Handley, W.} and {Helou, G.} and {Herranz, D.} and {Hildebrandt, S. R.} and {Hivon, E.} and {Huang, Z.} and {Jaffe, A. H.} and {Jones, W. C.} and {Keih\"anen, E.} and {Keskitalo, R.} and {Kiiveri, K.} and {Kim, J.} and {Kisner, T. S.} and {Krachmalnicoff, N.} and {Kunz, M.} and {Kurki-Suonio, H.} and {Lasenby, A.} and {Lattanzi, M.} and {Lawrence, C. R.} and {Le Jeune, M.} and {Levrier, F.} and {Liguori, M.} and {Lilje, P. B.} and {Lilley, M.} and {Lindholm, V.} and {L\'opez-Caniego, M.} and {Lubin, P. M.} and {Mac\'{\i}as-P\'erez, J. F.} and {Maino, D.} and {Mandolesi, N.} and {Marcos-Caballero, A.} and {Maris, M.} and {Martin, P. G.} and {Mart\'{\i}nez-Gonz\'alez, E.} and {Matarrese, S.} and {Mauri, N.} and {McEwen, J. D.} and {Meinhold, P. R.} and {Mennella, A.} and {Migliaccio, M.} and {Mitra, S.} and {Molinari, D.} and {Montier, L.} and {Morgante, G.} and {Moss, A.} and {Natoli, P.} and {Paoletti, D.} and {Partridge, B.} and {Patanchon, G.} and {Pearson, D.} and {Pearson, T. J.} and {Perrotta, F.} and {Piacentini, F.} and {Polenta, G.} and {Rachen, J. P.} and {Reinecke, M.} and {Remazeilles, M.} and {Renzi, A.} and {Rocha, G.} and {Rosset, C.} and {Roudier, G.} and {Rubi\~no-Mart\'{\i}n, J. A.} and {Ruiz-Granados, B.} and {Salvati, L.} and {Savelainen, M.} and {Scott, D.} and {Sirignano, C.} and {Sirri, G.} and {Spencer, L. D.} and {Suur-Uski, A.-S.} and {Svalheim, L. T.} and {Tauber, J. A.} and {Tavagnacco, D.} and {Tenti, M.} and {Terenzi, L.} and {Thommesen, H.} and {Toffolatti, L.} and {Tomasi, M.} and {Tristram, M.} and {Trombetti, T.} and {Valiviita, J.} and {Van Tent, B.} and {Vielva, P.} and {Villa, F.} and {Vittorio, N.} and {Wandelt, B. D.} and {Wehus, I. K.} and {Zacchei, A.} and {Zonca, A.}},
	title = {Planck intermediate results - LVII. Joint Planck LFI and HFI data processing},
	DOI= "10.1051/0004-6361/202038073",
	url= "https://doi.org/10.1051/0004-6361/202038073",
	journal = {A\&A},
	year = 2020,
	volume = 643,
	pages = "A42",
}

@ARTICLE{LL_2014,
       author = {{Lowery}, Bradley R. and {Langou}, Julien},
        title = "{Stability Analysis of QR factorization in an Oblique Inner Product}",
      journal = {arXiv e-prints},
         year = 2014,
        month = jan,
          eid = {arXiv:1401.5171},
        pages = {arXiv:1401.5171},
archivePrefix = {arXiv},
       eprint = {1401.5171},
 primaryClass = {math.NA},
       adsurl = {https://ui.adsabs.harvard.edu/abs/2014arXiv1401.5171L}
}

\appendix

\section{Pre-defined template classes.}
\label{sect:appA}

Here we describe some of the template classes which are directly implemented in the proposed framework. These are constructed following previous CMB data analysis experience, and selected given potential needs of forthcoming and future experiments. We also highlight the orthogonalization kernel structure these templates imply.

\subsection*{$\bullet$ Atmosphere and $1/f$ instrumental noise}
In a typical experiment the noise is correlated due to low frequency excess noise of instrumental origin, and the atmospheric emission in the case of ground experiments. We model this low frequency noise by a set of Legendre polynomials, and hence build a filtering operator which deprojects polynomial trends from the data. This can be seen as the most straightforward generalization of the standard destriper. 

We define a set of time intervals, $\bigcup\,\left[\,t_i,\,t_{i+1}\,\right]$, over which the amplitudes of the polynomial trends for each detector are considered to be constant. In practice, these time intervals can span the duration of each sweep in the constant elevation scans (CESs) of a ground telescope for example. The number of polynomial orders, $n_{poly}$, is fixed. We do not consider here common detector modes, which would model correlations between detectors. We can define each template as a compact support function, which represents one polynomial order for a given detector during one time interval $\left[\,t_i,\,t_{i+1}\,\right]$ (e.g., one sweep). In summary, for a detector $j$, there exists a set, $(\alpha_{j,k})_{k \in \llbracket 0,  n_{poly}-1\rrbracket}$ of real amplitudes, such as the $1/f$ noise component in the $i-$th time interval would be given by:
\begin{eqnarray}
n_j(t) = \sum_{k=0}^{n_{poly}-1}\,\alpha_{j,k}\,P_k(t),
\label{eqn:polytemplate}
\end{eqnarray}
where $P_k$ is the Legendre polynomial of degree $k$ rescaled so it is orthonormal to others on the relevant time interval. Each of these templates represents one column of the global template matrix $\mathbf{T}$. By virtue of this construction and the orthogonality of Legendre polynomials, the columns would be orthogonal, making the polynomial templates kernel $\left(\,\mathbf{T^\mathrm{T}}\,\mathbf{W^{-1}}\,\mathbf{T}\,\right)^{-1}$ diagonal, simplifying the computational cost of the filtering operator.
This approach to modeling the atmospheric signal explicitly ignores correlations between contributions to data registered by different detectors. Such common modes can be straightforwardly modeled in the proposed framework, though may come at additional computational cost. Whether such an extension is justified and beneficial will depend on other factors such as relative calibration errors between the detectors and should be decided on case-by-case.

\subsection*{$\bullet$ Scan synchronous signal}
These templates model any spurious signals which are synchronous with the scanning strategy. For ground experiments, these would be dominated by ground pickup sourced by the far side lobes of the beam which sense the surrounding terrain~\cite{BBpaper2020}. We model this signal with a template parametrized by the azimuth angle of the boresight pointing.

We define an azimuth grid, $\bigcup_{i=1}^{n_{az}}\,\left[\,\varphi_i,\,\varphi_{i+1}\,\right]$ for each CES, fixing \textit{a priori} a number of azimuth bins $n_{az}$. Then for each detector $j$ we consider a ground structure, $(g_{j,i})_{i \in \llbracket 1,  n_{az}\rrbracket}$,  which is stable over the duration of the full CES. The template marginalization performance is in such a case contingent upon having enough redundancy in the scanning strategy to separate between the ground and sky signal over the duration of the CES. The scan synchronous signal is given by the following,
\begin{eqnarray}
\forall t: \varphi(t) \in \left[\,\varphi_i,\,\varphi_{i+1}\,\right] \Rightarrow SSS_j(t) = g_{j,i}.
\label{eqn:groundtemplate}
\end{eqnarray}
In the template matrix formulation as in Eq.~\eqref{eqn:dataModel0} the amplitude vector stores the bin amplitudes, $g_{j,i}$ and there is one column of the corresponding scan-synchronous template matrix for each azimuthal bin. The columns have consequently non-zero values only in rows corresponding to times at which the telescope azimuth falls into the corresponding bin. As only one bin contributes to the overall measurment at any given time the template matrix has only one non-zero element per row and is therefore very sparse and the columns are by construction orthogonal. The non-zero values of the matrix can be typically set to unity, however, they may be replaced by some uniquely defined function of time, such as one describing gain drifts. 
For the diagonal noise weights, the corresponding diagonal block of the orthogonalization kernel, $\left(\,\mathbf{T^\mathrm{T}}\,\mathbf{W^{-1}}\,\mathbf{T}\,\right)^{-1}$, will be diagonal. However, the scan-synchronous templates may not be and, in general, will not be orthogonal to the other templates, such as the polynomial templates discussed earlier. As discussed in~\cite{Poletti_2017} in general the scan-synchronous templates will not be also orthogonal to the sky signal, leading to near-degeneracies in the system matrix broken only by the assumed sky pixelization.

We note that more involved models of the scan synchronous signal are readily possible. For instance, for a given grid of azimuths, $\{\varphi_i\}$, rather than binning we can linearly interpolate between the signal estimated in the grid nodes. This could be helpful whenever the scan-synchronous signal is changing rapidly with the azimuth and when the binning may require a very large number of azimuth bins in order to minimize the signal variation across each bin. In this case, the interpolation may be more efficient. There will be two non-zero values per each row of the corresponding template matrix corresponding to two azimuth grid points, $\varphi_k, \varphi_{k+1}$ bracketing the telescope azimuth, $\varphi(t)$, and the two non-zero elements would be given by,
\begin{eqnarray}
\mathbf{T}_{t\, k} \; = \; \frac{\varphi(t)-\varphi_k}{\varphi_{k+1}-\varphi_k},\ \ \ 
\mathbf{T}_{t\,k+1} \; = \; \frac{\varphi_{k+1}-\varphi(t)}{\varphi_{k+1}-\varphi_k}.
\end{eqnarray}
The correspond kernel in such a case would be tridiagonal and therefore could be inverted efficiently.

\subsection*{$\bullet$ Half-wave plate synchronous signal}

Many of the modern CMB experiments feature a half-wave plate allowing to modulate the incoming polarized signal. While they play a key role in reducing the noise correlations, reducing beam effects, and minimizing  leakages, they also tend to introduce their own specific effects, which are generally synchronous with the position of the HWP itself~\cite{Maxipol_2008, Kusaka_2014}. They arise from differential transmission
or reflection of unpolarized light as it propagates through the optical chain, or from polarized emission of the HWP itself and can be modulated at different harmonics of its rotation speed. We refer to them as HWP synchronous signals. 

We could define templates for these signals in a model-independent way as in the previous subsection this time relying on binning in the HWP angle. As the HWP synchronous signals tend to vary rather abruptly with the HWP angle, this may however require a large number of bins even if some interpolation is invoked. Alternately we can model these spurious signals as a sum of harmonics of the HWP rotation frequency. 
This is the approach we have chosen in this work.

We define a set of time intervals, $\bigcup\,\left[\,t_i,\,t_{i+1}\,\right]$, with an \textit{a priori} fixed length $\Delta t$, except for the remainder interval at the end of each CES. The amplitudes of the HWP harmonics in each detector are considered to be constant over each $\Delta t$. The number of HWP harmonics, $n_{HWP}$ is fixed. In a similar manner to the polynomial templates, we can define each template as a compact support function, which represents either the cosine or the sine part of HWP harmonic for a given detector on time interval $\left[\,t_i,\,t_{i+1}\,\right]$. So, for a detector $j$, there exists two sets, $(\alpha_{j,k})_{k \in \llbracket 1,  n_{HWP}\rrbracket}$ and $(\beta_{j,k})_{k \in \llbracket 1,  n_{HWP}\rrbracket}$ of real amplitudes, such as the HWP synchronous signal in the $i$-th time interval would be given by:
\begin{eqnarray}
\mathcal{H}_j(t) = \sum_{k=1}^{n_{HWP}}\,\alpha_{j,k}\,\cos(k\phi(t))\,+\,\beta_{j,k}\,\sin(k\phi(t)),
\label{eqn:hwptemplate}
\end{eqnarray}
where $\phi(t)$ is the HWP angle as a function of time. The interval length $\Delta t$ can be chosen so that we get an integer number of rotations of the HWP, in the ideal case where the rotation frequency is not fluctuating. This would simplify the structure of the HWP segment of the templates kernel. This is because each of the templates as we defined them above represents a column of the HWP segment of the templates matrix, and the proposed choice of the time intervals would make the columns orthogonal or near-orthogonal (except the last one in each CES span) by virtue of the orthogonality of cosines and sines in an integer number of periods (HWP rotations). Therefore the corresponding part of the kernel $\left(\,\mathbf{T^\mathrm{T}}\,\mathbf{W^{-1}}\,\mathbf{T}\,\right)^{-1}$ would be nearly diagonal (last column and row of each CES excluded), again speeding up its inversion and improving the performance of the filtering operator.

We note that one could further generalize Eq.~\eqref{eqn:hwptemplate} by introducing time dependent coefficients $\alpha$ and $\beta$ accounting on potential gain drifts. This would typically result in a full block-diagonal structure of the kernel, with each block corresponding to a different time interval and different detectors, and would therefore require a full numerical inversion.

\end{document}